\documentclass[journal=jacsat,manuscript=article,email=false]{achemso}
\usepackage{chemformula} % Formula subscripts using \ch{}
\usepackage[T1]{fontenc} % Use modern font encodings
\usepackage{amsmath}
\usepackage{amsfonts}
\usepackage{mathtools}
\usepackage{bm}
\usepackage{float}
\usepackage{algorithm}
\usepackage[noend]{algpseudocode}
\usepackage{graphbox,graphicx}
\usepackage{caption}
\usepackage{subcaption}
\usepackage{import}
\usepackage{xifthen}
\usepackage{pdfpages}
\usepackage{transparent}
\usepackage{colortbl}
\usepackage[flushleft]{threeparttable}
\author{David C. Williams}
\email{dwilliams@nobiastx.com}
\author{Neil Inala}
\email{inala@nobiastx.com}
\affiliation[Nobias Therapeutics]
{Nobias Therapeutics}

\title{Physics-informed generative model for drug-like molecule conformers}

\begin{document}

%\begin{tocentry}
%    \noindent
%    \includegraphics*[width=3.3in]{figs/table_of_contents.png}\par
%\end{tocentry}

\begin{abstract}
We present a diffusion-based, generative model  
for conformer generation.
Our model is focused on the reproduction of bonded structure and is constructed from
the associated terms traditionally found in classical force fields
to ensure a physically relevant representation. Techniques in 
deep learning are used to infer atom typing and geometric parameters
from a training set. Conformer sampling is 
achieved by taking advantage of recent advancements
in diffusion-based generation. 
By training on large, 
synthetic data sets of diverse, drug-like molecules optimized with the
semiempirical GFN2-xTB method, high accuracy is achieved for bonded
parameters, exceeding that of conventional, knowledge-based methods. 
Results are also compared to experimental
structures from the Protein Databank (PDB) and Cambridge Structural Database (CSD).
\end{abstract}

\section{Introduction}
Conformer generation is the process of identifying a valid and useful set
of atomic coordinates for a given molecule. Since so many tools
in computational chemistry rely on atomic coordinates, it plays an 
important role in structure-based drug-discovery~\cite{schaduangrat_towards_2020}.
As such, several different methods for 
conformer generation have been developed and refined over the years, 
each with their own advantages
and disadvantages, but all with the general goal of providing a tool of sufficient 
quality for down-stream computation work such as 
protein docking~\cite{gasteiger_automatic_1990,vainio_generating_2007,
corbeil_docking_2009,riniker_better_2015,hawkins_conformation_2017,
friedrich_benchmarking_2017,wang_improving_2020,leite_frog_2007,lagorce_dg-ammos_2009}.

In a broad sense, we can define a valid conformer for a target molecule as a 
local minima in potential energy. As a practical matter, we could also insist
that any such local energy minima be somewhat close in energy to a global 
minimum. There are still ambiguities, however, since the energy of a molecule
is influenced by its environment, such as whether the molecule is solvated,
in some type of solid form, or bound to a protein. In addition, from some
perspectives, the (Gibbs) free energy is a more appropriate measure.

\begin{figure}[H]
    \centering
    \def\svgwidth{\columnwidth}
    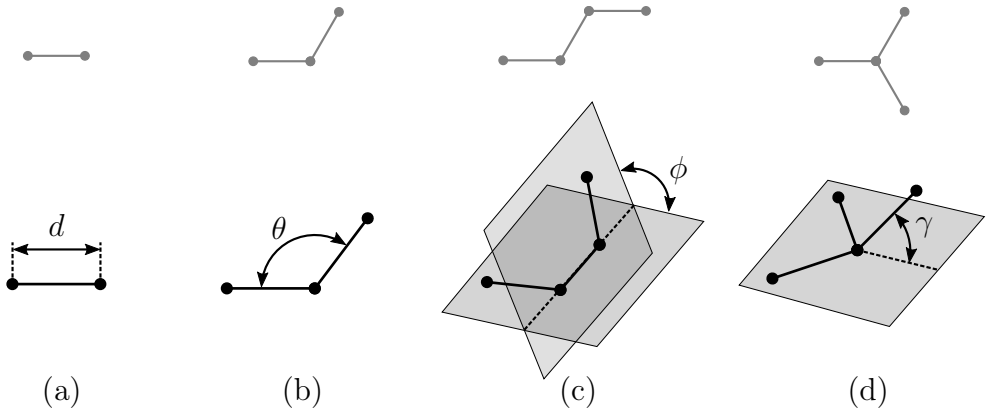

    \caption{Force fields typically include bonded terms associated with
    (a) bond lengths, (b) bond angles, (c) proper torsions, and (d)
    improper torsions. Each term has an associated subgraph topology
    and a single characteristic property.}
    \label{fig:bonded}
\end{figure}

To help address these ambiguities, consider the bonded parameters of a
molecule, such as bond lengths, bond angles, and
torsions (Fig.~\ref{fig:bonded}). We generally expect such parameters to be
weakly dependent on environment,
a tendency reflected in the "1-4" exclusion for pair energies found in many classical
force field 
parameterizations~\cite{dauber-osguthorpe_biomolecular_2019,case_amber_2005,halgren_merck_1996}.
As such, an accurate reproduction of these bonded parameters could be
considered a defining
characteristic of a valid molecule conformer, independent of environment.

Note that proper torsions are typically parameterized as periodic, such that
each associated term has an equally spaced set of favored values of torsion angle $\phi$.
Thus, it is possible for a molecule to exhibit multiple conformers with  
identical bonded parameters 
by sampling among a set of favored values of $\phi$, a property we shall
refer to as {\it torsional freedom}. Many
docking algorithms take advantage of torsional freedom to
sample ligand poses by manipulating
proper torsion angles along with overall
translation and rotation~\cite{corbeil_docking_2009,guo_comparison_2014,
kitchen_docking_2004,pagadala_software_2017,corso_diffdock_2022}.

From this point of view,
when developing a conformer generation algorithm, one could proceed with the
assumption that a reproduction of bonded terms is the primary goal, with
the sampling of torsional freedom as secondary, the latter being
dependent on the molecular environment and target application.
This approach is an advantage for models that 
rely on data sets for training,
because the availability of experimental data on molecular conformers 
is limited. If we instead rely on synthetic data, we are limited to the 
assumptions used in constructing such data sets. For example, the public
data sets used to train the model described in this work were generated
under conditions of a vacuum, an environment that is unnatural
for a drug-like molecule.

It should be emphasized, however, that sampling torsional freedom is a stated
goal of most conformer solutions~\cite{hawkins_conformation_2017,riniker_better_2015},
since many applications, especially in drug development, 
are primarily concerned with a specific molecule environment 
(water solvent) and many 
do not perform their own independent, torsional sampling.
For such applications, the model presented here will likely require additional
downstream processing to be useful.

The atoms associated with bonded parameters are separated by no more than three
bonds (Fig.~\ref{fig:bonded}). The distance between atoms separated by four or
more bonds is therefore not constrained by a single bonded term. Such atom pairs
will be referred to as {\it nonbonded}. In general, torsional freedom allows the
distances between nonbonded atom pairs to vary.

Rings can pose several challenges. Because a cycle of atoms must be closed,
constraints are imposed on bond angles. 
For non-aromatic systems, these
constraints can often be satisfied in multiple ways, introducing different
categories of corner folding (such as the ``chair'' and ``twist boat'' conformations
of cyclohexane~\cite{nelson_toward_2011}). The situation becomes more 
complicated for polycyclic systems. A successful conformer generator must
be able to sample from the various possibilities, within some reasonable
energy window.

Macrocycles impose constrains on torsional freedom that can be challenging
to accommodate algorithmically~\cite{wang_improving_2020}.

In addition to locating energy minima, a useful conformer 
generator needs to be
able to distinguish between stable isomers. The two
most important are chirality~\cite{jamali_enantioselective_1989,brooks_significance_2011}
and cis/trans isomerism~\cite{dugave2003cis}.

Several algorithmic approaches to conformer generation
have been developed over the years. CORINA~\cite{gasteiger_automatic_1990},
one of the first commercial offerings,
uses a simple ansatz for bond lengths and bend angles combined with 
geometric rules and backtracking, with a particular focus on the difficult
problem of ring systems. FROG~\cite{leite_frog_2007} employs
a template library for ring systems interconnected using canonical
bond lengths and angles, followed by torsional freedom sampling
via Monte Carlo. DG-AMMOS~\cite{lagorce_dg-ammos_2009} employs
a hybrid Krylov solver for distance geometry~\cite{chastine2005ammp}
followed by force-field minimization, and makes no attempt at sampling
torsional freedom. In contrast, the
OMEGA toolkit~\cite{hawkins_conformation_2017,hawkins_conformer_2010} 
specializes in targeted sampling of torsional freedom, and
employs a fragment library combined with rules-based sampling
for bonded parameters.
Balloon~\cite{vainio_generating_2007} is a conformer generator
based on a multiobjective genetic algorithm. 
ETKDGv3~\cite{riniker_better_2015,wang_improving_2020} is
a knowledge-based generator provided by RDKit~\cite{rdkit} 
based on distance geometry, with additional heuristics targeted at macrocycles.
EKTDGv3 is commonly followed by force-field
optimization to improve the accuracy of bonded parameters~\cite{riniker_better_2015},
an option conveniently provided by RDKit.

The problem of conformer generation has been an attractive target for
the machine-learning community, due to the broad availability of 
cheminformatics libraries, the approachable, geometric nature of the
problem, and relevance in drug discovery. 
Many attempts, however, have focused on technological
advancement at the detriment of physical viability and utility,
perhaps due to lack of appropriate domain knowledge.
Several strategies have been employed, such as energy 
gradients~\cite{luo_predicting_2021,shi_learning_2021,xu_learning_2021},
Gibbs sampling~\cite{mansimov_molecular_2019},
and conditional variational 
encoders~\cite{simm_generative_2020,xu_end--end_2021}.
The drawback of these approaches is that the energy of 
disordered molecule systems is difficult to characterize 
directly due to singularities and large energy barriers.
GeoMol~\cite{ganea_geomol_2021} learns local structure and 
applies incremental construction. Since incremental construction
is poorly suited to cycles, it fails to reproduce all but the
simplest ring systems.

GeoDiff~\cite{xu_geodiff_2022} is a stochastic diffusion model.
It follows conventions most closely related to  
``denoising diffusion probabilistic models''~\cite{ho_denoising_2020} 
(an approach developed for images),
employs around 800,000 independent parameters, 
and relies on an uncharacteristically large number of steps for generation. 
GeoDiff attempts to 
learn explicit values for the distances between nonbonded atom pairs,
a physically ambiguous quantity due to torsional freedom.
This requirement likely contributes to a high level of
computational complexity.

Torsional diffusion~\cite{jing_torsional_2022} is a generative
model designed to explore torsional freedom. It relies entirely on
the ETKDGv3~\cite{wang_improving_2020} algorithm for the difficult task
of establishing the bonded parameters. To make the task more
tractable, macrocycles are ignored, and molecules are limited
to no more than seven free torsion angles. The authors 
trained this model to reproduce the torsional freedom of 
a synthetic data set of drug-like
molecule conformers selected arbitrarily and
generated in a vacuum~\cite{axelrod_geom_2022}, 
an objective of questionable physical relevance.

\subsection{Diffusion-based generation}

Diffusion is a machine-learning technique introduced in 2015~\cite{sohl-dickstein_deep_2015}
and more recently the subject of groundbreaking
research in image generation.~\cite{ho_denoising_2020,croitoru_diffusion_2023}. 
The results have been impressive enough
to spawn several commercial 
endeavors~\cite{noauthor_dall-e_2023,noauthor_midjourney_2023}
and capture the imagination of the general public~\cite{schaul_ai_nodate}.

The methodology of diffusion-based generative modeling is described
in detail elsewhere~\cite{karras2022elucidating,luo_understanding_2022}
and only the general principles are summarized here.
By applying noise of varying amounts to a suitable data set, it is possible
to train a "denoising" model that can take a system with
noise and predict the original state. If properly prepared, a model
of this type can be applied in a series of sequential steps to
extract a random sample from pure noise. If the data used for training are
labeled in some fashion, and if those labels are incorporated in the model
during training, 
a diffusion model can be instructed
to bias generation to match a given set of labels, producing a
result corresponding to instructions.

Because of its success, diffusion-based models have been the subject
of considerable study~\cite{nichol_improved_2021,kingma_variational_2021,
song_score-based_2022} and many applications outside of image generation
have been 
developed~\cite{luo_diffusion_2021,kong2021diffwave,daniels2022scorebased,chen2023equidiff,wang2023nadiffuse}.
Various approaches to diffusion have been proposed,
from DDPM (denoising diffusion probabilistic models)~\cite{ho_denoising_2020},
VDM (variational diffusion models)~\cite{kingma_variational_2021},
score-based modeling~\cite{song_score-based_2022}, and 
LDM (latent diffusion models)~\cite{rombach_high-resolution_2022},
to name a few. The approach used in this work is based on
recent developments on formulating a universal
framework for describing diffusion-based models~\cite{karras2022elucidating}
around the concept of denoising
score matching~\cite{vincent_connection_2011}.

Extending diffusion-based techniques developed for images
to molecule conformer generation 
appears simple on the surface, however they are important differences.
In images, each
element $\bm{x}_i$ of the model space
corresponds to the color of a pixel, and the size of $\{\bm{x}_i\}$
depends on the number of pixels in the image to be generated.
For a molecule, the size of $\{\bm{x}_i\}$ corresponds to the number of atoms,
and each element corresponds to the projection of a given atom
position onto an arbitrary Cartesian reference system.
Whereas the solution $\{\bm{x}_i\}$ for an image is bounded to the color
space of the problem (with a universal origin and scale), 
the solution $\{\bm{x}_i\}$ of a molecule need only be internally
consistent and should otherwise be
translationally and rotationally invariant.

Described in this article is a novel method 
of conformer generation using a physics-informed, 
denoising model (PIDM). By taking advantage of established
methods employed in classical force fields, the intent is to construct a model
that is accessible, transferable and robust. Suitable training sets 
and benchmark criteria are chosen with these same concepts in mind. 
Building upon recent theoretical
advancements~\cite{karras2022elucidating}, both deterministic and
stochastic methods of generation are explored. A form of guided
generation is demonstrated as a means of exploring torsional freedom.

\subsection{Limitations}

Neglecting to appropriately sample torsional freedom will produce
conformers that are inadequate for applications that do not
perform their own sampling, such as rigid ligand docking or 3D pharmacophore
modeling, unless
additional processing of some kind is applied.

Conformer generation for molecules with chemical groups
or atom types
outside the training set may perform poorly or fail. 
Molecules with certain challenging
topologies, such as a central ring with several large branches, may
perform poorly (see Fig.~\ref{fig:atorvastatin}).
Conformer quality is expected to degrade as molecules grow 
in size beyond {$\sim$}200 heavy atoms.

\section{Methods}

Our goal is to provide a method that can
be used to generate realistic conformers, independent of
torsional freedom, for any drug-like
molecule when provided with just the atom composition,
connectivity, and isomer (chirality, cis/trans) identity.
Emphasis is placed on the reproduction of bonded terms
and the preservation of given chirality
and cis/trans isomerism.

\subsection{Denoising Function}

As discussed earlier, diffusion-based models can be generalized
around the concept of denoising. To that end,
we may represent our model as a denoising function $D$ that yields
an estimate of the actual coordinates $\bm{x}$ of the atoms of a molecule
when provided with coordinates that have been perturbed by a centered, 
uncorrelated Gaussian of width $\sigma$
\begin{equation}
\bm{x} \approx D(\mathcal{N}(\bm{x};0,\sigma^2\bm{I}),\sigma;\bm{\zeta}) \,,
\label{eq:D}
\end{equation}
where $\bm{\zeta}$ represents the 
composition of the molecule ({\it i.e.}
its atom types, connectivity, and isomer identity). The behavior of $D$ is controlled
by a set of internal parameters that are optimized during training
using standard deep-learning techniques.

\begin{figure}[H]
    \fbox{\includegraphics[width=\textwidth]{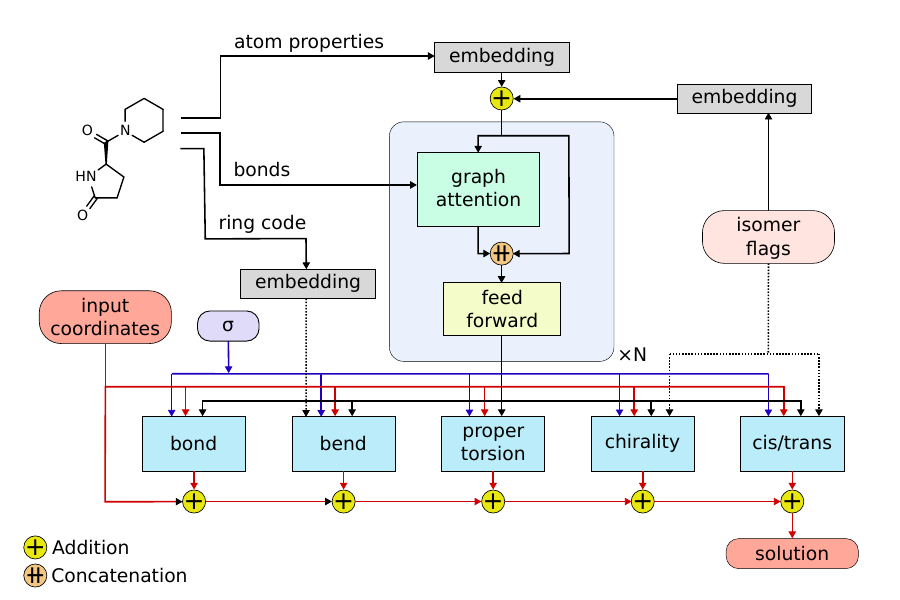}}
    \caption{A schematic of the denoising model $D$.}
    \label{fig:block}
\end{figure}

The overall structure of the model for $D$ is shown in Fig.~\ref{fig:block}
and consists of two major components: a graph transformer network
to build a form of atom typing and a series of bonded subcomponents 
whose outputs are summed together for coordinate prediction.

The purpose of the graph transformer network is to place the
atoms of each molecule into a suitably descriptive embedding space 
that can be employed by the bonded components. It begins by
assigning an initial embedding by enumerating
by atomic number, formal charge,
and hybridization. The latter is arrived using the algorithm
built into the RDKit cheminformatics library~\cite{rdkit}.
All hydrogens are treated as explicit. Chiral and cis/trans
atoms are flagged by the addition of global vectors reserved
for this purpose.

The initial atom embedding is refined by multiple
layers of a graph transformer network, based on the GATv2
algorithm~\cite{brody_how_2022}. The final output 
are the atoms of the molecule represented in a new embedding 
space that captures information on connectivity.
No information on atom coordinates is used at this stage.

The GATv2 graph attention network contained in each layer is
constructed by associating each molecular bond with a graph edge.
Self edges are not included since they would not reflect
an appropriate physical analog (a self edge would be
equivalent to a bond to an identical atom). Instead,
the input to the graph network is concatenated to the
output in order to preserve a form
of atom self-identity.
This concatenation is fed into a feed-forward network
to form the output of each layer. 

It should be noted that the edges used in the graph layers
include no labeling, such as bond order. Such labeling
was not deemed necessary since connectivity alone in
combination with atomic number, formal charge, and hydridization 
should provide sufficient context to describe relevant 
chemistry~\cite{glendening_natural_1998}.

Each of the bonded components have their own challenges
and are individually described in what follows.

The purpose of the ``bond'' component is to calculate
a correction to atom positions to account for 
the expected distance $|\bm\delta|$ between bonded atoms:
\begin{equation}
    \bm{\delta_{ij}} \doteq \bm{x}_j - \bm{x}_i \,.
\end{equation}
In this notation, we are using the subscript to refer to the
Cartesian coordinates of the corresponding atom.
The correction $\bm\Delta^d$, a vector in Cartesian space, is
calculated as a displacement along $\bm\delta$ separately for each 
of the two atoms associated with the bond.
The magnitudes of the displacements
are implemented, using a multilayer perceptron (MLP), as an arbitrary 
function of the current distance, the characteristic Gaussian width
of the noise, and the associated atom embeddings $\bm a$. The
correction applied to each atom of a bond can thus be described,
under a certain convention, as the following:
\begin{equation}
    \left[-\bm\Delta^{d}_i,\bm\Delta^{d}_j\right] = \frac{1}{2}
    \mathrm{MLP}\left(
        |\bm\delta_{ij}|, \sigma; \bm{a}_i, \bm{a}_j
    \right) \hat{\bm\delta}_{ij} \,,
    \label{eq:bond}
\end{equation}
where $\hat{\bm\delta}_{ij}$ is the normal vector along the direction
of $\bm\delta_{ij}$.
In the implementation, the input to the MLP is a concatenation
of the function variables and parameters, after suitable weighting.
A similar method is used for the other components described below.

For the ``bend'' component, a complication is that the target
bend angle depends on whether the corresponding three atoms
are a part of a ring, and if so, the size of the ring. 
One of the unfortunate
weaknesses of message-passing, graph convolution networks, 
of which GATv2 is a member, is the inability to detect 
cycles~\cite{xu_how_2019}.

To work around this weakness, a bit-encoded integer is 
used to enumerate the size and number of rings which belong to all three of the
atoms of the bend, for ring sizes up to and including 6 
(it is possible for a bend to be embedded in more than one ring). 
This integer is mapped to an
embedding $\bm{c}_{ijk}$. The bend angle
correction $\bm\Delta^\theta$ applied to the two outer atoms then follows a scheme
similar to the bond correction:
\begin{equation}
    \left[-\bm\Delta^{\theta}_i,\bm\Delta^{\theta}_k\right]=\frac{1}{2}
    \mathrm{MLP}\left(
        |\bm\delta_{ik}|, \sigma; \bm{a}_i, \bm{a}_j, \bm{a}_k, \bm{c}_{ijk}
    \right) \hat{\bm\delta}_{ik} \,.
    \label{eq:mlp_bend}
\end{equation}

\begin{figure}[H]
    \centering
	\begin{subfigure}{240pt}
    \def\svgwidth{\columnwidth}
    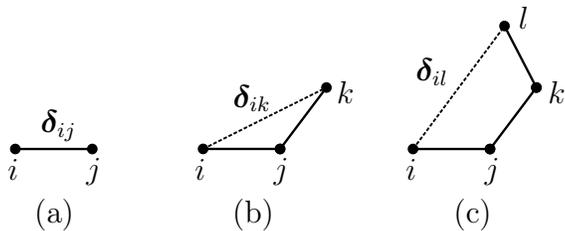

    \end{subfigure}
    \caption{Characterizing bonded terms using atom distances $|\bm\delta|$.
    Shown are (a)~bonds, (b)~bends, and (c)~proper torsions.}
    \label{fig:deltas}
\end{figure}

Notice that Eq.~\ref{eq:mlp_bend} uses the distance between the outer
two atoms instead of the bend opening angle (Fig.~\ref{fig:deltas}). 
This is a deliberate choice.
Unlike the opening angle, the distance is unbounded, and thus more 
numerically tractable. Eliminating the dependence on the position
of the central atom also removes some noise, especially when $\sigma$
approaches the associated bond lengths. For similar reasons,
the correction for
atom position is calculated along the vector between the two outer atoms, 
rather than as a rotation around the central atom.

A similar strategy is used for the ``proper torsion'' component
\begin{equation}
    \left[-\bm\Delta^{\phi}_i,\bm\Delta^{\phi}_l\right] =\frac{1}{2}
    \mathrm{MLP}\left(
        |\bm\delta_{il}|, \phi_{ijkl}, \sigma; \bm{a}_i, \bm{a}_j, \bm{a}_k, \bm{a}_l
    \right) \hat{\bm\delta}_{il} \,,
    \label{eq:mlp_proper}
\end{equation}
where the proper torsion angle $\phi$ is included as an argument 
to the MLP. The angle is useful because proper torsions, as discussed earlier, 
typically have multiple, favored values of $\phi$, each of which will be
characterized by a corresponding
distance $|\bm\delta_{il}|$ (see Fig.~\ref{fig:bonded}c). 
To avoid discontinuities in the modulo of $\phi$,
the concatenation for the input of the MLP
uses $[\cos\phi,\sin\phi]$ in place of $\phi$.

\begin{figure}[H]
    \centering
	\begin{subfigure}{240pt}
    \def\svgwidth{\columnwidth}
    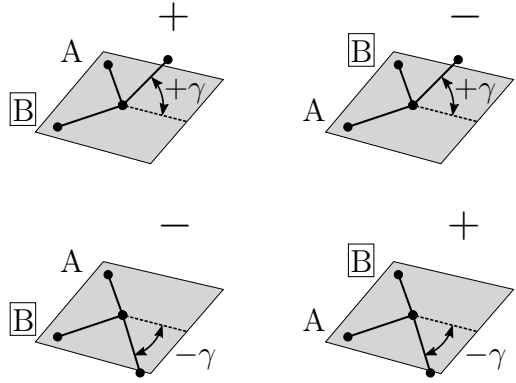

    \end{subfigure}
    \caption{The sign of chirality in terms of an improper torsion
    angle. The sign depends on whether a neighboring atom is above or below the
    plane formed by the other three atoms. The sign changes if any two 
    neighboring atoms are swapped.}
    \label{fig:tetra}
\end{figure}

Improper torsions, used by force-field parameterizations primarily
as a means to enforce planarity in conjugated systems, are already
constrained by bond lengths and bend angles. There is, however,
an important connection with chirality and the sign of the
improper torsion angle $\gamma$ (Fig.~\ref{fig:tetra}).

Consider an improper torsion centered on atom $i$, with a plane
defined in combination with atoms $j$ and $k$, and a forth atom $l$
for which we calculate $\gamma$. We can define a distance $d_\parallel$
along the plane and a signed distance $d_\perp$ out of the plane:
\begin{equation}
    \begin{aligned}
        d_{\parallel,ijkl} &= \bm\delta_{il} \cdot \hat{\bm\delta}_{\parallel,ijk} \\
        d_{\perp,ijkl} &= \bm\delta_{il} \cdot \hat{\bm\delta}_{\perp,ijk} \,,
    \end{aligned}
\end{equation}
where
\begin{equation}
    \begin{aligned}
        \bm\delta_{\parallel,ijk} &\doteq 
            -\left(\bm\delta_{ij}+\bm\delta_{ik}\right) \\
        \bm\delta_{\perp,ijk} &\doteq \chi_{ijkl}
            \left(\bm\delta_{ij}\times\bm\delta_{ik}\right) \,,
    \end{aligned}
\end{equation}
and where $\chi_{ijkl} = \pm 1$ is the target chirality.
The ``chirality'' bonded component of the model uses the above to enforce a given 
signed distance out of the plane, by displacing atoms $i$ and $l$:
\begin{equation}
    \left[-\bm\Delta^{t}_i,\bm\Delta^{t}_l\right] =\frac{1}{2}
    \mathrm{MLP}\left(
        d_{\perp,ijkl}, d_{\parallel,ijkl}, \sigma; \bm{a}_i, \bm{a}_j, \bm{a}_k, \bm{a}_l
    \right) \hat{\bm\delta}_{\perp,ijk} \,.
    \label{eq:mlp_tetra}
\end{equation}
The above term is used for all three combinations of the atoms in an improper torsion,
such that the total correction for the central atom is a vector sum
of the three. For a chiral atom with four bonded atoms, there are four improper
torsions, and thus twelve terms in total.

The ``cis/trans'' component is the last of five and is designed to enforce
explicit cis/trans isomerism associated with double bonds.
The planer structures typically associated with double bonds allow two possible
configurations for the associated proper torsions, corresponding
to $\phi\simeq 0$ and $\phi\simeq\pm\pi$. Rather than use the angle $\phi$
directly, we introduce, for a more robust solution,
the average vector between the inner and outer atoms of a proper
torsion:
\begin{equation}
    \bm\delta_{ijkl} \doteq 
     \frac{1}{2}\left(
        \bm{x}_j + \bm{x}_l - \bm{x}_i - \bm{x}_k 
    \right) \,,
\end{equation}
which is equivalent to the vector between the center of the
segment connecting the outer atoms and the center of the bond
connecting the inner atoms. The implementation of the component
can be expressed as:
\begin{equation}
    \left[-\bm\Delta^{c}_i,\bm\Delta^{c}_j\right] = \left[-\bm\Delta^{c}_l,\bm\Delta^{c}_k\right] =
    \frac{1}{2}
    \mathrm{MLP}\left(
        |\bm\delta_{ijkl}|, \chi_{ijkl}, \sigma; \bm{a}_i, \bm{a}_j, \bm{a}_k, \bm{a}_l
    \right) \hat{\bm\delta}_{ijkl} \,,
    \label{eq:mlp_cistrans}
\end{equation}
where $\chi_{ijkl}$ is equal to $-1$ ($+1$) for the target cis (trans) state
and the same correction is applied to pairs of atoms.
Depending on the chemistry, each cis/trans group can be associated with up to four
proper torsions, in which case each outer atom receives a vector sum of two corrections 
and the two inner atoms each receive four.

The final output of the model is an estimate of the unsmeared
coordinates and can be expressed symbolically as the sum of all contributions:
\begin{equation}
    D(\bm x,\sigma) = \bm x + \sum_{\rm bonds} \bm\Delta^d
    + \sum_{\rm bend} \bm\Delta^\theta
    + \sum_{\rm proper} \bm\Delta^\phi
    + \sum_{\rm chirality} \bm\Delta^t
    + \sum_{\rm cis{\slash}trans} \bm\Delta^c \,.
    \label{eq:modelsummary}
\end{equation}
Here the various arguments and indices have been suppressed for clarity.
By training on the vector sum of the five components, the model can learn
to adapt to correlated behavior.

As consistent with our goals, 
the denoising function of Eq.~\ref{eq:modelsummary} makes no attempt to
predict the distance between nonbonded pairs of atoms.

For the models reported here, an atom embedding of dimension 50 is used
throughout. Four graph transformer layers were employed.
The result is a model with a total of 135,080 parameters (weights), with 63,480
reserved for the molecule graph and 71,240 in the geometry components. 

\subsection{Training}

For the purposes of training, we choose to evenly sample from a canonical
set $\left\{\sigma_1\ldots\sigma_N\right\}$:
\begin{equation}
    \sigma_i = \begin{cases}
        \left(
            \sigma_{\vphantom{\rm min}\rm max}^{1/\rho} + \frac{i-1}{N-1}\left(
                \sigma_{\rm min}^{1/\rho} - \sigma_{\vphantom{\rm min}\rm max}^{1/\rho}
        \right)\right)^\rho & 1 \le i < N \\
        0 & i = N \\
   \end{cases}     \,.
\end{equation}
The total loss is calculated as the weighted sum of the contribution from each 
sample $\sigma_i$
\begin{equation}
\mathcal{L} = \sum_{i=1}^N \frac{1}{\sqrt{\sigma_i^2 + \epsilon^2}} \mathcal{L}(\sigma_i)
\end{equation}
with
\begin{equation}
    \mathcal{L}(\sigma) =
    \mathbb{E}_{\bm x \sim {\rm data},\bm n \sim \mathcal{N}(0,\sigma^2 \bm I)}
    \| \bm{D}(\bm x + \bm n,\sigma) - \bm x\|_2^2 \,,
\end{equation}
and where $\mathbb{E}$ is the expected value obtained from the training data
after applying random Gaussian sampling.
Standard techniques such as
the AdamW algorithm~\cite{loshchilov_decoupled_2017} and
mini batches were used to establish model weights that minimize $\mathcal{L}$.
Hyperparameter values of $N=100$, $\sigma_{\rm max}=8$\AA,
$\sigma_{\rm min} = 10^{-5}$\AA, $\epsilon=10^{-5}$\AA, and $\rho=6$
were found to produce satisfactory results.

\begin{figure}[H]
    \centering
    \includegraphics[width=\textwidth]{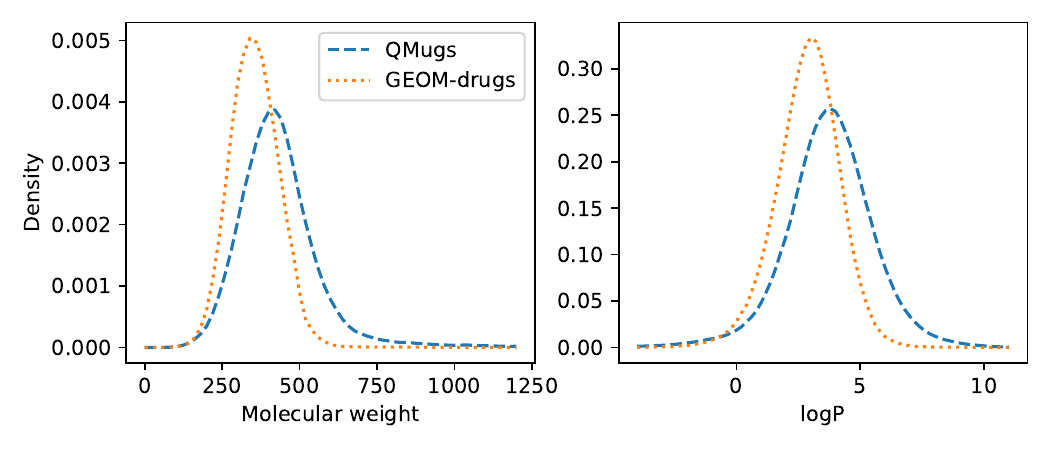}
    \caption{The distribution of molecular weight (left)
    and logP (right) for two datasets. The logP value is estimated
    using the Crippen algorithm~\cite{wildman_prediction_1999}.}
    \label{fig:datasets}
\end{figure}

Concerning data sources for training, a large, representative sample
of drug-like molecules would be ideal. For a meaningful benchmark,
a well-quantified baseline is also desirable. Two publically available,
synthetic data sets  come to mind: QMugs~\cite{isert_qmugs_2022} and 
GEOM-drugs~\cite{axelrod_geom_2022}. Both contain samples of
several hundred thousand drug-like molecules with conformers
optimized (in vacuum) using the GFN2-xTB  semiempirical quantum mechanical
method~\cite{grimme_robust_2017,bannwarth_gfn2-xtbaccurate_2019}.
Statistics on both sets are shown in Table~\ref{tab:datasets}.
Molecular weight and estimated LogP distributions are 
shown in Fig.~\ref{fig:datasets}.

To measure the similarity of two compounds, we employ the ECFP6 (extended-connectivity) 
fingerprint, as implemented by RDKit~\cite{rdkit} and folded to 1024 bits.
Using a relatively permissive Tanimoto threshold of 0.9, we find that only 5.6\%
of the compounds in QMugs overlap with those in GEOM-drugs. This increases
to 6.3\% if the threshold is lowered to 0.8.

\begin{table}[H]
    \centering
    \caption{Statistics on the QMugs~\cite{isert_qmugs_2022}
    and GEOM-drugs~\cite{axelrod_geom_2022} data sets, after quality filtering.}
    \label{tab:datasets}
    \begin{tabular}{lrr}
        Quantity & QMugs & GEOM-drugs \\
        \hline
        Molecules &  665,911 & 301,821 \\
        Molecule atoms & 36,679,641 & 13,378,196 \\
        Molecule bonds & 38,640,204 & 13,981,212 \\
        Molecule angles & 67,526,717 & 24,090,418 \\
        Proper torsions & 98,098,868 & 33,819,616 \\
        Improper torsions & 32,187,240 & 11,220,749 \\
        Proper torsions with cis/trans isomerism & 237,879 & 149,831 \\
        Improper torsions with chirality & 2,864,349 & 540,979 \\
        \hline 
        Conformers & 1,992,984 & 30,906,135 \\
        Conformer atoms (nodes) & 110,044,367 & 1,599,467,582 \\
        Conformer bonds (edges) & 202,017,554 & 1,658,998,071 \\
        \hline
        Average molecular weight & 433.3 & 355.5 \\
        Average number of heavy atoms & 30.6 & 24.8 \\
    \end{tabular}
\end{table}

The processing used by their authors
to create the two data sets are somewhat different.
For the QMugs data set~\cite{isert_qmugs_2022}, 
molecules extracted from the ChEMBL database~\cite{mendez_chembl_2019}
were charge neutralized, and a limit
of three conformers are selected for each molecule. Radical species
were also removed. The result is a total of 28 distinct atom types,
as enumerated by element, formal charge, and hydridization.
The total number of edges, when counting by conformer, is approximately
200 million.

For the GEOM-drugs data set~\cite{axelrod_geom_2022}, 
molecules were taken from the AI Cures conference 
open challenge~\cite{aicures,aicures_data}
and MoleculeNet~\cite{wu_moleculenet_2018}.
No molecule charge neutralization was attempted, which is 
somewhat unnatural, given that the conformers were generated
and optimized in vacuum. The result is a total of 64 distinct atom
types, a superset of those found in the QMugs data set. Conformer 
generation was more liberal, resulting in an average
of around 100 conformers per molecule, or a total of nearly 1.7
billion edges.

Performing quantum-level optimization of charged species in
vacuum can increase the likelihood of artifacts, as bonds 
are broken and created.
For the purposes of this study, the conformers provided by GEOM-drugs 
for each molecule was 
checked for strict consistency at the graph level, and if any
discrepancy was detected, the molecule was removed from consideration. 
This affected about 2\% of the data set.

It should be noted that the conformers in the QMugs data set are stored
in standard SDF format. This limits coordinate resolution to $10^{-4}$\AA.
The GEOM-drugs data set does not suffer from this limitation.

\begin{figure}[h]
    \centering
    \includegraphics[width=0.47\textwidth]{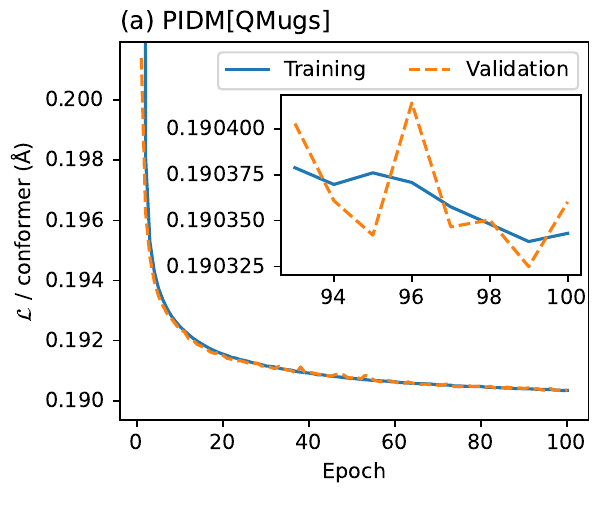}
    \includegraphics[width=0.47\textwidth]{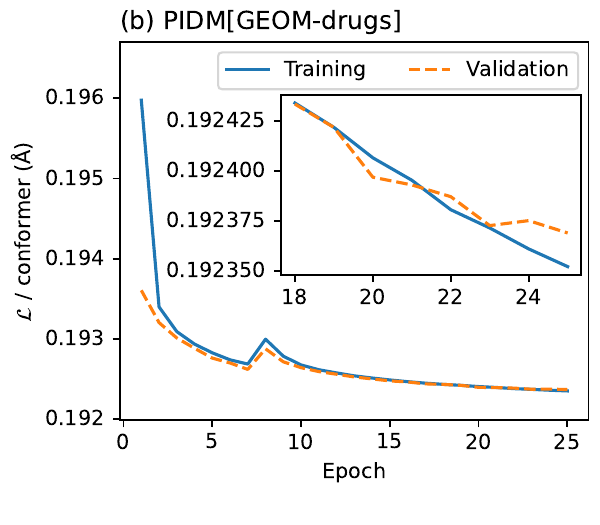}
    \caption{Loss per conformer as calculated during model training
    for (a) QMugs and (b) GEOM-drugs. Plotted are losses 
    calculated for the training subset and for an 
    independent validation set of $1/8$ the size.}
    \label{fig:train}
\end{figure}

Both data sets are randomly divided into training (80\%), validation (10\%), 
and test (10\%) subsets. One version of the model is trained on the 
QMugs training subset using a fixed schedule
of 100 epochs, corresponding to approximately 1.6 million steps,
with no evidence of overtraining (Fig.~\ref{fig:train}a). The loss
as calculated independently for the validation set during training is
remarkably consistent with the training loss.

A second version of the model is trained on the GEOM-drugs
training subset using a fixed schedule
of 25 epochs, corresponding to approximately 6.2 million steps.
There is also little evidence of overtraining and no apparent 
difference between training and validation losses
(Fig.~\ref{fig:train}b).

\subsection{Probing the models}

Perhaps the most direct way to judge the quality of
a trained model is to apply it in generation and
inspect the resulting conformers. That task will be the subject
of the following sections. Before doing so, it is instructive
to probe the model structure using example compounds and infer
characteristics of its learned behavior. 
Some examples are described below.
Further examples are available in the Supporting Information.

\begin{figure}[H]
    \centering
    \includegraphics[width=\textwidth,align=t]{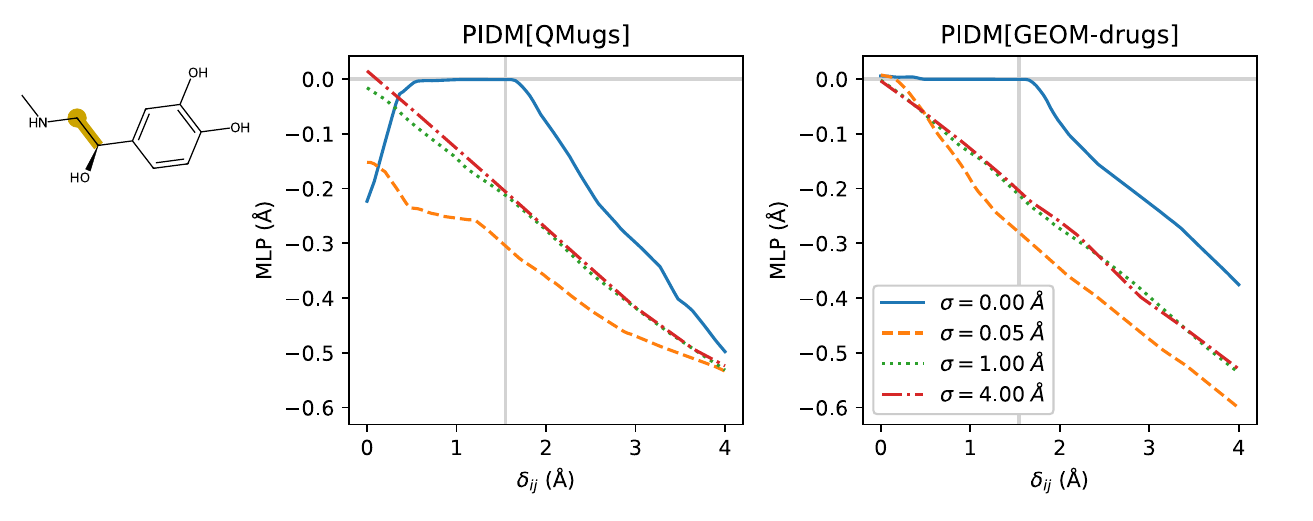}
    \caption{Example bond correction for the two models
    for an atom involved in an alkane bond in 
    adrenaline. Corrections for various values of $\sigma$ are plotted.
    The vertical gray line is the expected bond distance from a GFN2-xTB
    optimization.
    }
\label{fig:bondexample}
\end{figure}

Shown in Fig.~\ref{fig:bondexample} is the output of the
bond component for an example alkane bond as a function
of $|\bm\delta_{ij}|$, calculated
for various values of $\sigma$ (see Eq.~\ref{eq:bond}). 
As $|\bm\delta_{ij}|$
increases, the model predicts larger corrections. This
has the overall tendency of pulling bonded atoms closer
together. As $\sigma$ approaches zero, the correction
vanishes once the correct bond length is achieved.

Notice that the correction for the bond is smaller
than required for that bond alone (that is, the correction
falls well inside a line of unit slope). This is characteristic
of the individual corrections learned by the model
because the final result is the sum of all corrections.
As a consequence, bonded components tend to act in concert.

\begin{figure}[H]
    \centering
    \includegraphics[width=\textwidth,align=t]{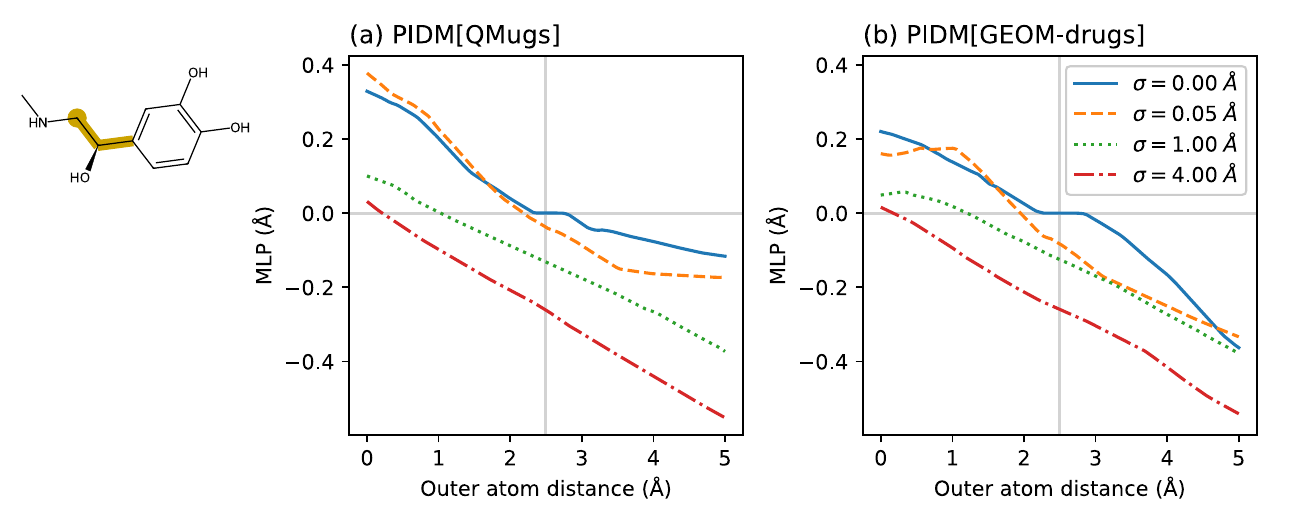}
    \caption{Example bend correction for the two models 
    for an atom involved in the ethanol group of 
    adrenaline. Corrections for various values of $\sigma$ are plotted.
    The vertical gray line is the expected atom distance from a GFN2-xTB
    optimization.
    }
\label{fig:bendexample}
\end{figure}

An example of the output of the bend component is shown in
Fig.~\ref{fig:bendexample}. The example has a similar behavior
to the bond, except there is more of a tendency to push
atoms apart if they are too close. Presumably this is part of a 
compensating mechanism for the tendency of the bond component to pull
atoms together.

Probes of the proper torsion, chirality, and cis/trans components
are also revealing. Details can be found in the Supplemental Information.

\subsection{Generation}

We adopt a score-based, probability flow framework~\cite{song_score-based_2022} 
in order to generate conformers from our trained model. 
As is typical in this approach,
we consider a multidimensional Wiener process applied
to molecule coordinates $\bm x$ over a time interval $t \in [0,1]$:
\begin{equation}
    p_t(\bm y(t)|\bm x;\sigma(t)) = \mathcal{N}(\bm y(t); \bm x, \sigma(t)^2{\bm I}) \,,
\end{equation}
where $\bm y(t)$ are the resulting random coordinates and where
$\sigma(t)$ is a width schedule 
we are free to choose to suit our task with the only requirement
that $\lim_{t\rightarrow 0} \sigma(t) = 0$.
For generation,
we start by sampling from a random Gaussian distribution 
$\mathcal{N}(0,\sigma(1)^2\bm I)$ as an approximation for $\bm y(1)$
and solve for the corresponding reverse process (denoising)
to obtain $\bm y(0)$ as a candidate solution for $\bm x$.

To construct a solution for the reverse process, we 
identify the marginal distribution 
$p(\bm y;\sigma)$ as
\begin{equation}
p(\bm y;\sigma) = \int p_t(\bm y|\bm x;\sigma) \; p(\bm x) \; d \bm x \,,
\end{equation}
where the $t$ dependence is implicit and
$p(\bm x)$ represents the marginal distribution of the training data.
We can use $p(\bm y;\sigma)$ to express the
time dependence of $\bm y$ as a {\it probability flow} 
ODE~\cite{karras2022elucidating}:
\begin{equation}
d\bm y = -\frac{d\sigma}{dt} \sigma\; \nabla_y \log p(\bm y; \sigma) \; dt \,,
\label{eq:flow}
\end{equation}
where $\nabla_y\log p(\bm y)$ is the {\it score function}.
In a score-based framework, there is a 
direct relationship between the score function and our denoising model 
$D$~\cite{karras2022elucidating,vincent_connection_2011}:
\begin{equation}
    \nabla_y \log p_t(\bm y;\sigma) \approx \frac{1}{\sigma^2}\left(
        D(\bm y, \sigma;\bm\zeta) - \bm y
    \right) \,,
\label{eq:score}
\end{equation}
where $\bm\zeta$ represents the molecular structure of interest.
This important relation connects our denoising model to the 
conformer generation process.

In our implementation, we have selected a
linear function $\sigma(t) = \alpha t$, where
$\alpha$ is a scale parameter in units of \AA. Applying this selection to
Eq.~\ref{eq:flow} and \ref{eq:score} results in
a simple form for the probability flow ODE:
\begin{equation}
\frac{d\bm y}{dt} =
\left( \bm y - D(\bm y;\alpha t;\bm\zeta) \right) / t\,.
\label{eq:ode}
\end{equation}
Our conformer generation process is the numerical solution to this equation,
calculated in steps of $t$ in reverse, and using as initial conditions
$\bm y(1) \sim \mathcal{N}(0,\alpha^2\bm I)$.

Inspired by work elsewhere~\cite{karras2022elucidating}, 
we solve Eq.~\ref{eq:ode} using  
Heun's 2\textsuperscript{nd}-order method, augmented
by a form of backtracking (Algorithm~\ref{algo:gen}). 
The backtracking provides
an option to add additional noise to the generation process.
We can begin by dividing the interval $[0,1]$ into a fixed
set of sequentially diminishing steps $\{t_i\}$ over 
which we iterate, to calculate a set of intermediate
solutions $\{\bm y_i\}$. Instead of relying on 
solving along the connected, nonoverlapping intervals $t_{i+1} \le t < t_i$,
we can substitute modified values for the upper bound $t_i$ of each interval:
\begin{equation}
\begin{aligned}
\tilde{t}_i &= \beta t_i   \\
\tilde{\bm y}_i &\sim \mathcal{N}(\bm y_i; 0, \lambda^2\alpha^2 t_i^2 (\beta^2-1) )
\end{aligned}
\end{equation}
where $\beta \ge 1$ and $\lambda \ge 0$ are free parameters. 
This has the effect of introducing 
Gaussian noise at each step of the solution.
For $\lambda = 1$, the amount of added noise compensates for the
change in interval size.

\begin{algorithm}
    \linespread{1.1}\selectfont
    \caption{Conformer generation.}
    \label{algo:gen}
    \begin{algorithmic}[1]
    
    \Procedure{Generate}{$D(\bm y,\sigma)$, $\{t_i\}$, $\alpha$, $\beta$, $\lambda$}
        \State $\bm y \gets \mathcal{N}(0,\alpha^2\bm I)$ \Comment{Prepare random initial state}
        \For{$i\gets 1$ to $|t|$}
            \State $\tilde{t} \gets \beta t_i$ \Comment{Widen effective interval}
            \State $\tilde{\bm y} \gets 
                \mathcal{N}(\bm y; 0, \lambda^2\alpha^2 t_i^2 (\beta^2-1)\bm I )$ \Comment{Add noise}
            \State $\bm d_1 \gets 
                \left(\tilde{\bm y} - D(\tilde{\bm y}, \alpha\tilde{t}) \right)/\tilde{t}$
                    \Comment{Evaluate $d\bm y/dt$}
            \State $\bm y \gets 
                \tilde{\bm y} + \left(t_{i+1} - \tilde{t}\right) \bm d_1$
                    \Comment{Solve}
            \If{$t_{i+1} > 0$}
                \State $\bm d_2 \gets \left(\bm y - D(\bm y, \alpha t_{i+1}) \right)/t_{i+1}$
                    \Comment{Apply 2\textsuperscript{nd}-order correction}
                \State $\bm y \gets \tilde{\bm y}_i 
                    + \frac{1}{2}\left( t_{i+1} - \tilde{t} \right)\left( \bm d_1 + \bm d_2 \right)$
            \EndIf
            \State $\bm y \gets \bm y - \langle\bm y\rangle$ \Comment{Remove center of mass}
        \EndFor

        \State \Return{$\bm y$}
    \EndProcedure
    
    \end{algorithmic}
\end{algorithm}

For reasons of convenience, we remove an overall center-of-mass
during each generation step. This prevents solutions from slowing walking
in coordinate space and helps with inspecting results. The correction is small
and quality of output is not affected.

If we generate using $\lambda = 0$, no noise is added during the
intermediate steps. In the language of diffusion-based models, we call
this a ``deterministic'' approach, even though we still begin
with a random initial state $y(1)$. Combined with $\beta > 0$, 
the algorithm is equivalent to pretending 
that each intermediate value $\bm y_i$ belongs to a solution
sampled from a larger value of $\sigma$. This has the effect
of overcorrecting, which improves accuracy in our case.

If we generate using $\lambda > 0$ and $\beta > 0$, we inject
noise during each step of generation. This is referred to as a
``stochastic'' approach. Both stochastic and deterministic 
approaches have been used for image generation, with impressive
results~\cite{ho_denoising_2020,nichol_improved_2021,
song_score-based_2022,karras2022elucidating,song_denoising_2022,
bansal_cold_2022,kingma_variational_2021,song_consistency_2023}.

To apply our algorithm, we use an exponentially decreasing set $\{t_1\ldots t_N\}$
of a given size $N$ and final step size $t_\epsilon$:
\begin{equation}
    t_i = \begin{cases}
        t_\epsilon^{i/(N-1)}& 1 \le i < N  \\
        0 & i = N \\
   \end{cases}     \,.
\end{equation}
The quality of generated output is reasonably stable for a large range
of parameter values. The results reported here 
use $t_\epsilon = 0.0006$, $\alpha = 2.5$~\AA, and $\beta = 5$.
Quality improves marginally if the solution is calculated
using more steps at a proportional cost in processing time.
To quantify this trade off, we report on results for $N = 100$, 200,
and 500. We also report results for both deterministic ($\lambda = 0$)
and stochastic ($\lambda = 1$) generation. 

\subsection{Results}

\begin{figure}[H]
    \centering
    \includegraphics[width=\textwidth]{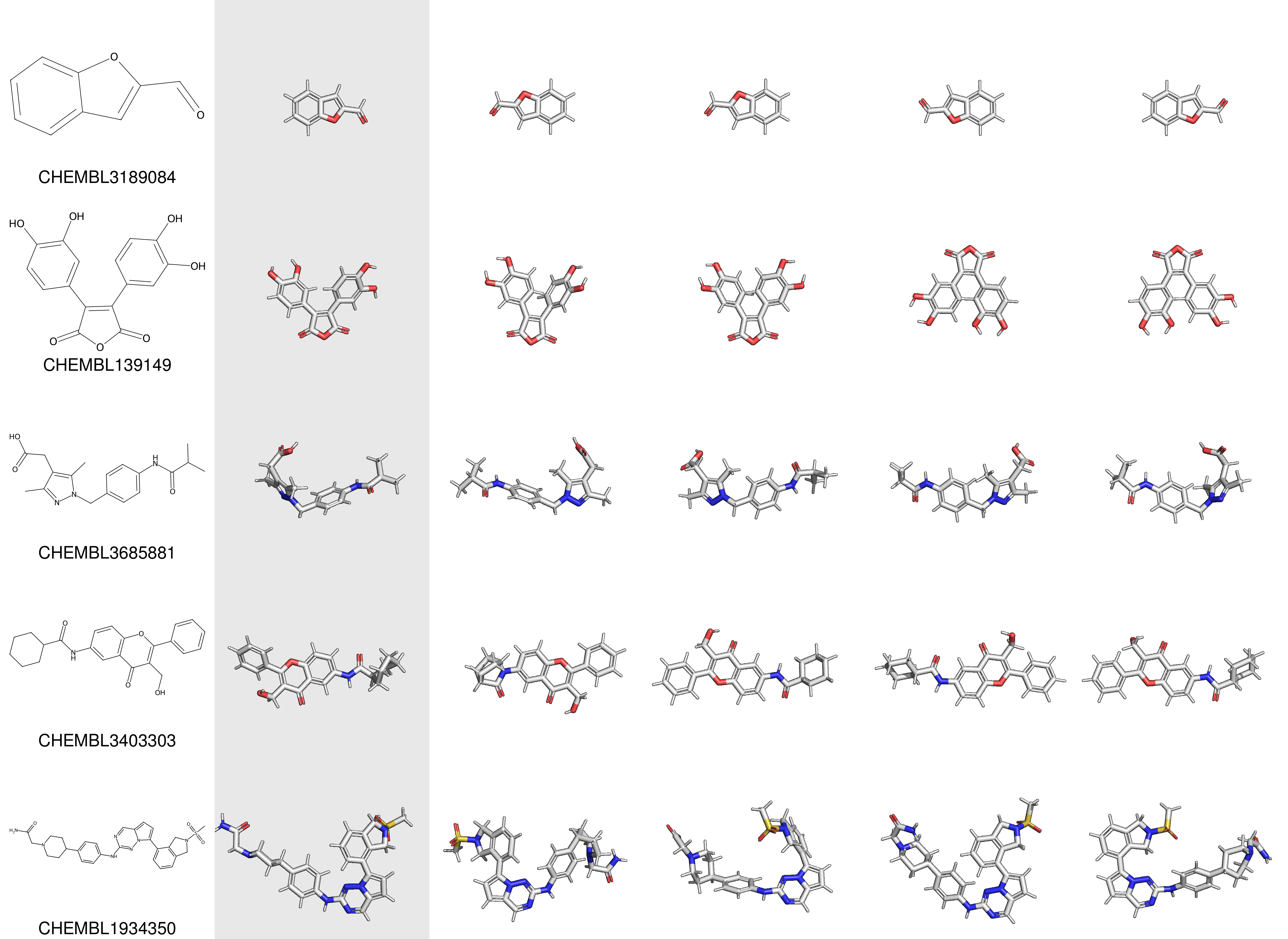}
    \caption{Example conformer output, for the model trained on QMugs,
    using deterministic generation and 500 steps, and for molecules randomly selected
    from an independent benchmark molecule set
    (see text for details). Shown in the second column 
    from left (grey background) is the first conformer from QMugs. Shown on
    the right are four conformer outputs, selected randomly.
    All molecule renderings are oriented by principal component.
    }
\label{fig:render01}
\end{figure}

Shown in Fig.~\ref{fig:render01} are random examples of
generated conformers,
using deterministic generation, 500 steps, and the model
trained on QMugs (PIDM[QMugs]). Similar figures, generated
under different conditions, such as stochastic generation
or using PIDM[GEOM-drugs], are available in the Supporting Information,
along with corresponding molecular structures files. 
Visual inspection alone does
not reveal any apparent difference in conformer quality.

\begin{figure}[H]
    \centering
    \includegraphics[width=\textwidth]{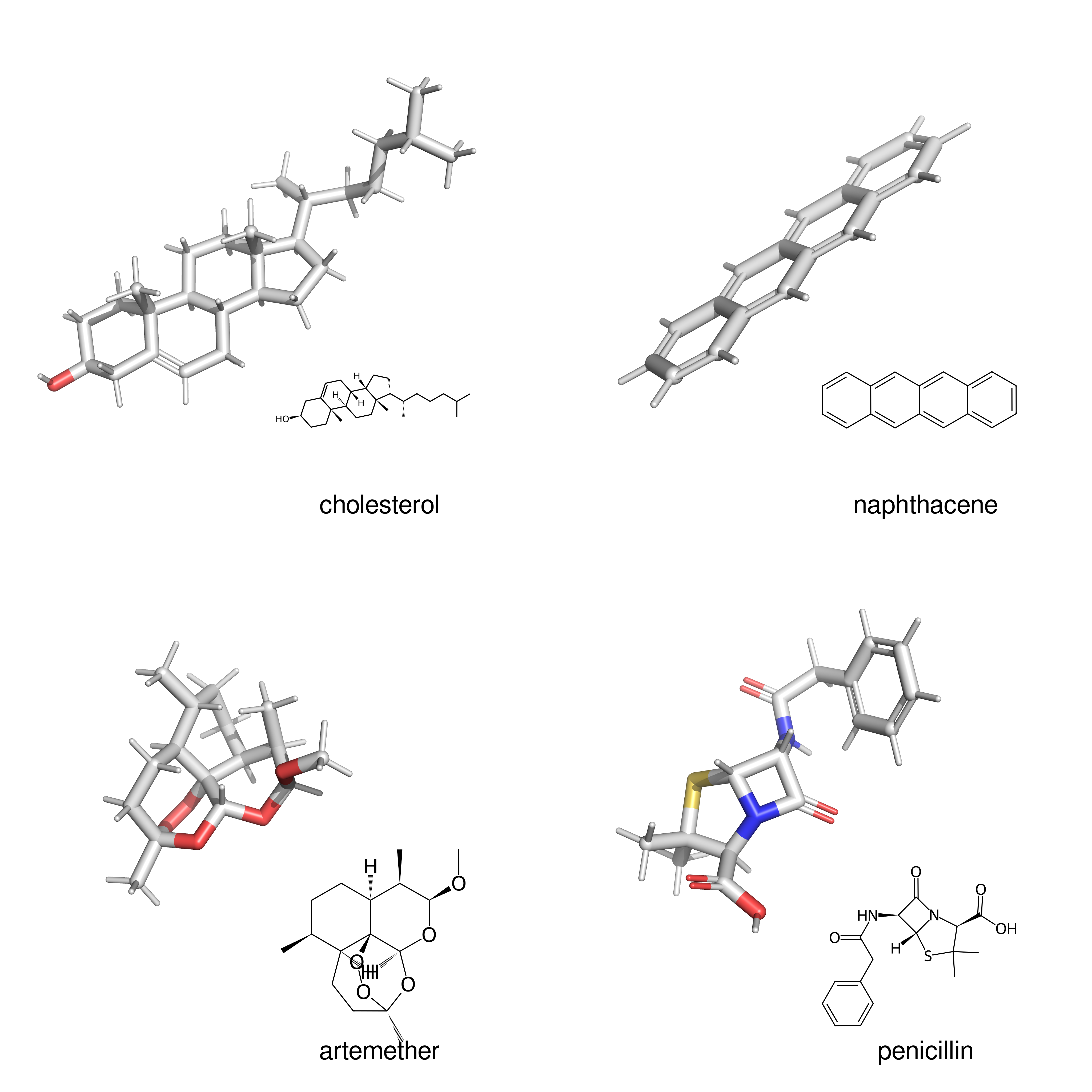}
    \caption{Select examples of conformer generation.
    These examples were generated using PIDM[QMugs] 
    with 500 steps in the deterministic scheme.
    }
\label{fig:examples}
\end{figure}

Shown in Fig.~\ref{fig:examples} are some selected examples of conformer
generation. For each of these molecules, good quality conformers are generated
the majority of the time. Sterols such as cholesterol contain fused ring
systems whose conformation could not have been reliably generated without
faithfully reproducing multiple chiralities. Large aromatic
systems, such as naphthacene, generate as reasonably flat, even though
our model does not contain the associated improper torsion terms. 
Complex, fused ring systems
are also reliably reproduced, even those with bridges, such as artemether.

\begin{figure}[H]
    \centering
    \includegraphics[width=\textwidth]{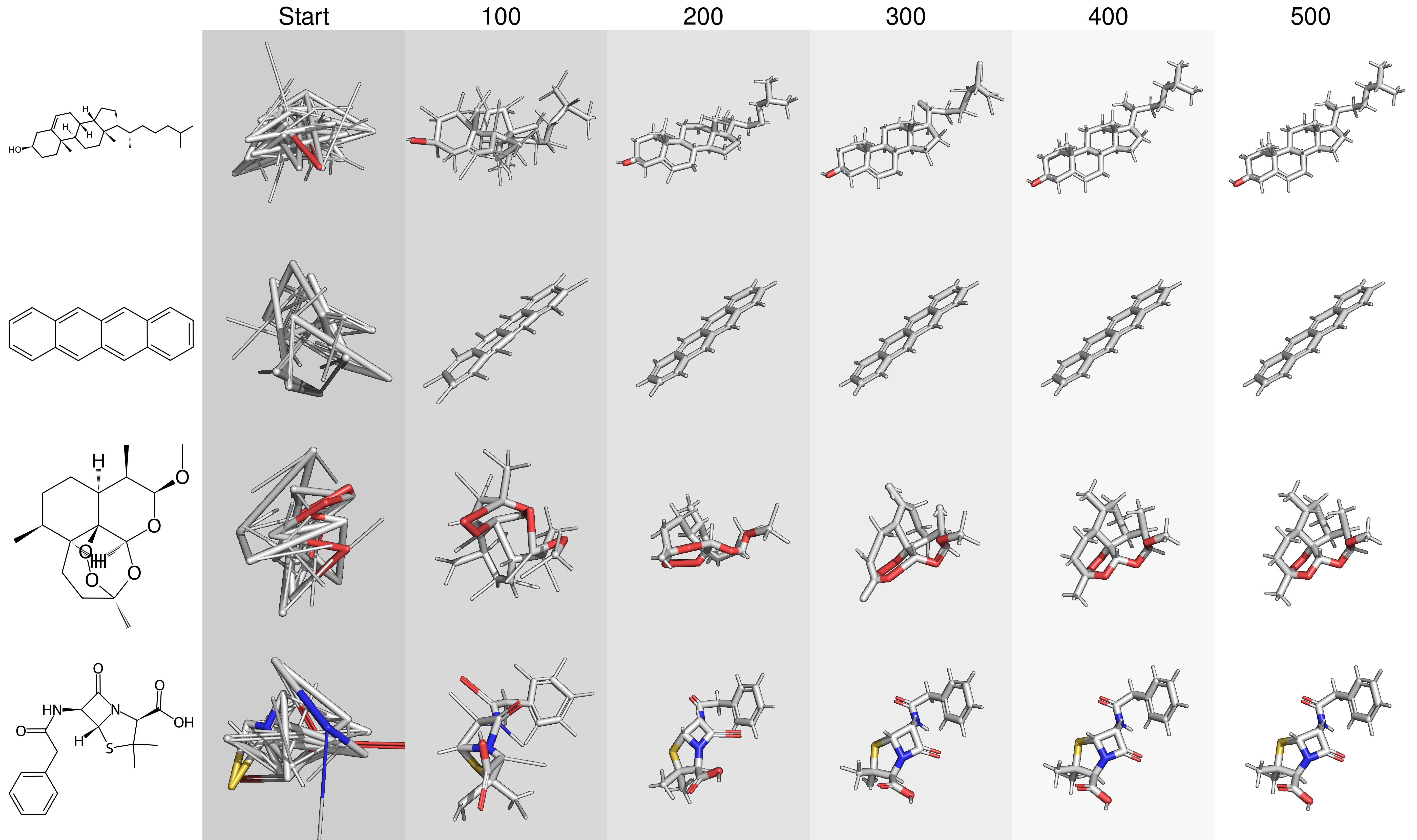}
    \caption{Steps in the generation of the example conformers
    illustrated in Fig.~\ref{fig:examples}
    }
\label{fig:frames}
\end{figure}

One of the characteristics of deterministic generation is that the frames
transition smoothly toward the final solution. This behavior is most
apparent when generation is presented as an animation, samples of which are
included in the Supporting Information. 
A static depiction of stages in the generation process
is shown in Fig.~\ref{fig:frames} for the four examples shown 
in Fig.~\ref{fig:examples}. The overall structure of the molecule conformer
arises early in the generation process, with the remainder focused
on refinement.

\begin{figure}[H]
    \centering
    \includegraphics[width=0.7\textwidth]{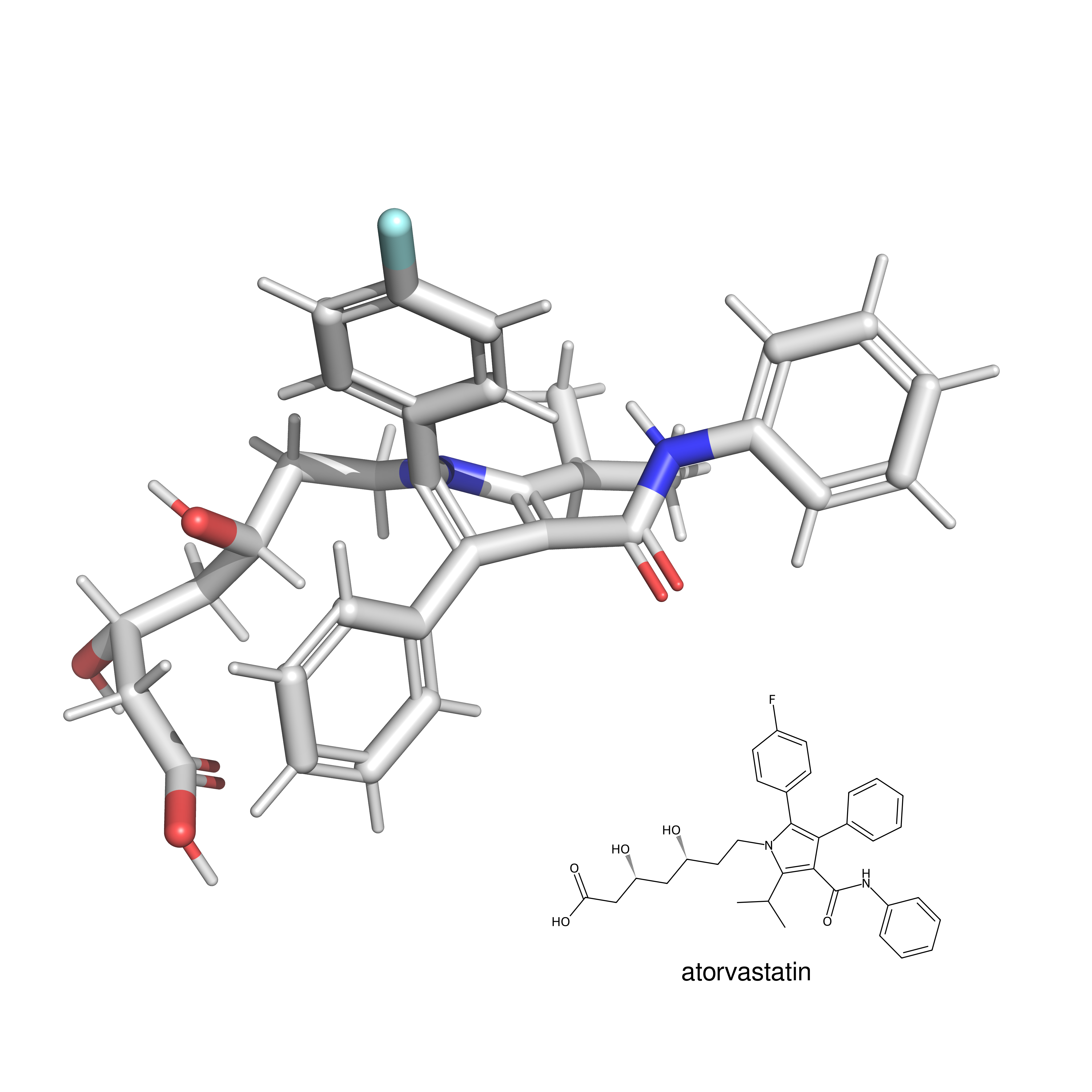}
    \caption{Example of a challenging molecule. Atorvastatin (Lipitor)
    contains a central, aromatic ring that consistently fails to 
    generate as planar.
    }
\label{fig:atorvastatin}
\end{figure}

There are some systems that are challenging to generate.
An example is atorvastatin (Fig.~\ref{fig:atorvastatin}). This
molecule has a central, aromatic ring connected to four large
substitutions. This aromatic ring usually fails to generate as
flat, probably because planarity imposes tight constraints on 
the orientation of two of the attached phenyl groups
that are difficult to satisfy.

\subsection{Benchmarks}

As discussed earlier, 10\% of the QMugs data set was set aside as
a test subset, corresponding to a random selection of 66,591 molecules. 
This subset will form the basis for our first set of benchmarks.
Because drug molecules are
often synthesized as part of a family of closely related compounds
during drug development, 
this randomly selected subset likely contains
compounds similar in structure to molecules in the training subset.
To remove structural overlap (at the level of the entire molecule)
and thus guard against data leakage, 
Tanimoto similarity is calculated
for each molecule 
against the contents of the QMugs training subset
and the entire GEOM-drugs data set. Any 
molecule with similarity exceeding 0.7
is discarded. The same threshold is also used 
within the test set to reduce its size and
ensure some level of diversity. 

We are interested in reproducing the annotated chirality
and cis/trans isomerism of our test compounds. To independently
verify this information, each test molecule is checked against the public 
PubChem database~\cite{kim2023pubchem}, and the annotations
available there are retrieved for this purpose. In additional,
we queried PubChem for a copy of the first 10 of their 
generated conformers~\cite{bolton_pubchem3d_2011}
for later analysis. Molecules that could not be validated
or did not have a generated PubChem conformer were 
discarded. 

The final result is an independent
set of 15,763 fully annotated test molecules reserved for benchmarks.
The molecular weight is an average 15\% smaller than the full
QMugs data set, but otherwise has a similar shaped distribution.

Although it was not mentioned earlier, the random sample
of compounds shown in Fig.~\ref{fig:render01}
were drawn from this independent benchmark set, as are the
other examples included in the Supporting Information.

We are interested in establishing conformer accuracy by
measuring the reproduction of bonded parameters
such as bond length ($d$), bond angle ($\theta$),
and proper torsion angle ($\phi$). Proper torsions
require special attention because, as discussed earlier,
they can have multiple favored values.
To limit our statistics to proper torsion angles associated with the 
same favored angle, we only consider angles that are 
generated within $\pm 30$\textdegree{} of the angle found in the
reference conformer.

We are also interested in measuring how often a generated
structure fails to reproduce the desired chirality
and cis/trans isomerism. To do so, we check each related
improper torsion and cis/trans bond for a
geometry that is consistent with given annotations.
Failures are recorded as a fraction of total occurrence
of improper torsion atom or cis/trans atom pair.

Because our model makes no attempt to address torsional freedom,
the resulting conformers may have atoms that overlap in
position (clashes). Although less important for applications which 
introduce their own torsional sampling (such as flexible ligand docking),
clashes nevertheless represent an unphysical molecular state
that is explicitly excluded in conventional conformer generators.
To measure their occurrence, we count the fraction of generated
conformers that include any nonbonded atom pair within a distance of less
than 1.5\AA{}.

To provide overall benchmark statistics, we generate ten
random conformers for each of the molecules in the
benchmark set. Each generated conformer is compared against
all the corresponding conformers for that molecule as provided by 
QMugs. Overall errors in
$d$, $\theta$, and $\phi$ (within cutoffs) are measured
using the mean absolute deviation (MAD), to avoid
sensitivity to tails. 

Results are shown in Table~\ref{tab:benchmark1}.
In the same table are results taken from 
other published conformer generation
solutions. In all cases, ten conformers are requested, although some 
conformer solutions by design provide less than the requested number 
under certain circumstances. Results are 
briefly summarized below.

\begin{table}[H]
    \centering
    \caption{Benchmark data for the models presented here under
    various different running conditions and
    compared against other conformer generation methods.
    Results for models trained on the QMugs and GEOM-drug
    data sets are shown for generation steps of size 100, 200 and 500,
    and using both deterministic and stochastic schemes.
    Best values in each category are highlighted.
    }
    \begin{tabular*}{\linewidth}{@{}ll@{\extracolsep{\fill}}rrr@{\extracolsep{\fill}}rr@{}r}
\hline
& & \multicolumn{3}{c}{Mean absolute deviation} & \multicolumn{2}{c}{Inconsistency rate} & Clash rate\\\cline{3-5}\cline{6-7}
& & $d$ (\AA)& $\theta$ (rad) & $\phi$ (rad) & chirality & cis-trans & $<1.5$\AA\\
\hline\\[\dimexpr-\normalbaselineskip+2pt]
\multicolumn{7}{@{}l}{PIDM[QMugs]} \\
 Deterministic & 100 &
$0.0042$ &
$0.015$ &
$0.036$ &
$0.031$ &
$0.033$ &
$0.753$ \\
 Deterministic & 200 &
$0.0038$ &
$0.013$ &
$0.027$ &
$0.022$ &
$0.009$ &
$0.681$ \\
 Deterministic & 500 &
\multicolumn{1}{>{\columncolor[gray]{.92}[3pt]}r}{$0.0036$} &
\multicolumn{1}{>{\columncolor[gray]{.92}[3pt]}r}{$0.012$} &
$0.023$ &
$0.013$ &
\multicolumn{1}{>{\columncolor[gray]{.92}[3pt]}r@{}}{$0.002$} &
$0.576$ \\
 Stochastic & 100 &
$0.0051$ &
$0.021$ &
$0.079$ &
$0.112$ &
$0.027$ &
$0.587$ \\
 Stochastic & 200 &
$0.0047$ &
$0.019$ &
$0.069$ &
$0.081$ &
$0.013$ &
$0.571$ \\
 Stochastic & 500 &
$0.0045$ &
$0.018$ &
$0.062$ &
$0.057$ &
$0.004$ &
$0.549$ \\
\hline\\[\dimexpr-\normalbaselineskip+2pt]
\multicolumn{7}{@{}l}{PIDM[GEOM-drugs]} \\
 Deterministic & 100 &
$0.0044$ &
$0.015$ &
$0.034$ &
$0.031$ &
$0.034$ &
$0.761$ \\
 Deterministic & 200 &
$0.0040$ &
$0.013$ &
$0.027$ &
$0.023$ &
$0.015$ &
$0.705$ \\
 Deterministic & 500 &
$0.0037$ &
$0.012$ &
$0.023$ &
$0.015$ &
$0.005$ &
$0.601$ \\
\hline\\[\dimexpr-\normalbaselineskip+2pt]
\multicolumn{2}{@{}l}{Pubchem3D (OMEGA)} &
$0.0075$ &
$0.020$ &
$0.020$ &
$0.020$ &
$0.014$ &
$0.000$ \\
\multicolumn{2}{@{}l}{ETKDGv3} &
$0.0183$ &
$0.039$ &
\multicolumn{1}{>{\columncolor[gray]{.92}[3pt]}r@{}}{$0.019$} &
\multicolumn{1}{>{\columncolor[gray]{.92}[3pt]}r}{$0.000$} &
$0.017$ &
$0.000$ \\
\multicolumn{2}{@{}l}{ETKDGv3+MMFF94} &
$0.0081$ &
$0.017$ &
$0.021$ &
$0.000$ &
$0.017$ &
$0.000$ \\
\multicolumn{2}{@{}l}{Balloon} &
$0.0082$ &
$0.018$ &
$0.032$ &
$0.002$ &
$0.011$ &
$0.000$ \\
\multicolumn{2}{@{}l}{GeoMol} &
$0.0125$ &
$0.030$ &
$0.042$ &
$0.032$ &
$0.087$ &
$0.495$ \\
\multicolumn{2}{@{}l}{GeoDiff} &
$0.0051$ &
$0.017$ &
$0.170$ &
$0.500$ &
$0.263$ &
$0.032$ \\
\hline
\end{tabular*}
    \label{tab:benchmark1}
\end{table}

The public PubChem API~\cite{kim2023pubchem}
provides access to a set of conformers
calculated under the PubChem3D scheme~\cite{bolton_pubchem3d_2011}.
This scheme is based on the OMEGA toolkit~\cite{hawkins_conformer_2010}
using parameters selected by the authors. Of particular
note is the choice to apply the MMFF94s classical 
forcefield~\cite{halgren_merck_1996} minus long-distance
charged interactions. The results, according to our benchmarks,
are robust conformer prediction, particularly for
proper torsion angles.

The 
ETKDGv3~\cite{riniker_better_2015,wang_improving_2020,rdkit}  
algorithm produces the most accurate proper torsion
values, but is less accurate with bond lengths and angles.
Following
conformer generation with MMFF94 force field optimization
produces conformers of quality similar to
PubChem3D, which relies on a similar force field for parameterization.
No chirality inconsistencies were detected.

Balloon produces bond distance
and bend angle accuracies consistent with other conformer
generation solutions that take advantage of the MMFF94 force field.

Like the model presented here, GeoMol
focuses on the bonded components of molecules. As mentioned
earlier, message passing networks are incapable of detecting
cycles~\cite{xu_how_2019}, and so it is not surprising
the GeoMol has difficulty accurately representing them,
despite incorporating
ad-hoc corrections to compensate for this weakness. 
Tested here is the
version of the model trained on the GEOM-drugs data set. Performance
is poor by all metrics presented here, especially for bond lengths.

\begin{figure}[H]
    \centering
    \includegraphics[width=\textwidth]{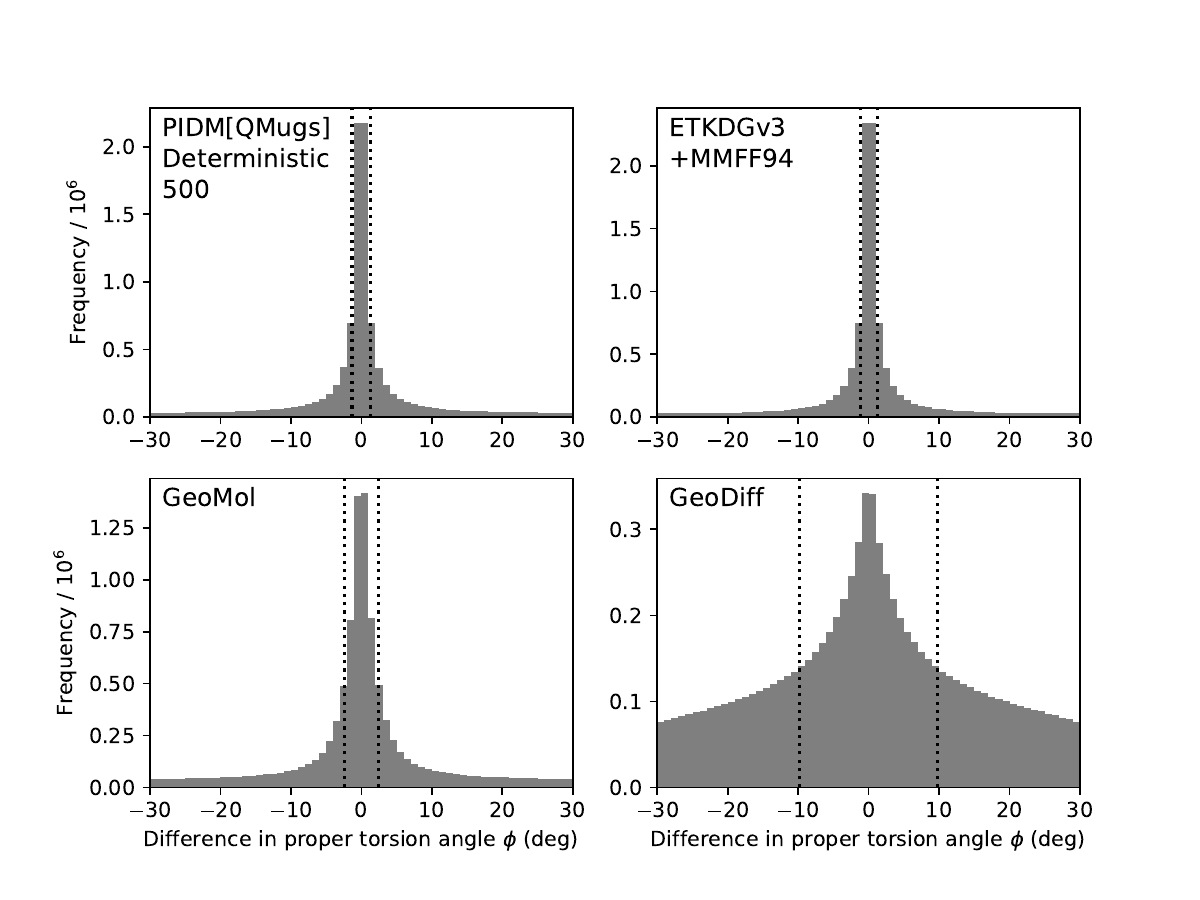}
    \caption{The difference in generated and ground truth for
    proper torsion angle $\phi$ for four different generation
    methods applied to the set of benchmark conformers. 
    The vertical dashed lines indicate the
    median absolute deviation, calculated for differences
    with $\pm 30$\textdegree.}
\label{fig:benchproper}
\end{figure}

Tested here is the version of GeoDiff~\cite{xu_geodiff_2022} 
trained on the GEOM-drugs data set, as provided by the authors.
Although this model is capable of accurate bond distance and 
bend angle prediction, proper torsion angles are poorly 
reconstructed (Fig.~\ref{fig:benchproper}).
GeoDiff has no mechanism for enforcing chirality nor
cis/trans isomerism. Although this limits the usefulness 
for drug discovery, the omission appears to be an 
oversight of the authors rather
than a fundamental limitation of their approach.

\begin{figure}[H]
    \centering
    \includegraphics[width=\textwidth]{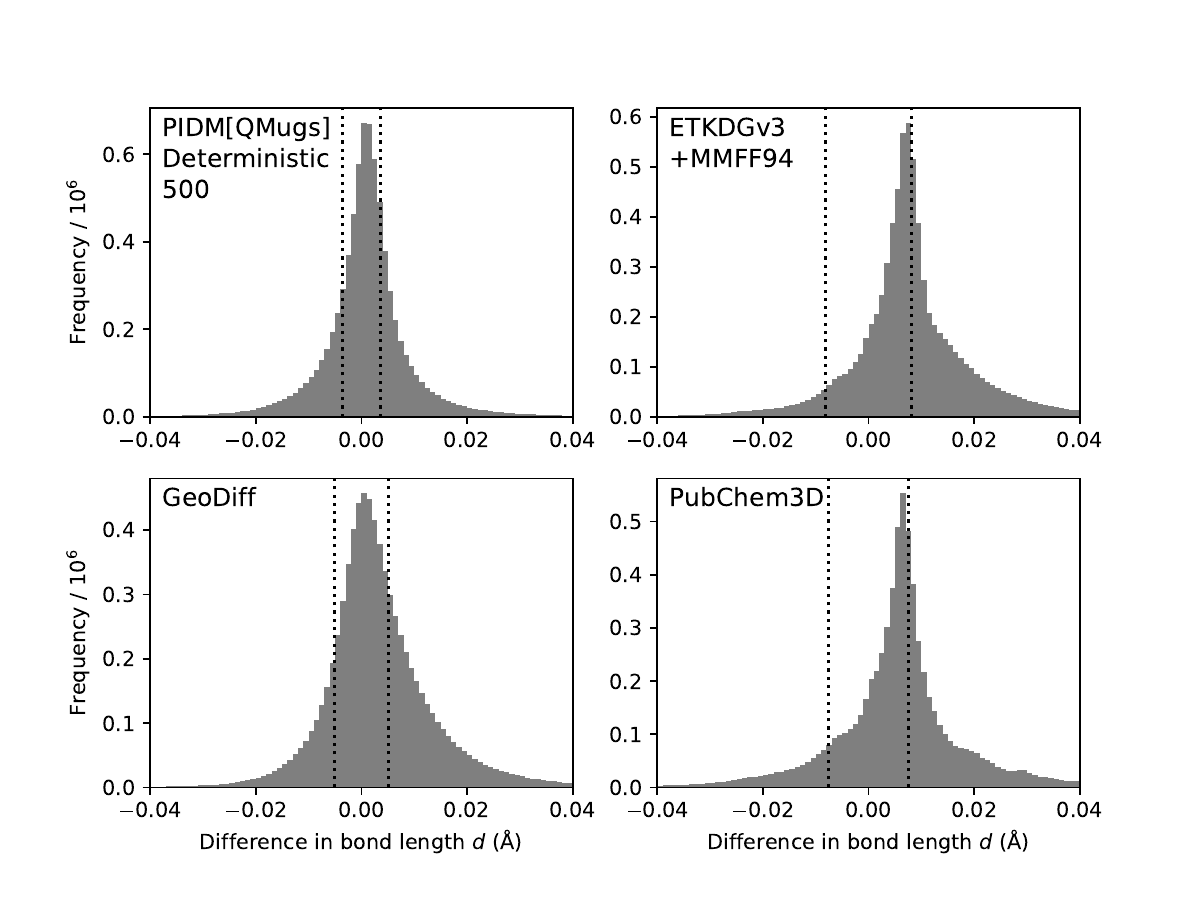}
    \caption{The difference in generated and ground truth for
    bond distance $d$ for four different generation
    methods applied to the set of benchmark conformers. 
    The vertical dashed lines indicate the
    median absolute deviation. The methods in the left
    column were trained on conformers optimized in the
    same fashion as the benchmark conformers (GFN2-xTB).
    The methods in the right column rely on some variation 
    of the MMFF94 force field.}
\label{fig:benchbond}
\end{figure}

Several of the conformer methods described above rely on the
MMFF94 force field. When compared against benchmark conformers
optimized by the more realistic GFN2-xTB 
semiempirical quantum mechanical method, an overall
bias is apparent in bond lengths (Fig.~\ref{fig:benchbond}).
If the bond length parameters in the MMFF94 force field were
refit, it's possible that conformer methods such
as PubChem3D, RDKit followed by MMFF94, and Balloon would
outperform the generative model presented here.

\begin{figure}[H]
    \centering
    \includegraphics[width=240pt]{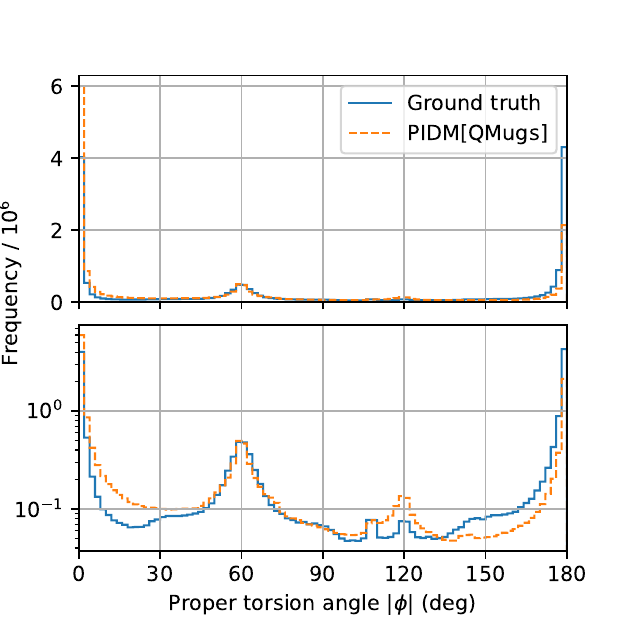}
    \caption{The distribution of the magnitude of proper torsion
    angle $\phi$ for the benchmark set of conformers
    compared to the distribution generated
    from the same set of molecules using PIDM.}
\label{fig:genproper}
\end{figure}

The statistics on $\phi$ accuracy shown
in Table~\ref{tab:benchmark1} were limited to
those cases where the generated proper angle aligned
within $\pm30$\textdegree{}. 
Accuracy aside, we can check to see how well the
overall distribution is reproduced. Interestingly enough,
the PIDM models tend to favor $\phi = 0$
more than the conformers provided by QMugs (Fig.~\ref{fig:genproper}).
The authors of QMugs used an elaborate procedure involving
molecular dynamics and clustering via RMSD to select the
conformers in their data set, and a uniformity in proper torsion
angles is likely a natural consequence of this process.

For ligand-protein docking methods that explore torsional freedom during
pose optimization, dihedral angle sampling for input conformers
should have little relevance. For the benefit of applications that do 
not independently sample dihedral angles, we can perform further
analysis.

The QMugs and GEOM-drugs
data sets provide a sample of conformers taken from
states of favorable energy calculated in
vacuum. These samples are selected 
to represent some amount of diversity,
as measured using the root-mean-squared deviation
(RMSD)~\cite{isert_qmugs_2022,axelrod_geom_2022}.
Although these data sets were constructed with care,
they remain synthetic, and the strategies used
to enforce diversity somewhat arbitrary.
Thus, we see little value in comparing the dihedral sampling 
of our conformer model to these data sets.

In place of synthetic data sets, we can
rely on experimental data. To measure conformer generation
performance, the authors
of the OMEGA toolkit selected two small
experimental sets~\cite{hawkins_conformer_2010}: 
480 molecules from the Cambridge
Structural Database (CSD) and 197 ligands 
from the PDB.
We will use the same experimental data here.

Both the CSD and PDB data sets
are derived from X-ray data in which only
heavy atoms are reliably resolved. The CSD
data is for crystalline solids of the molecules
either by themselves or with salts. The PDB data set
is for ligands bound to proteins. Both are typically
resolved in the solid state. 

\begin{figure}[H]
    \centering
    \includegraphics[width=\textwidth]{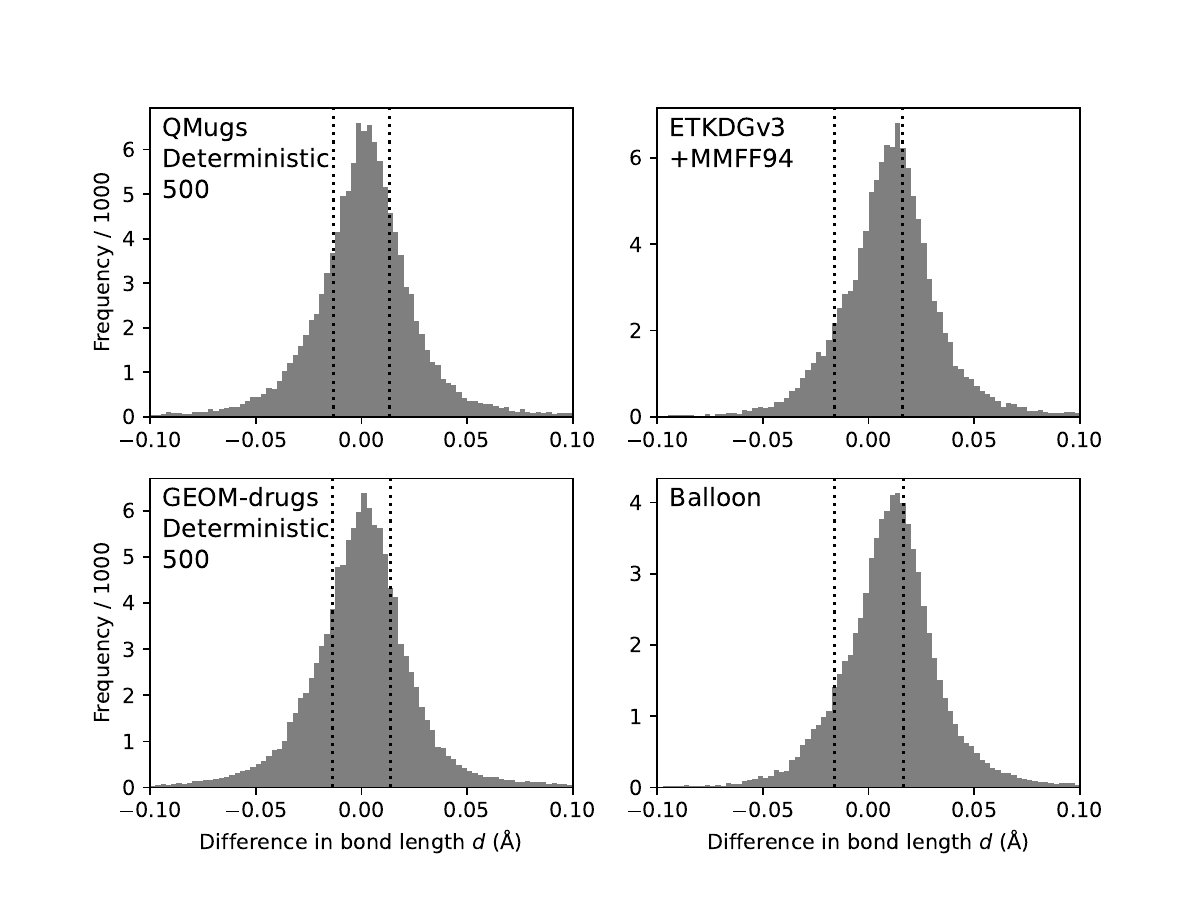}
    \caption{The difference in generated and experimental 
    bond distance $d$ for 480 structures from the CSD.
    Only bonds involving two heavy atoms are included. 
    The vertical dashed lines indicate the
    median absolute deviation. Both models on the right
    rely on some variation of the MMFF94 force field.}
\label{fig:csdbenchbond}
\end{figure}

The X-ray structures in the PDB have limited resolution
and their solutions are reconstructed, in part, based
on assumed force field parameters~\cite{deller_models_2015}. As such,
PDB files are not useful for testing the accuracy
of bonded parameters. The atom coordinates
in the CSD data set, however, are not as  
constrained by such assumptions.
A comparison of the bonded parameters in generated conformers 
for the CSD data set show the same
trends as observed from the QMugs data set,
although with lower resolution, presumably due to 
experimental uncertainties (Table~\ref{tab:benchmark2}).
A bias in MMFF94 bond length is confirmed
(Fig.~\ref{fig:csdbenchbond}). Generated results show
little bias, presumably reflecting the accuracy (on average) of
GFN2-xTB, used in the training data.

\begin{table}[H]
    \centering
    \caption{Statistics on generated conformers compared
    to experimental data for 480 structures extracted
    from the CSD.
    Only bonded terms entirely involving heavy atoms are included.
    }
    \begin{tabular*}{\linewidth}{@{}ll@{\extracolsep{\fill}}rrr@{\extracolsep{\fill}}rr@{}r}
\hline
& & \multicolumn{3}{c}{Mean absolute deviation} & \multicolumn{2}{c}{Inconsistency rate} & Clash rate\\\cline{3-5}\cline{6-7}
& & $d$ (\AA)& $\theta$ (rad) & $\phi$ (rad) & chirality & cis/trans & $<1.5$\AA\\
\hline\\[\dimexpr-\normalbaselineskip+2pt]
\multicolumn{7}{@{}l}{PIDM[QMugs]} \\
Deterministic & 500 &
$0.0134$ &
$0.030$ &
$0.032$ &
$0.026$ &
$0.003$ &
$0.581$ \\
Stochastic & 500 &
$0.0141$ &
$0.035$ &
$0.061$ &
$0.112$ &
$0.016$ &
$0.492$ \\
\hline\\[\dimexpr-\normalbaselineskip+2pt]
\multicolumn{7}{@{}l}{PIDM[GEOM-drugs]} \\
 Deterministic & 500 &
$0.0136$ &
$0.031$ &
$0.033$ &
$0.039$ &
$0.027$ &
$0.584$ \\
\hline\\[\dimexpr-\normalbaselineskip+2pt]
\multicolumn{2}{@{}l}{ETKDGv3+MMFF94} &
$0.0161$ &
$0.026$ &
$0.033$ &
$0.000$ &
$0.042$ &
$0.000$ \\
\multicolumn{2}{@{}l}{Balloon} &
$0.0163$ &
$0.029$ &
$0.037$ &
$0.000$ &
$0.041$ &
$0.000$ \\
\multicolumn{2}{@{}l}{GeoDiff} &
$0.0140$ &
$0.030$ &
$0.153$ &
$0.488$ &
$0.242$ &
$0.008$ \\
\hline
\end{tabular*}
    \label{tab:benchmark2}
\end{table}

We will use the RMSD 
to measure how well experimental coordinates of the entire
molecule are reproduced,
calculated on heavy atoms after solid-body
alignment. To do so, atoms between generation and experiment 
are paired by graph matching, where atoms are 
distinguished by element, formal charge, and number of hydrogens, and bond
orders are ignored. The graph matching may produce more
than one solution (due to symmetries). The match
that produces the smallest RMSD value is used.

Our goal is to use RMSD to measure how close a generator is able
to mimic the torsional freedom experimentally
observed for a molecule.
For our model, in which each conformer is randomly generated, 
the result will depend on how many attempts we allow.
Shown in Table~\ref{tab:exper} are statistics for the closest conformer out of 
10, 100, and 1,000 attempts. Also shown are published 
values~\cite{hawkins_conformer_2010} for the OMEGA toolkit
and to 1,000 conformers generated by RDKit followed by MMFF94 optimization.
Both OMEGA and RDKit clearly outperform the model presented here, even
after 1,000 attempts.

Our model does not consider the distances between nonbonded atom pairs, 
so it may be unrealistic to expect it to randomly sample torsional 
angles as effectively as algorithms like OMEGA and ETKDGv3, 
which are specifically designed to do so. This flaw is evident in
the clash rates (Table~\ref{tab:benchmark2}).

\begin{table}[H]
    \centering
    \caption{RMSD statistics on the closest conformer generated through
    various methods compared
    to experimental data 
    from the CSD and PDB.
    }
    \begin{threeparttable}
    \begin{tabular}{@{}lrrr@{\extracolsep{8pt}}rr@{}}
    & & \multicolumn{4}{c}{RMSD (\AA)} \\
    \cline{3-6}\\[\dimexpr-\normalbaselineskip+2pt]
    & & \multicolumn{2}{c}{CSD}&\multicolumn{2}{c}{PDB} \\
    \cline{3-4}\cline{5-6}\\[\dimexpr-\normalbaselineskip+2pt]
    Model & N & Mean & Median & Mean & Median \\
    \hline\\[\dimexpr-\normalbaselineskip+2pt]
    PIDM[QMugs]
    & 10
    & 1.24 & 1.38
    & 1.37 & 1.52 \\
    Deterministic 500
    & 100
    & 0.93 & 1.03
    & 1.04 & 1.16 \\
    & 1000
    & 0.74 & 0.84
    & 0.90 & 0.98 \\
    \hline\\[\dimexpr-\normalbaselineskip+2pt]
    PIDM[GEOM-drugs]
    & 10
    & 1.28 & 1.41
    & 1.37 & 1.59 \\
    Deterministic 500
    & 100
    & 0.97 & 1.10
    & 1.15 & 1.32 \\
    & 1000
    & 0.78 & 0.91
    & 0.96 & 1.10 \\
    \hline\\[\dimexpr-\normalbaselineskip+2pt]
    OMEGA\tnote{a}
    & --- & 0.51 & 0.44 & 0.67 & 0.51 \\
    \hline\\[\dimexpr-\normalbaselineskip+2pt]
    RDKit+MMFF94 
    & 10
    & 0.70 & 0.76
    & 0.92 & 1.06 \\
    & 100
    & 0.47 & 0.54
    & 0.67 & 0.80 \\
    & 1000
    & 0.43 & 0.48
    & 0.53 & 0.64 \\
    \hline
    \end{tabular}
    \begin{tablenotes}
        \item[a] Published statistics~\cite{hawkins_conformer_2010}
    \end{tablenotes}
    \end{threeparttable}
    \label{tab:exper}
\end{table}

\begin{figure}[H]
    \centering
    \includegraphics[width=0.28\textwidth]{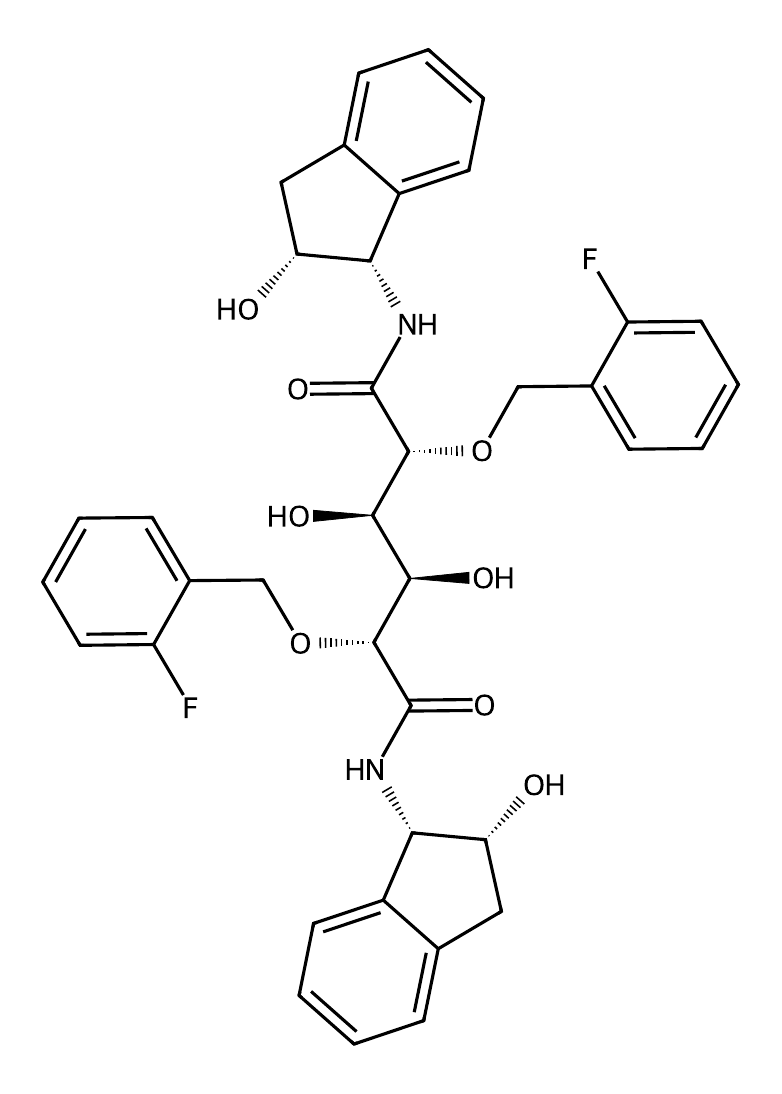}
    \hspace{-0.3in}
    \includegraphics[width=0.75\textwidth]{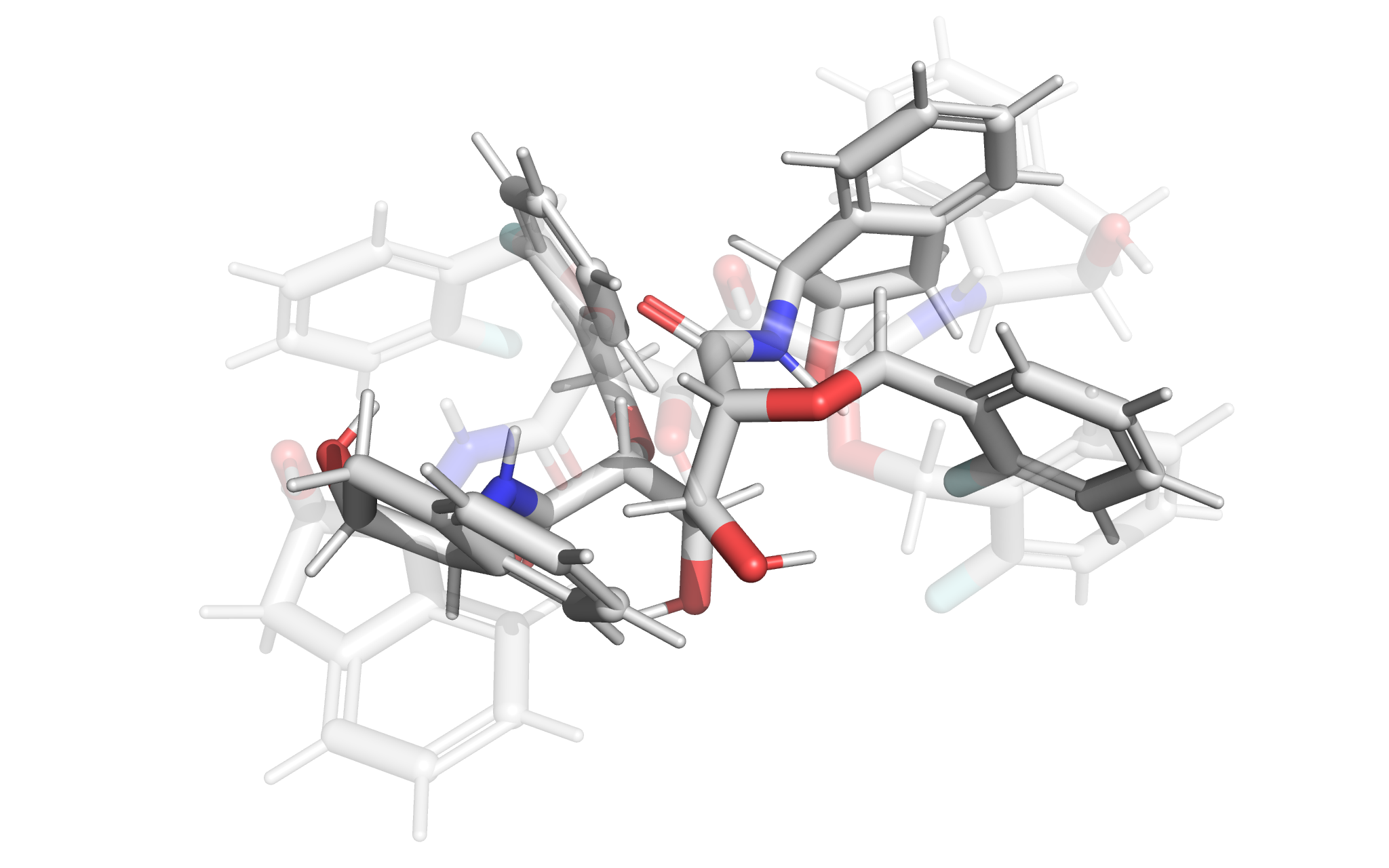}
    \caption{Conformer generation result for the ligand (BEA403)
    contained in PDB structure 1EC0. The
    experimental conformer is drawn as transparent. 
    Overlaid is the best RMSD
    result of 1,000 attempts at generation using PIDM[QMugs].}
\label{fig:1ec0_default}
\end{figure}

To better appreciate the kind of dihedral sample space that our 
generative model is failing to sample, we can investigate the
tail of the RMSD distribution. Shown in Fig.~\ref{fig:1ec0_default}
is the ligand from PDB structure 1EC0, which is the 
molecule with the worst RMSD result (2.9\AA). This ligand, a symmetric
system with four ring systems,
is observed in an extended conformation in the PDB.
The best generated pose is more confined. The failure to generate
a more compatible, extended conformation could be due to a 
lack of any explicit mechanism to repel nonbonded atom
pairs from each other.

\subsection{Guided Generation}

As discussed above, our model makes no attempt at predicting
the distance between nonbonded pairs of atoms. If we are
not concerned about torsional freedom, this is an acceptable
compromise. Even so, there are some drawbacks,
including the tendency to produce conformers that are more
tightly constrained in space than experimentally observed,
as discussed in the previous section. The introduction
of unphysical clashes may also be an issue, depending
on application.

If dihedral sampling is important, some method of
introducing bias during generation could be useful.
Given that our denoising model places no explicit constraint
on the distance between nonbonded atom pairs, one approach for 
dihedral sampling would be to add some type of bias on
that distance during generation. This section describes
such an addition as a proof of concept intended 
to prevent overlapping atoms.

Consider a modified probability flow ODE:
\begin{equation}
\frac{d\bm y}{dt} =
\left( \bm y - D(\bm y;\alpha t) - E(\bm y)\right) / t\,,
\label{eq:dode}
\end{equation}
where $E(\bm y)$ is a function external to the denoising
model that is introduced in
order to guide generation in a desired fashion. 
Solving for Eq.~\ref{eq:dode} in place of Eq.~\ref{eq:ode}
provides a mechanism for guided generation
where $E(\bm y)$ serves as a type of conditional 
score~\cite{luo_understanding_2022}.

For our proof of concept, consider a term that is analogous to a
repulsive force of strength $\bm\delta^{-10}$, 
similar to what is found in the repulsive portion 
of a Van der Waals interaction:
\begin{equation}
    \left[\bm\Delta^{n}_i,\bm\Delta^{n}_j\right] = -\frac{r_u^{11}}{2}
    \sum_{ij\in \mathrm{pairs}} 
        \mathrm{max}\left(\bm\delta_{ij}^2, r_c^2 \right)^{-5}
        \hat{\bm\delta}_{ij} \,,
    \label{eq:repuls}
\end{equation}
where $r_c$ is a clipping distance, nominally
set to a value of 0.7\AA, and $r_u$ is a unit distance
of 1\AA. Because this term is
not trained, it contains no atom embedding and is applied
to all nonbonded atom pairs equally. Using Eq.~\ref{eq:repuls}, 
we can introduce
a corresponding candidate for $E(\bm y)$ that is moderated
by an overall strength $\eta$:
\begin{equation}
    E(\bm y) = \eta\sum \bm\Delta^n \,.
    \label{eq:repul}
\end{equation}

\begin{figure}[H]
    \centering
    \includegraphics[width=\textwidth]{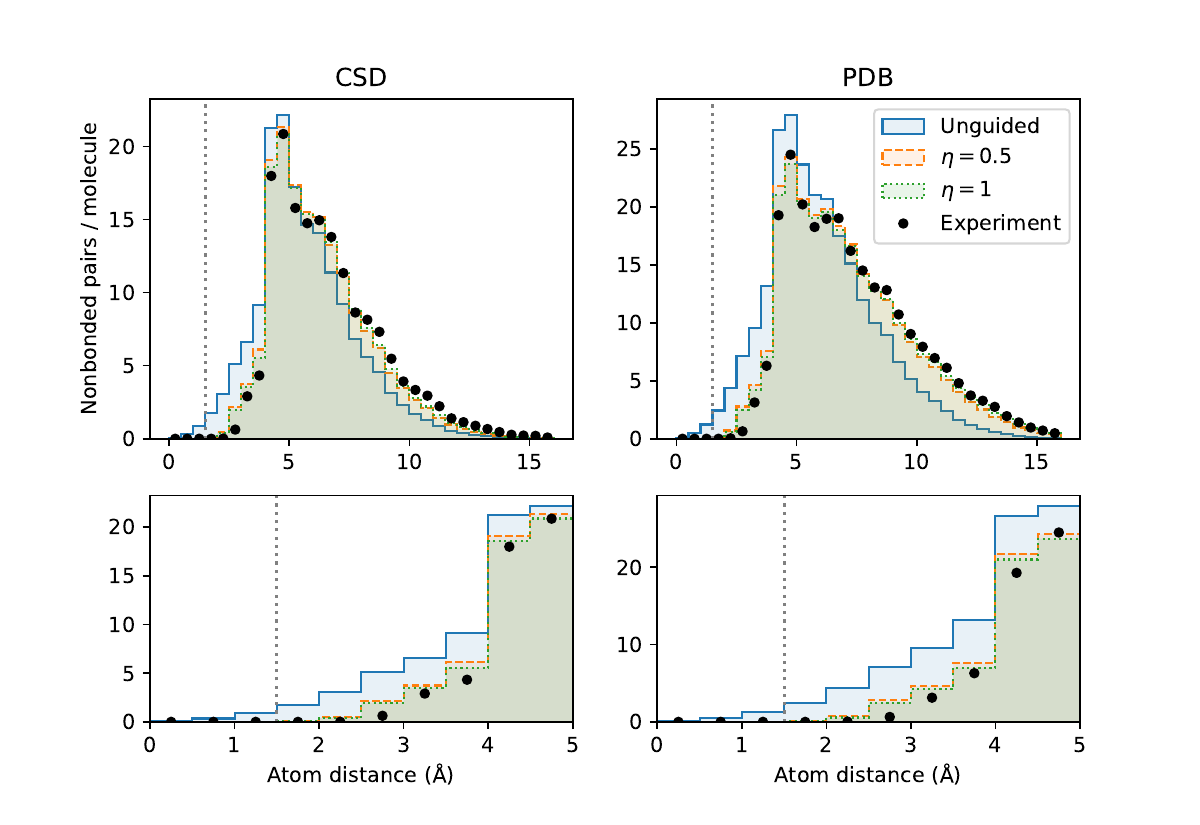}
    \caption{The distribution of distance between all nonbonded 
    atom pairs for the conformers in the CSD and PDB
    experimental data sets. Also shown are generated
    results for PIDM[QMugs] using 500
    steps in the deterministic scheme. Difference
    strengths of a repulsive term are applied.
    The vertical dashed line represents the distance used
    (1.5\AA) when reporting the clash rate.
    }
\label{fig:pairdist}
\end{figure}

Shown in Fig.~\ref{fig:pairdist}
is the distribution of distances between all atom pairs
in the CSD and PDB data sets. Also shown are the sames distances
sampled from ten conformers generated by PIDM[QMugs]
under various conditions and otherwise
using 500 steps in the deterministic scheme. With default 
generation, there is a tendency for atom pairs to overlap, 
which is unphysical. Adding a term of Eq.~\ref{eq:repul} with
strength $\eta = 0.5$
is sufficient to move the distribution closer to experiment.

\begin{figure}[H]
    \centering
    \includegraphics[width=\textwidth]{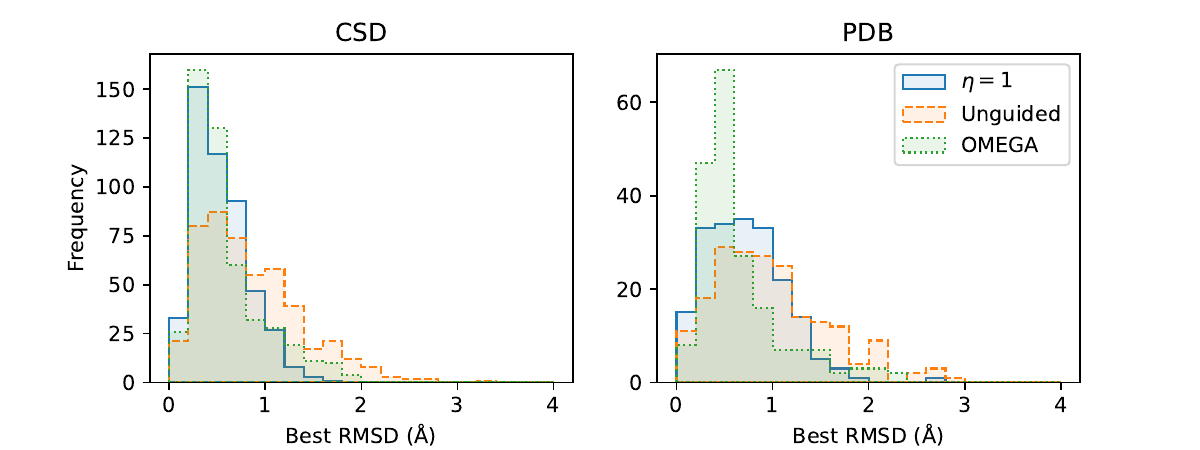}
    \caption{RMSD performance with and without
    a repulsive term for the CSD and PDB 
    experimental data sets. Shown for comparison are the
    published results from OMEGA~\cite{hawkins_conformer_2010}.}
\label{fig:exper.repul}
\end{figure}

\begin{figure}[H]
    \centering
    \includegraphics[width=0.75\textwidth]{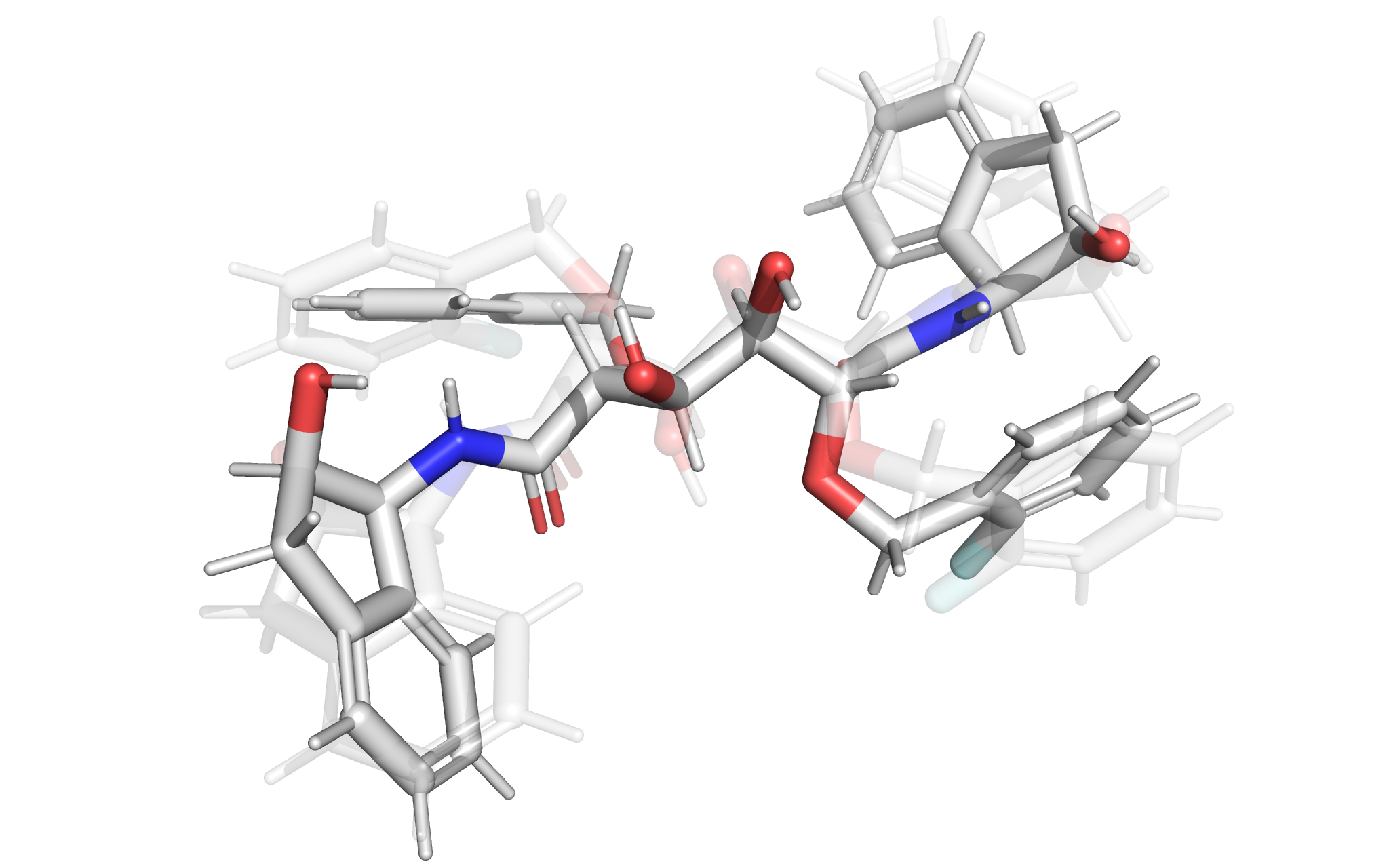}
    \caption{Improved conformer generation result after employing
    guided generation for the ligand
    contained in PDB structure 1EC0. 
    The experimenter conformer is presented as transparent.
    Overlaid is the best RMSD
    result of 1,000 attempts at generation using PIDM[QMugs] 
    in the deterministic scheme
    guided by a repulsive strength of $\eta=1$.}
\label{fig:1ec0_repul}
\end{figure}

The result is a modest improvement in RMSD statistics,
as shown in Fig.~\ref{fig:exper.repul} and Table~\ref{tab:exper.repul}.
If we revisit PDB system 1EC0, we observe an improvement in
matching the extended conformer observed in that structure
(Fig.~\ref{fig:1ec0_repul}).

\begin{table}[H]
    \centering
    \caption{RMSD statistics on the best out of 1,000
    conformers generated 
    with and without a repulsion term compared
    to experimental data 
    from the CSD and PDB. The model shown here
    is PIDM[QMugs] using the deterministic scheme.
    }
    \begin{threeparttable}
    \begin{tabular}{@{}lrr@{\extracolsep{12pt}}rr@{}}
    &\multicolumn{4}{c}{RMSD (\AA)} \\
    \cline{2-5}\\[\dimexpr-\normalbaselineskip+2pt]
    &\multicolumn{2}{c}{CSD}&\multicolumn{2}{c}{PDB} \\
    \cline{2-3}\cline{4-5}\\[\dimexpr-\normalbaselineskip+2pt]
    Conditions & Mean & Median & Mean & Median \\
    \hline\\[\dimexpr-\normalbaselineskip+2pt]
    Undirected 
    & 0.74 & 0.84
    & 0.90 & 0.98 \\
    $\eta=0.5$
    & 0.53 & 0.58
    & 0.70 & 0.75 \\
    $\eta=1$
    & 0.51 & 0.55
    & 0.67 & 0.73 \\
    \hline\\[\dimexpr-\normalbaselineskip+2pt]
    OMEGA\tnote{a}
    & 0.51 & 0.44 & 0.67 & 0.51 \\
    \hline
    \end{tabular}
    \begin{tablenotes}
        \item[a] Published statistics~\cite{hawkins_conformer_2010}
    \end{tablenotes}
    \end{threeparttable}
    \label{tab:exper.repul}
\end{table}

Including a simple repulsive term, however, comes with a cost. 
As shown in Table~\ref{tab:benchmark3}, error rates for  
chirality grow to as much as 6\%. Fortunately,
accuracy in bonded parameters such as bond length $d$ is only 
marginally affected. Interestingly enough, error rates for
cis/trans isomerism actually tend to slightly improve.
Clashes are almost entirely eliminated.

\begin{table}[H]
    \centering
    \caption{Bonded benchmark data for our models measured
    against the benchmark set of 15,763 molecules and generated
    with and without a repulsive term.
    Results for models trained on the QMugs and GEOM-drug
    data sets are shown for 500 steps in the deterministic scheme.
    }
    \begin{tabular*}{\linewidth}{@{}ll@{\extracolsep{\fill}}rrr@{\extracolsep{\fill}}rrr@{}}
\hline
& & \multicolumn{3}{c}{Mean absolute deviation} & \multicolumn{2}{c}{Inconsistency rate} & Clash rate\\\cline{3-5}\cline{6-7}
& & $d$ (\AA)& $\theta$ (rad) & $\phi$ (rad) & chirality & cis/trans & $<1.5$\AA \\
\hline\\[\dimexpr-\normalbaselineskip+2pt]
\multicolumn{2}{@{}l}{PIDM[QMugs]} &&&&\\
 Unguided & &
$0.0036$ &
$0.012$ &
$0.023$ &
$0.013$ &
$0.002$ &
$0.576$ \\
 $\eta=0.5$ & &
$0.0035$ &
$0.012$ &
$0.025$ &
$0.028$ &
$0.001$ &
$0.000$ \\
 $\eta=1$ & &
$0.0036$ &
$0.012$ &
$0.026$ &
$0.057$ &
$0.001$ &
$0.000$ \\
\hline\\[\dimexpr-\normalbaselineskip+2pt]
\multicolumn{2}{@{}l}{PIDM[GEOM-drugs]} &&&&\\
 Unguided & &
$0.0037$ &
$0.012$ &
$0.023$ &
$0.015$ &
$0.005$ &
$0.601$ \\
 $\eta=0.5$ & &
$0.0036$ &
$0.012$ &
$0.025$ &
$0.029$ &
$0.002$ &
$0.001$ \\
 $\eta=1$ & &
$0.0036$ &
$0.012$ &
$0.025$ &
$0.048$ &
$0.003$ &
$0.000$ \\
\hline
\end{tabular*}
    \label{tab:benchmark3}
\end{table}

\begin{figure}[H]
    \centering
    \includegraphics[width=0.7\textwidth]{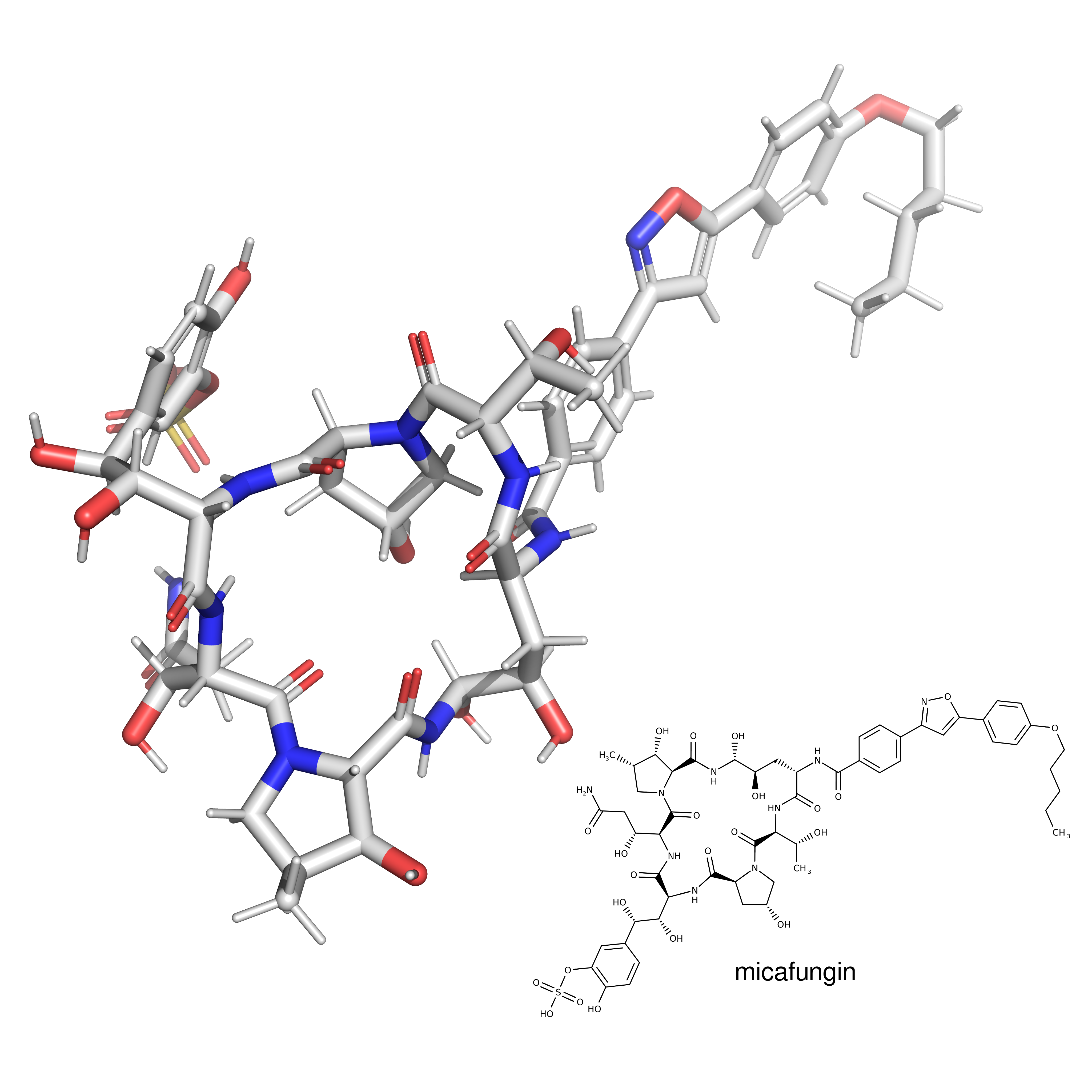}
    \caption{Micafungin is cyclic hexapeptide antifungal agent with a molecular
    weight of 1,270. A reasonable
    conformation can be generated if a repulsive force ($\eta$ = 1)
    is employed.
    }
\label{fig:micafungin}
\end{figure}

Adding a repulsive term can also helpful for generating conformers
for both long molecules
and macrocycles. One example is shown in Fig.~\ref{fig:micafungin}

\begin{figure}[H]
    \centering
    \includegraphics[width=\textwidth]{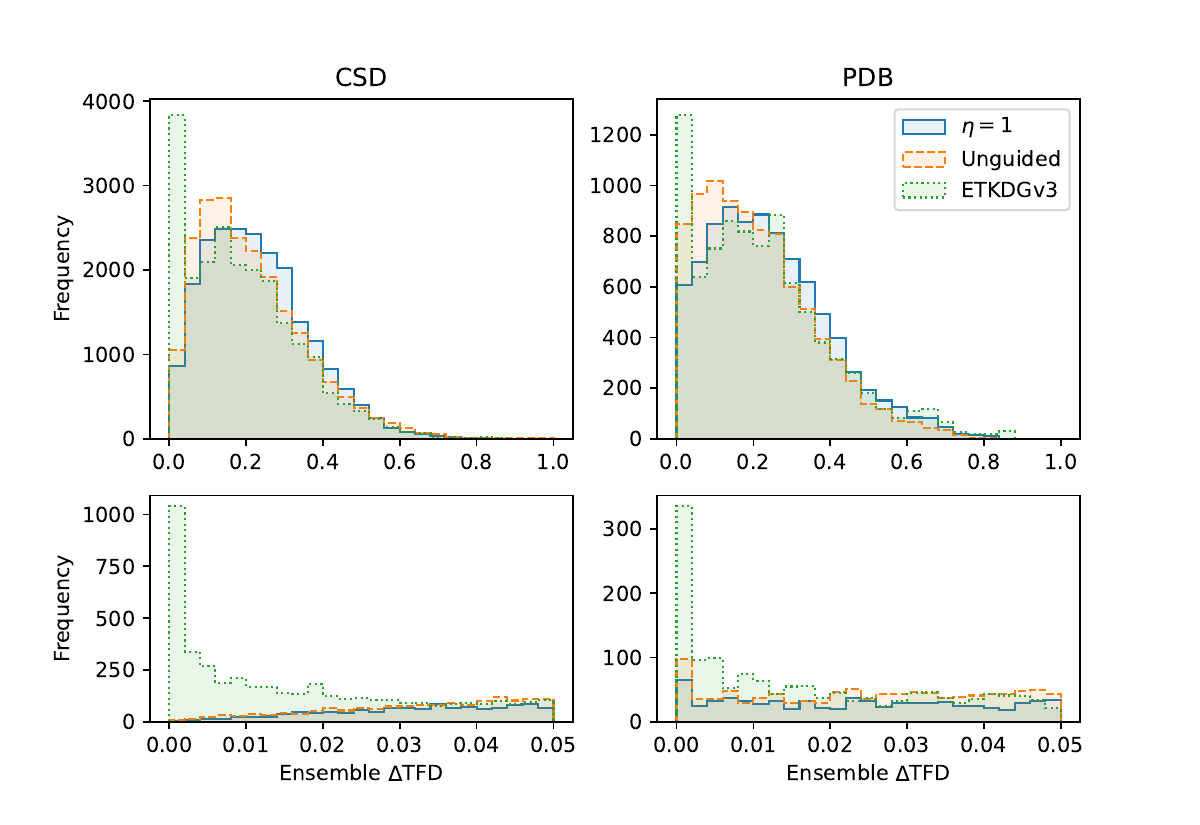}
    \caption{The difference in torsional fingerprint (TFD) values between 
    all pairs of ten conformers
    generated for the CSD and PDB experimental data sets.
    Results for PIDM[QMugs]
    are shown for 500 steps in the deterministic scheme
    and with or without a repulsive term. Results from
    RDKit are included for comparison.
    }
\label{fig:tfd}
\end{figure}

Besides largely eliminating overlapping atoms, an interesting question
is whether a repulsive term improves the overall sampling of
torsional freedom. To explore this question, one can use 
a torsional fingerprint such as TFD~\cite{schulz-gasch_tfd_2012}
to compare generated conformers against each other. Results using
the TFD implement of RDKit~\cite{rdkit}
are shown in Fig.~\ref{fig:tfd} for each pair of ten conformers
generated for the CSD and PDB data sets. Based on this criteria,
a repulsive term does
improve torsional sampling, but only marginally. 

The TFD results also indicate that it is rare for PIDM to generate
the same conformer twice, with or without guided generation. 
This is in contrast to ETKDGv3~\cite{riniker_better_2015}. The generation of
duplicate conformers by ETKDGv3 may not
be surprising since some of the molecules in the CSD and PDB data sets have a 
relatively small number of proper torsions, and in those cases torsional
space can be algorithmically exhausted even with just ten conformers.
Note that the RDKit implement of ETKDGv3 provides the option to remove duplicate
conformers based on a RMSD threshold, although
this option is disabled by default and was not used for the results in this work.

\section{Discussion}

Conformer generation is an established problem in computational,
structure-based drug discovery. Conventional solutions, which
have served the community well for decades, are based on
carefully tuned, hand-crafted algorithms. Machine learning
has made rapid advances recently, and so it
is natural to see how well the latest advancements in deep-learning 
can be applied to this space.
One of the goals of this current effort is to remain sensitive to 
prior work in computational chemistry, particularly in the
fields of cheminfomatics, classical force fields, and quantum simulation.

In this work, we combine some of the latest advancements in
diffusion-based, generative modeling with established techniques
used in classical force fields.
Classical force fields divide up energy contributions based
on subgraph topologies that naturally fit into graph convolution
models. They also conveniently divide energy contributions
between components that we value (bonded interactions) and
those that we deem less important (nonbonded interactions).

Our generative model is based on a denoising model constructed
from the sum of five separate, bonded terms. Each term is parameterized
by a simple MLP. By training all terms together, the denoising model
can account for correlated behavior. Accuracy is improved by
ensuring that each term is provided with sufficient information. 
For example, small rings
can influence the angle of any bends they contain. We also find
that parameterizing bonded terms as functions of atom distances is
more robust than using other geometric measures such as relative angles.

A critique of deep-learning is that the
internal structure of a trained model is often inscrutable (they are
so-called "black box" models)~\cite{rudin_stop_2019,blazek_explainable_2021}.
By constructing our denoising model from the sum of bonded components,
we can probe the response of each component separately and
explore how the model learned to solve the denoising problem.
By using different input molecules as a probe, the solutions
for bond lengths, bend angles, and torsion angles can be explored
for specific chemical groups. Some examples are presented above, and
further are available in the Supporting Information. This ability
to test each bonded component under controlled conditions
was invaluable during development.

We have used two separate data sets (QMugs and GEOM-drugs)
for training. Both include conformers
of drug-like molecules
optimized in vacuum using GFN2-xTB, a semiempirical quantum mechanical
calculation. This type of calculation 
is expected to provide bonded parameters more accurate
than available from classical force fields. During training, the losses
calculated on holdout validation data sets show no evidence of 
increasing, suggesting little overtraining. This loss also matches
the loss from the training subset, implying a level of
transferability. Remarkably, despite little overlap in the two
training sets, the resulting models perform nearly identically on
all benchmarks. Probes indicate that both models have learned
similar solutions.

For generation, we numerically solve for the probability flow ODE,
using a form of oversampling. Both deterministic and stochastic generation
are tested, and the former outperforms the latter for most criteria.
Generation in as little as 100 steps appears sufficient for conformers
of reasonable quality, but using more steps noticeably improves
performance.

To test the accuracy of generated conformers, we use an independent
subset of the QMugs data set. We used Tanimoto similarity
to ensure the molecules in this subset had little
structural overlap with the training subsets. Based on the
median absolute deviation, generated conformers have more
accurate bond lengths and bond angles than established solutions,
at least on average. Much of the reason is likely due to
training on structures optimized by GFN2-xTB, whereas established conformer 
solutions rely, in part, on force field parameterization, such as MMFF94.

Average performance on proper torsion angle is also competitive.
It is interesting that another published diffusion model, GeoDiff,
performs quite poorly on torsional angles. GeoDiff also requires
ten times the amount of generation steps and contains about six times
as many parameters.

Molecule conformers for drug-like compounds typically
achieve a large part of their diversity from torsional freedom. 
Outside of macrocycles, dihedral angles
are relatively easy to manipulate, which is why they are among the
degrees of freedom commonly optimized in ligand-protein docking 
algorithms. In such applications, the focus of a conformer
generator should be the production of quality bonded parameters,
such as bond lengths, bend angles, and ring conformations.

Conformer RMSD is sensitive to dihedral angles. This makes the
RMSD a poor metric for evaluating the quality of generated
conformers, unless the goal is to reproduce the selection of dihedral angles.
Gauging the quality of a conformer generator based on reproducing
the RMSD of synthetic data sets, such as QMugs and GEOM-drugs,
is merely testing if the generator can reproduce the somewhat
arbitrary choices the authors of these
data sets employed for dihedral angle sampling.

Measuring RMSD against experimental data does have value,
but data is limited in size and resolution and 
restricted to certain physical conditions, such as crystal solids. 
For training, limited statistics is a problem
for deep-learning techniques which rely on
a large and diverse sample of training data. 

We used experimental
data to explore the dihedral space sampled by our
conformer generation model and have uncovered some deficiencies.
This is a problem for applications that do not 
independently sample dihedral space. This deficiency can
be partially mitigated by introducing a form of guided
generation. We tested a simple technique that applied
a universal repulsion between nonbonded atom 
pairs. One could imagine more complex guided techniques
that are capable of sampling dihedral angles appropriate
for any number of molecule environments or conditions.

The atom encoding used in our model was generated
using a Graph Attention Transformer 
(GATv2)~\cite{brody_how_2022}. One could
substitute other types of comparable message-passing graph networks, 
such as a Graph Isomorphism Network (GIN)~\cite{xu2018powerful},
and probably obtain comparable results. We suspect that
using a type of graph network that can capture the same
quality of chemical information while also recognizing
cycles would be an improvement. Such a graph model 
would not only allow the removal of the ad-hoc ring embedding we 
use for the bend angle component, but also provide
a mechanism for correcting bond lengths and torsion angles
that are also impacted by rings.

Other challenges remain. There are some molecules that fail
to generate correctly, such as atorvastatin.
Because our denoising model is based on local molecular
structure (via bonded components), large scale movements, such
as thoses needed to swing an arm of a molecule to 
ensure planarity for an aromatic ring, are difficult to 
accommodate.

On average, the generative model presented here reproduces bonded
parameters with an accuracy comparable to conventional methods (such as OMEGA 
and ETKDGv3), with perhaps the exception of dihedral angle
sampling. This is accomplished without relying on preestablished constraints,
template libraries, or external parameterizations such as classical force fields.
Extending this model to additional atom types should only
require the construction of a suitably representative
molecule data set for training that can be prepared using GFN2-xTB. 
It should also be possible
to employ masking techniques, widely available
for images~\cite{liu2020rethinking,lugmayr_repaint_2022,kirillov_segment_2023}, 
to extend molecule conformer generation
to tasks such as fragment 
bridging~\cite{lauri_caveat_1994,thompson_confirm_2008,imrie_deep_2020} 
and fragment
screening~\cite{kumar_fragment_2012,jacquemard_bright_2019}.

\section{Conclusion}

Presented is a physics-inspired, diffusion-based model
for molecule conformer generation inspired by the
bonded components of classical force fields. Parameters
were trained on high-quality conformers from the QMugs
and GEOM-drugs data sets. Learning appears robust,
transferable, and explainable. Both deterministic and stochastic
generation schemes are demonstrated.
Average performance on the reproduction
of bonded parameters exceeds conventional
conformer generation tools. A simple example of guided
generation is successful at improving dihedral sampling
when compared to experimental data.

\section{Data and Software Availability}

A complete model implementation is available in source form
at https://github.com/nobiastx/diffusion-conformer. Included
are model checkpoints and scripts for training and inference.
The training and benchmark data (QMugs, GEOM-drugs, CSD, and PDB) are available
from public sources.

\begin{suppinfo}
    
    The following files are available free of charge:
    \begin{itemize}
      \item Figs.~S1--S17, additional model probes, Figs.~S18--S23, additional
            example conformer output.
      \item Video of example generation of cholesterol, using PIDM[QMugs] and 500 steps in the deterministic scheme.
      \item Video of example generation of naphthacene, same conditions.
      \item Video of example generation of artemether, same conditions.
      \item Video of example generation of penicillin, same conditions.
      \item Molecular structure file containing the conformer output for Fig.~\ref{fig:render01}
from the main text and Figs. S18--S23 from the Supporting Information.
    \end{itemize}
    
\end{suppinfo}

\bibliography{refs}

\section{Supporting Information}

\section{Additional model probes}

The bonded components of any molecule can be explored.
The following are selected as representative. It is also informative
to compare PIDM[QMugs] and PIDM[GEOM-drugs]. 
Interested readers are encouraged to perform their own experiments
and source code is provided for this purpose.

Each of the figures below include four plots arranged in a grid.
The left (right) column is reserved for PIDM[QMugs] (PIDM[GEOM-drugs]).
Each row corresponds to one of a pair of atoms for the bond,
bend, proper torsion, and chirality probes, or a pair of atoms
for the cis-trans probe. The identity of
the atoms are highlighted by the circle(s) 
in the molecule depiction on the left. 

Each plot includes corrections for various values of $\sigma$.
The vertical gray line is the expected value, from a GFN2-xTB
optimization.

\subsection{Bond component}

\begin{figure}[H]
    \centering
    \includegraphics[width=\textwidth]{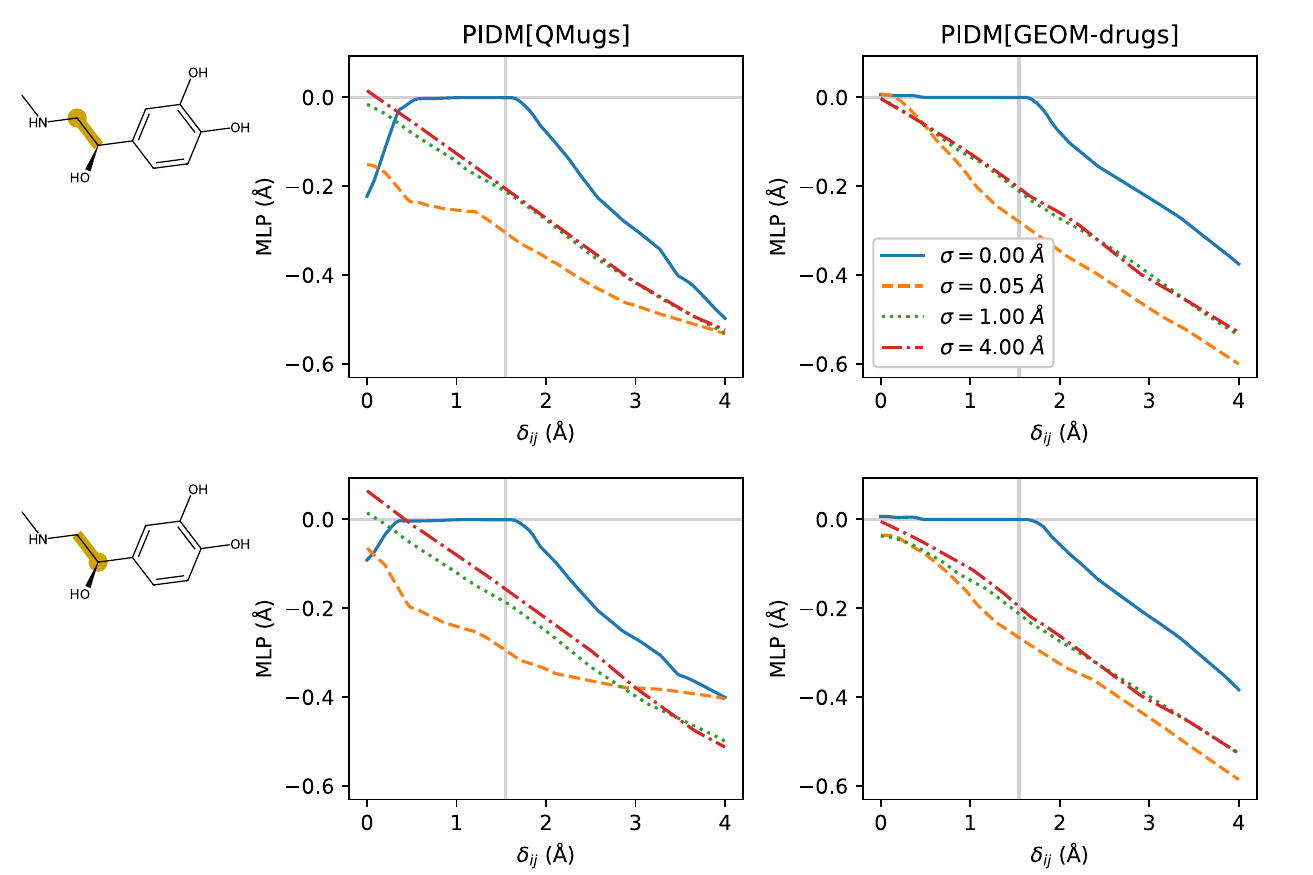}
    \caption{Example bond correction for the two atoms involved in an alkane 
    bond in adrenaline. 
    }
\end{figure}

\begin{figure}[H]
    \centering
    \includegraphics[width=\textwidth]{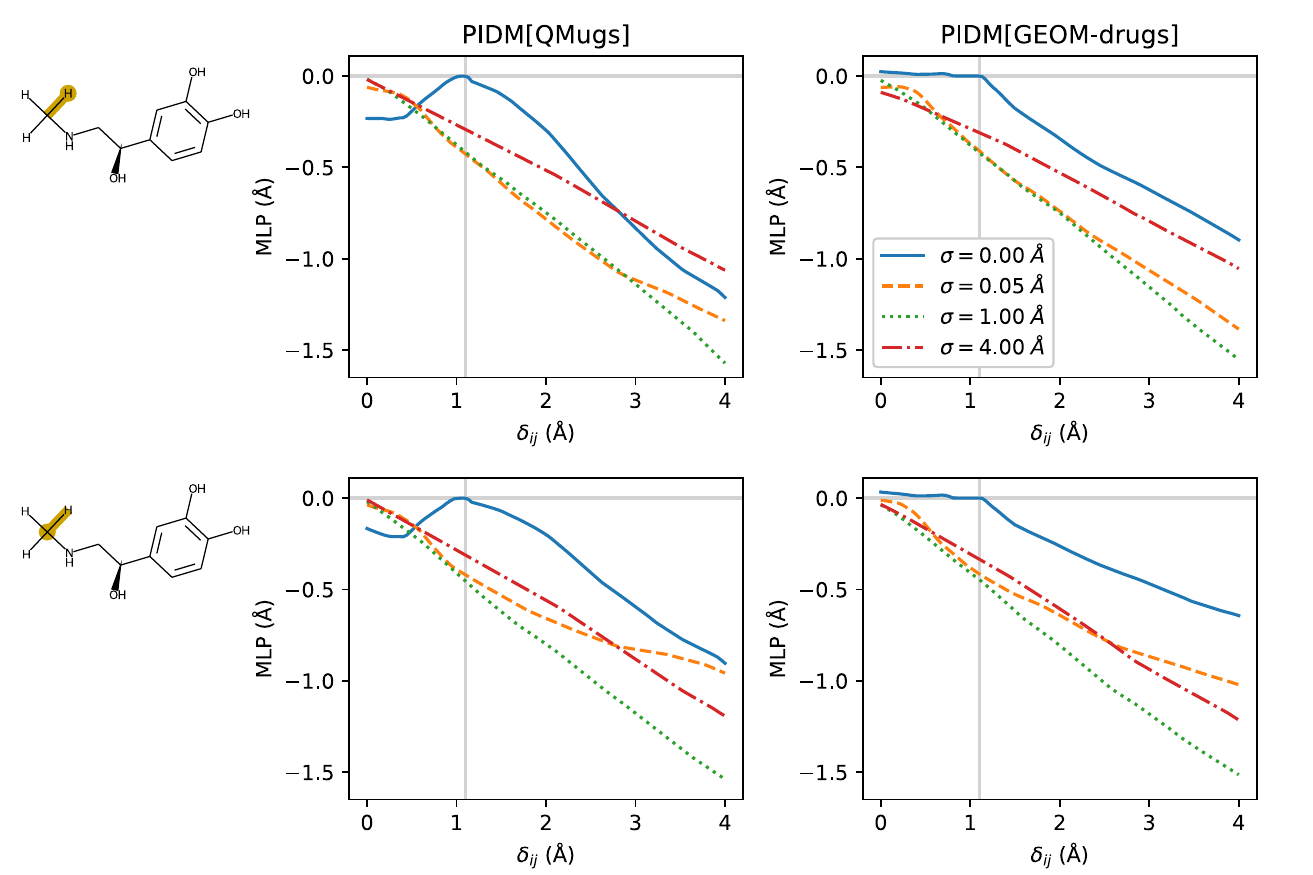}
    \caption{Example bond correction for one of the hydrogens (top row)
    on a primary carbon (bottom row) 
    in adrenaline. 
    }
\end{figure}

\begin{figure}[H]
    \centering
    \includegraphics[width=\textwidth]{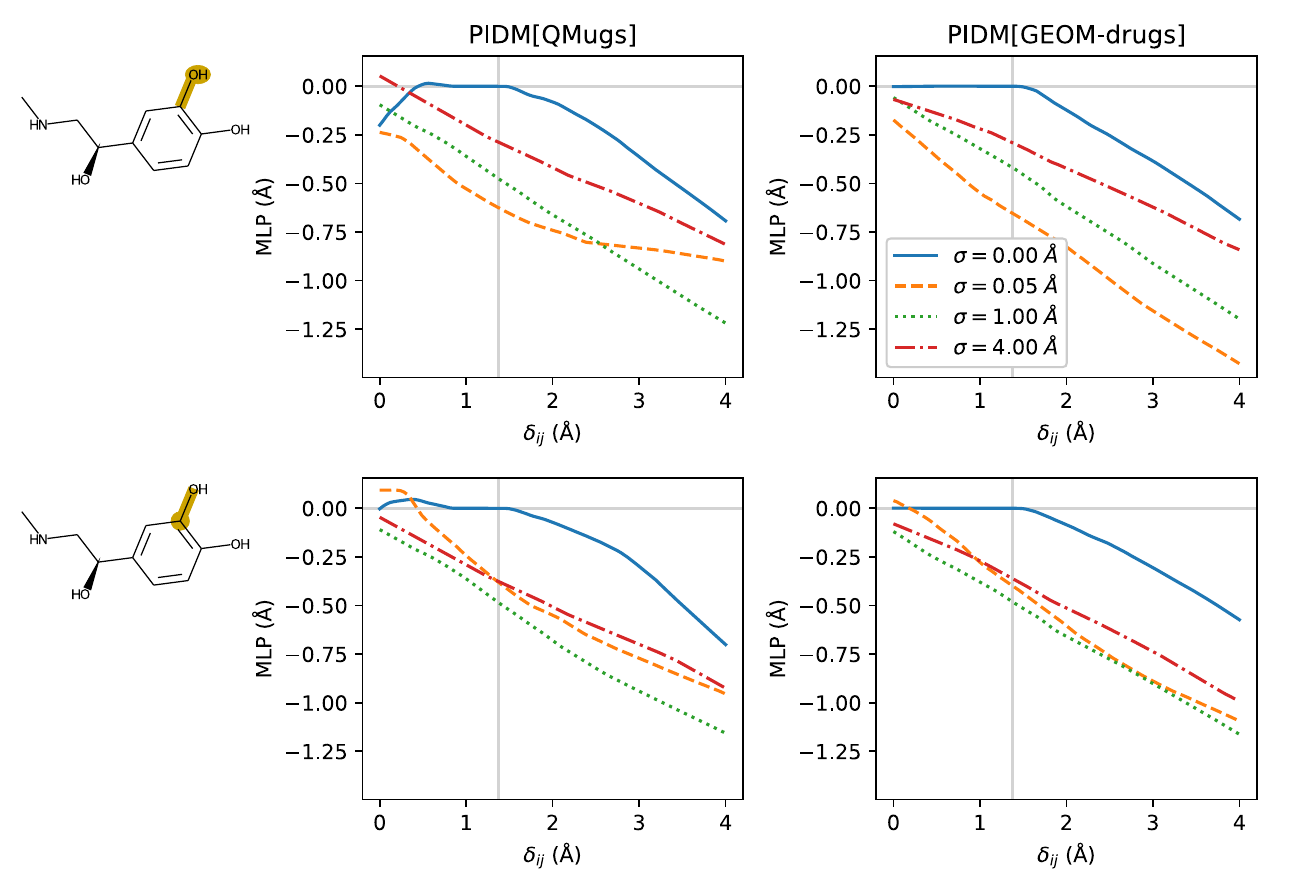}
    \caption{Example bond correction for the oxygen (top row) and carbon (bottom row)
    in a hydroxyl in adrenaline. 
    }
\end{figure}

\begin{figure}[H]
    \centering
    \includegraphics[width=\textwidth]{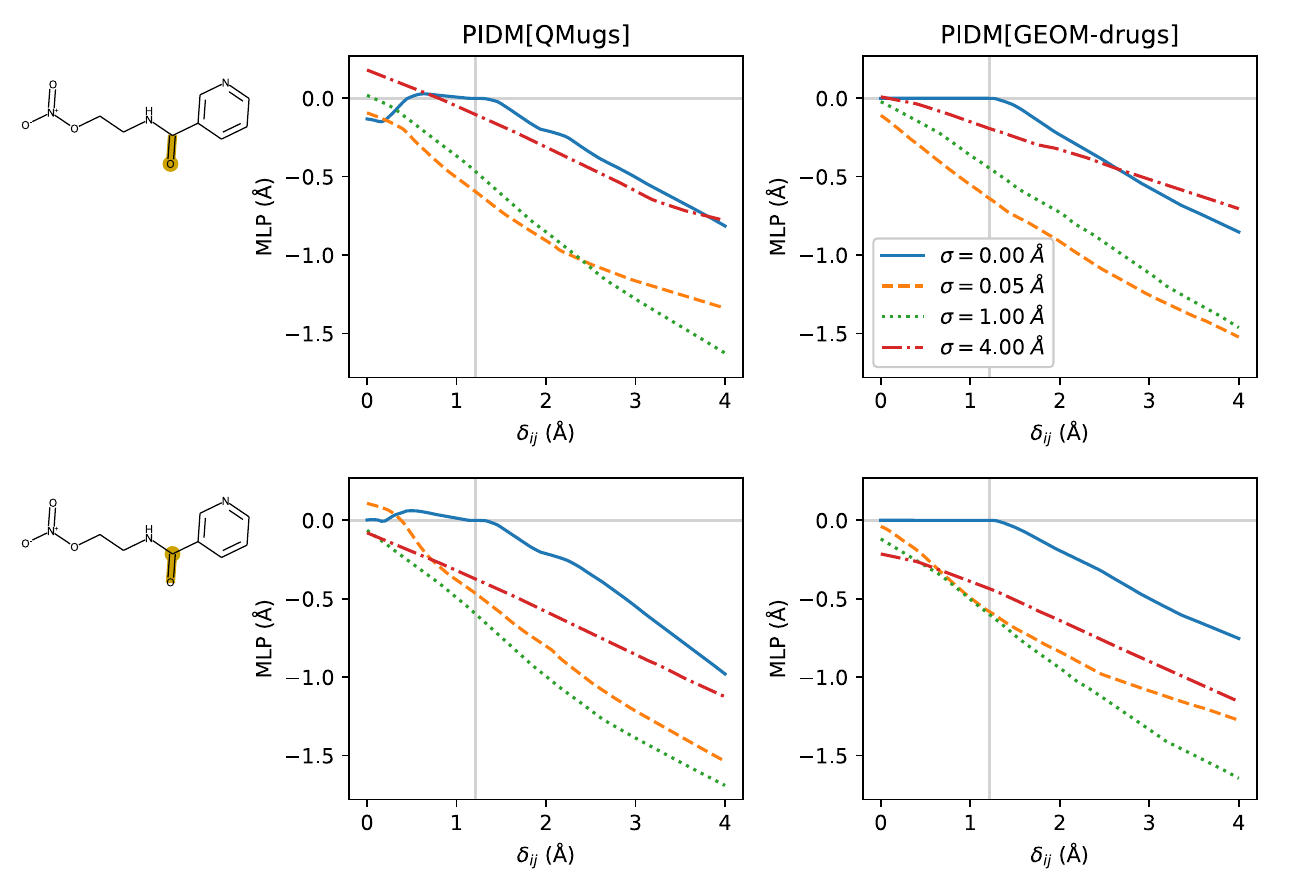}
    \caption{Example bond correction for the oxygen (top row) and carbon (bottom row)
    in an amide in nicorandil. 
    }
\end{figure}

\subsection{Bend component}

\begin{figure}[H]
    \centering
    \includegraphics[width=\textwidth]{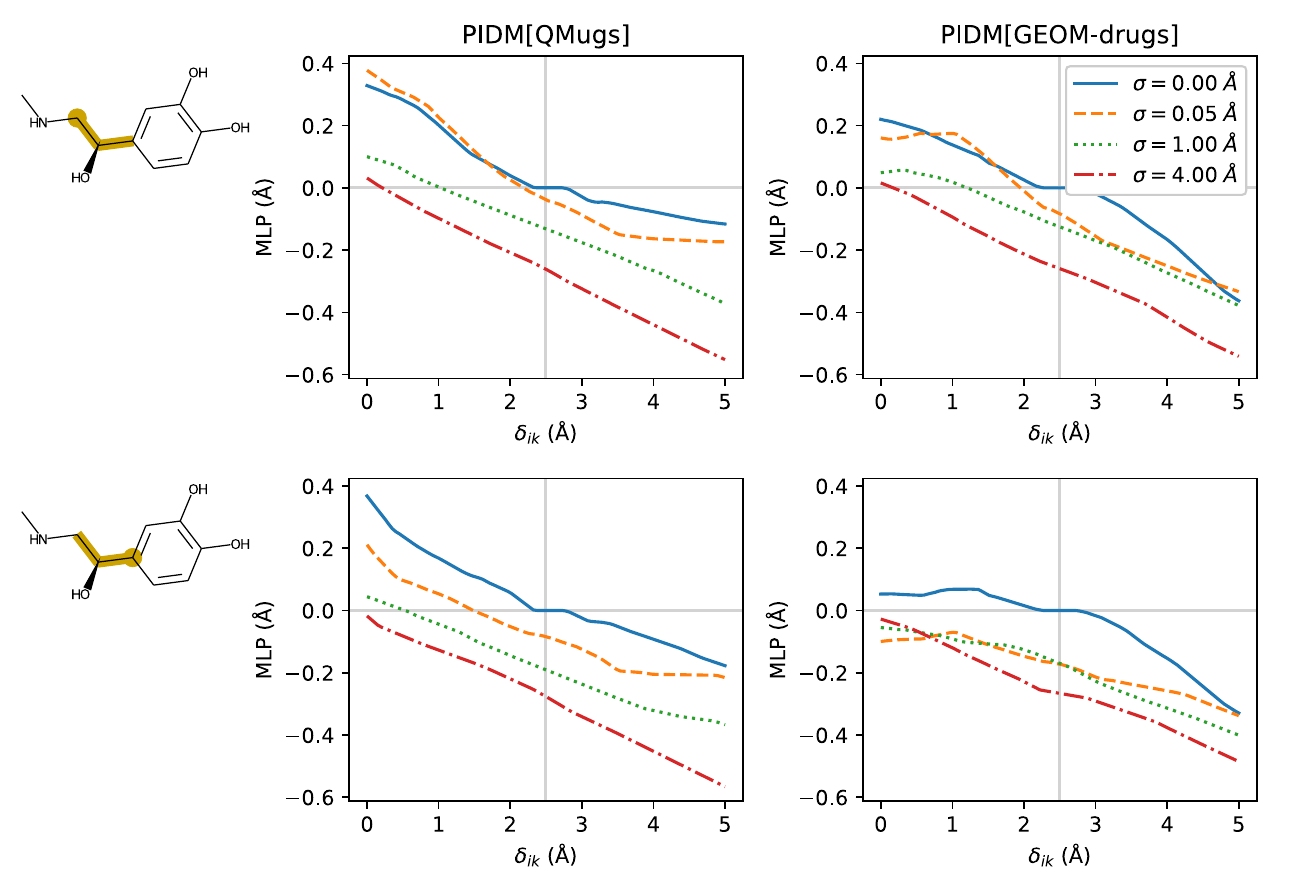}
    \caption{Example bend angle correction for an alkane in adrenaline. 
    }
\end{figure}

\begin{figure}[H]
    \centering
    \includegraphics[width=\textwidth]{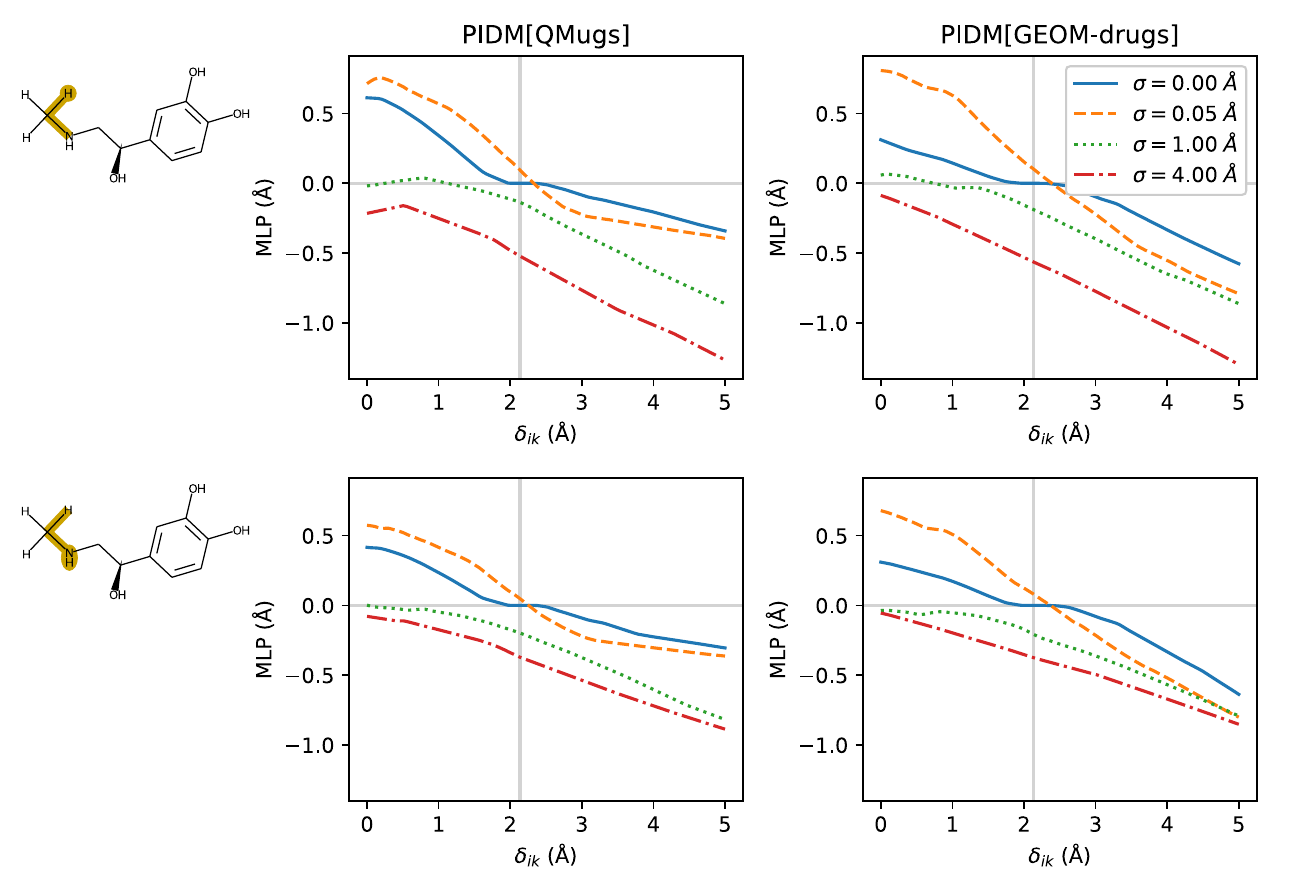}
    \caption{Example bend angle correction for hydrogen and nitrogen
    in the methylamine in adrenaline. 
    }
\end{figure}

\begin{figure}[H]
    \centering
    \includegraphics[width=\textwidth]{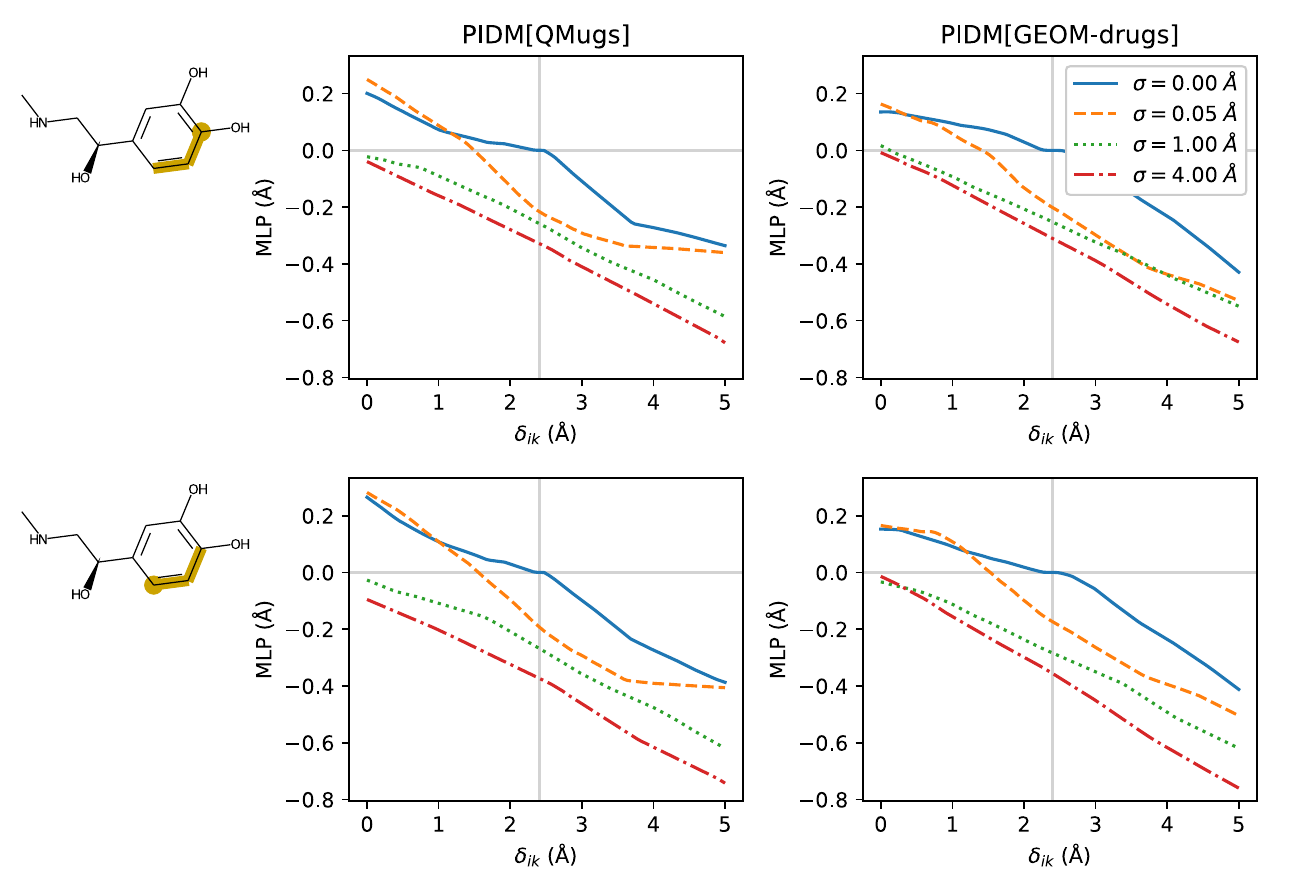}
    \caption{Example bend angle correction for carbons in the phenol
    ring in adrenaline. 
    }
\end{figure}

\begin{figure}[H]
    \centering
    \includegraphics[width=\textwidth]{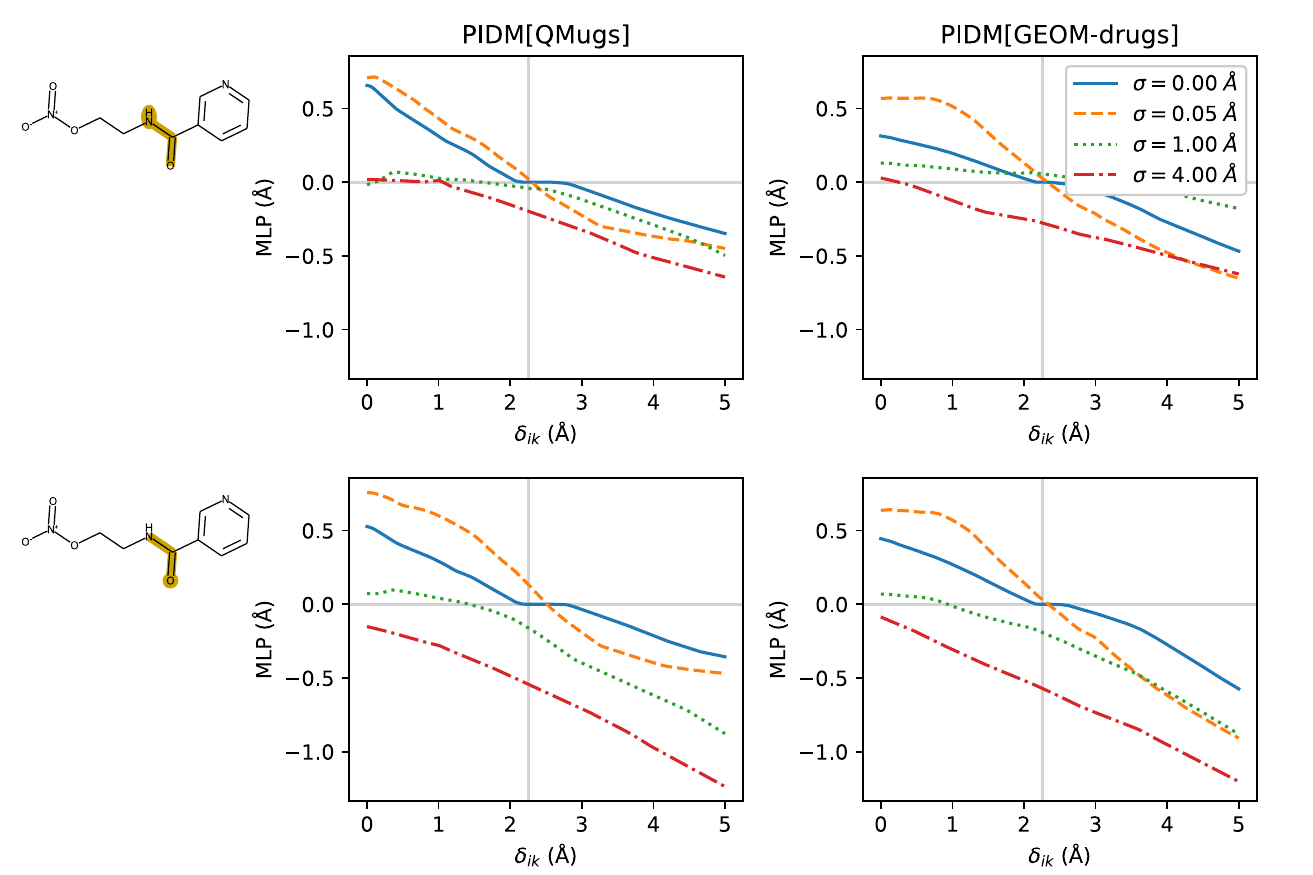}
    \caption{Example bend angle correction for the amide in nicorandil. 
    }
\end{figure}

\begin{figure}[H]
    \centering
    \includegraphics[width=\textwidth]{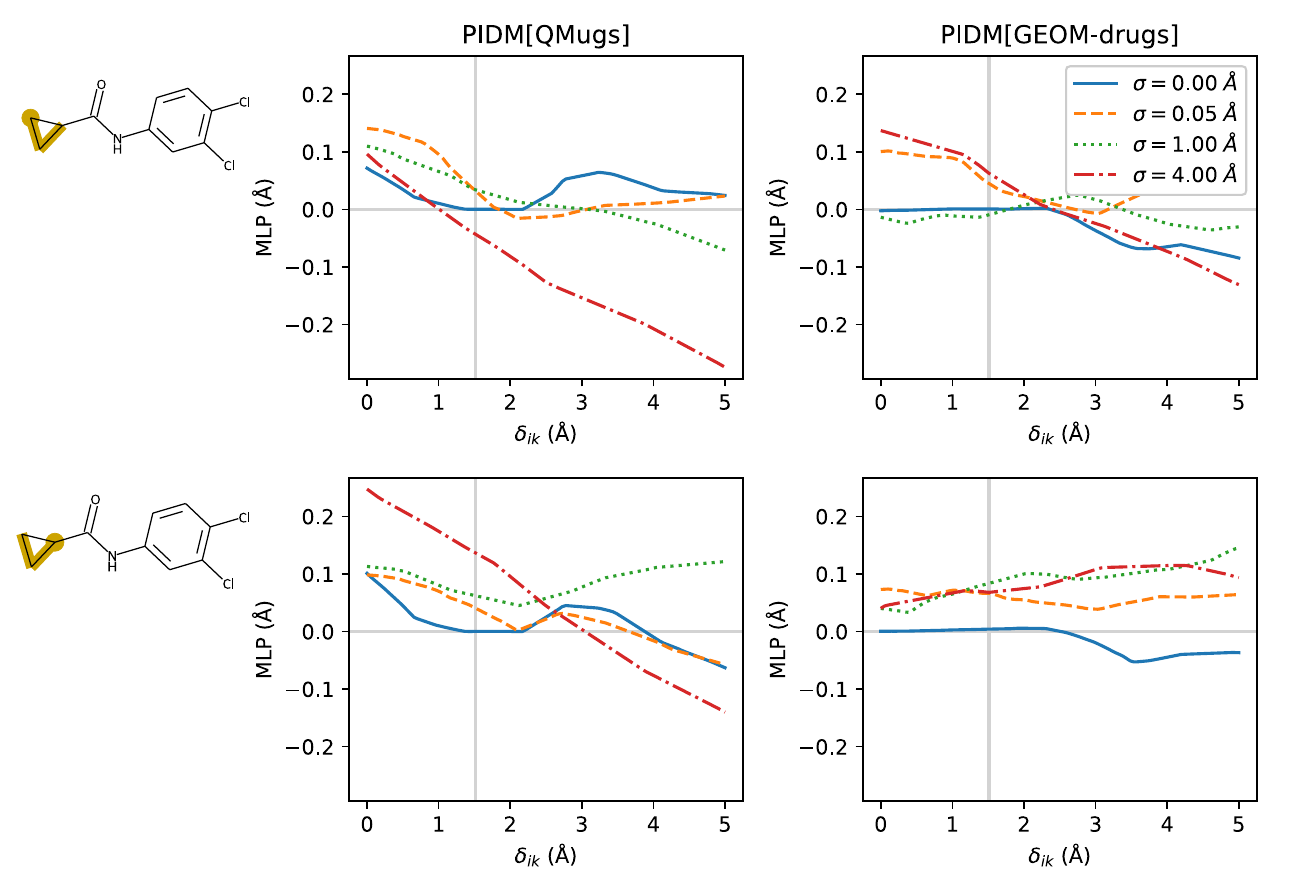}
    \caption{Example bend angle correction for carbons in the cyclopropyl
    of cypromid. 
    }
\end{figure}

\subsection{Proper torsion component}

\begin{figure}[H]
    \centering
    \includegraphics[width=\textwidth]{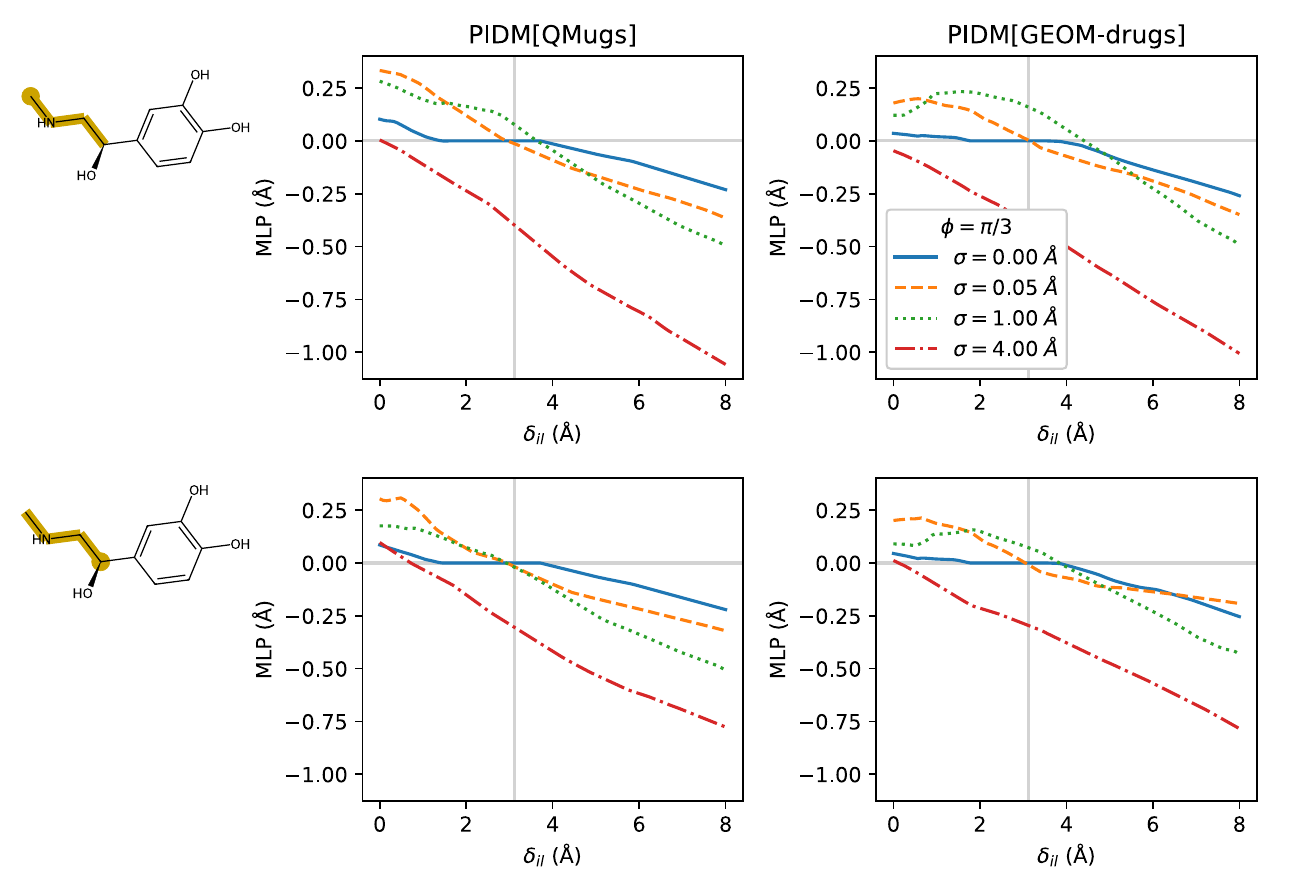}
    \caption{Example proper torsion angle correction for the methylethylamine in adrenaline. 
    }
\end{figure}

\begin{figure}[H]
    \centering
    \includegraphics[width=\textwidth]{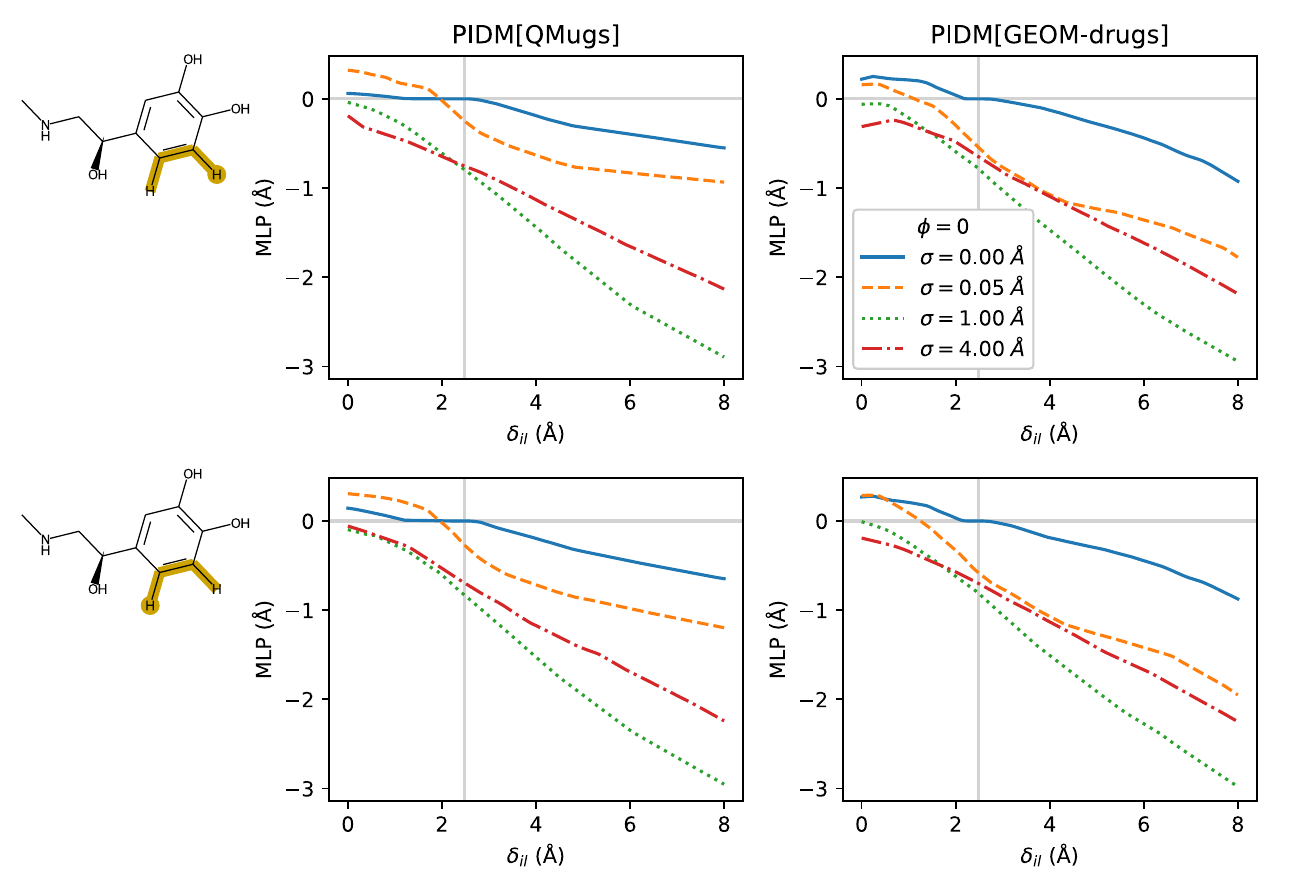}
    \caption{Example proper torsion angle correction for two hydrogens in the phenol
    ring in adrenaline. 
    }
\end{figure}

\begin{figure}[H]
    \centering
    \includegraphics[width=\textwidth]{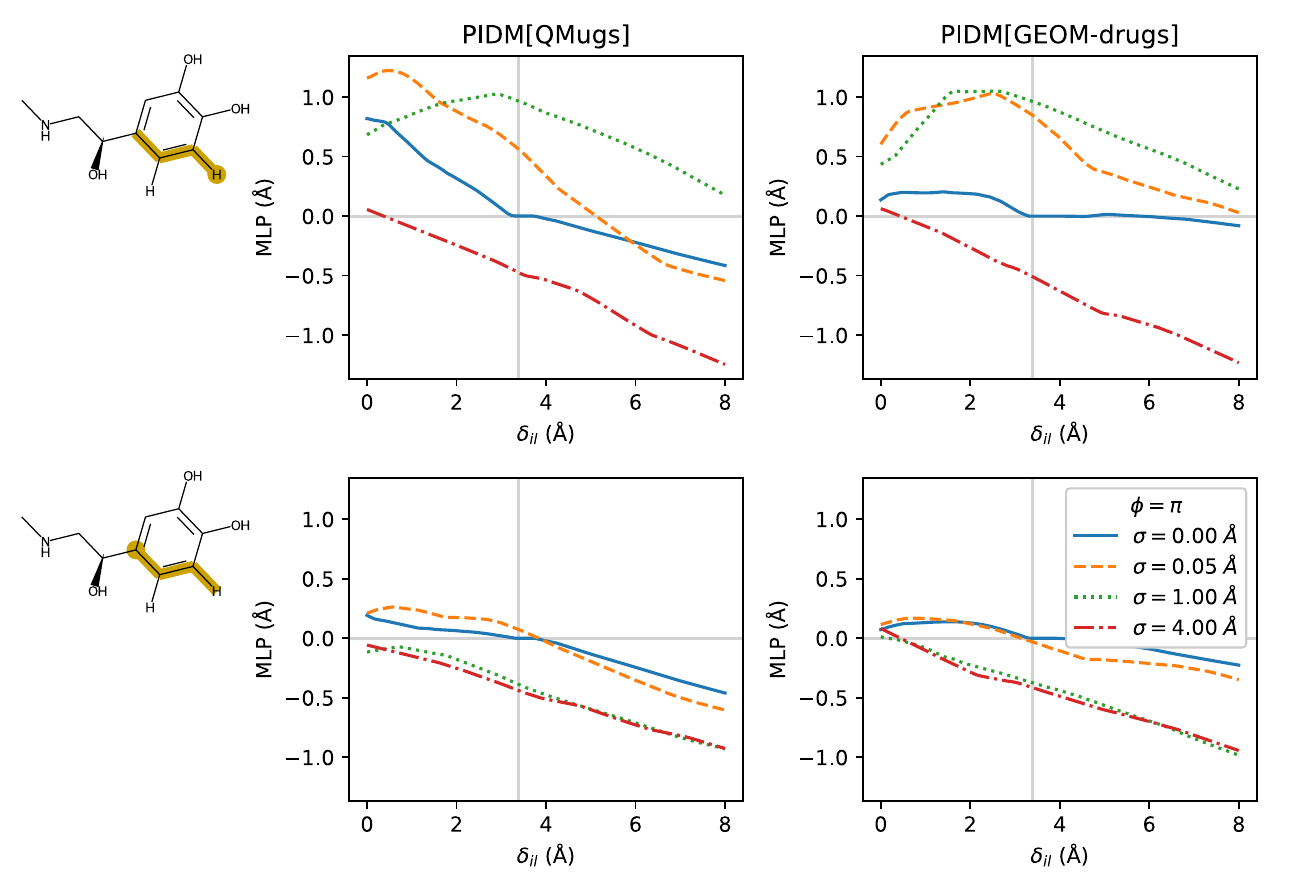}
    \caption{Example proper torsion angle correction for a hydrogen and a carbon in the phenol
    ring in adrenaline. 
    }
\end{figure}

\begin{figure}[H]
    \centering
    \includegraphics[width=\textwidth]{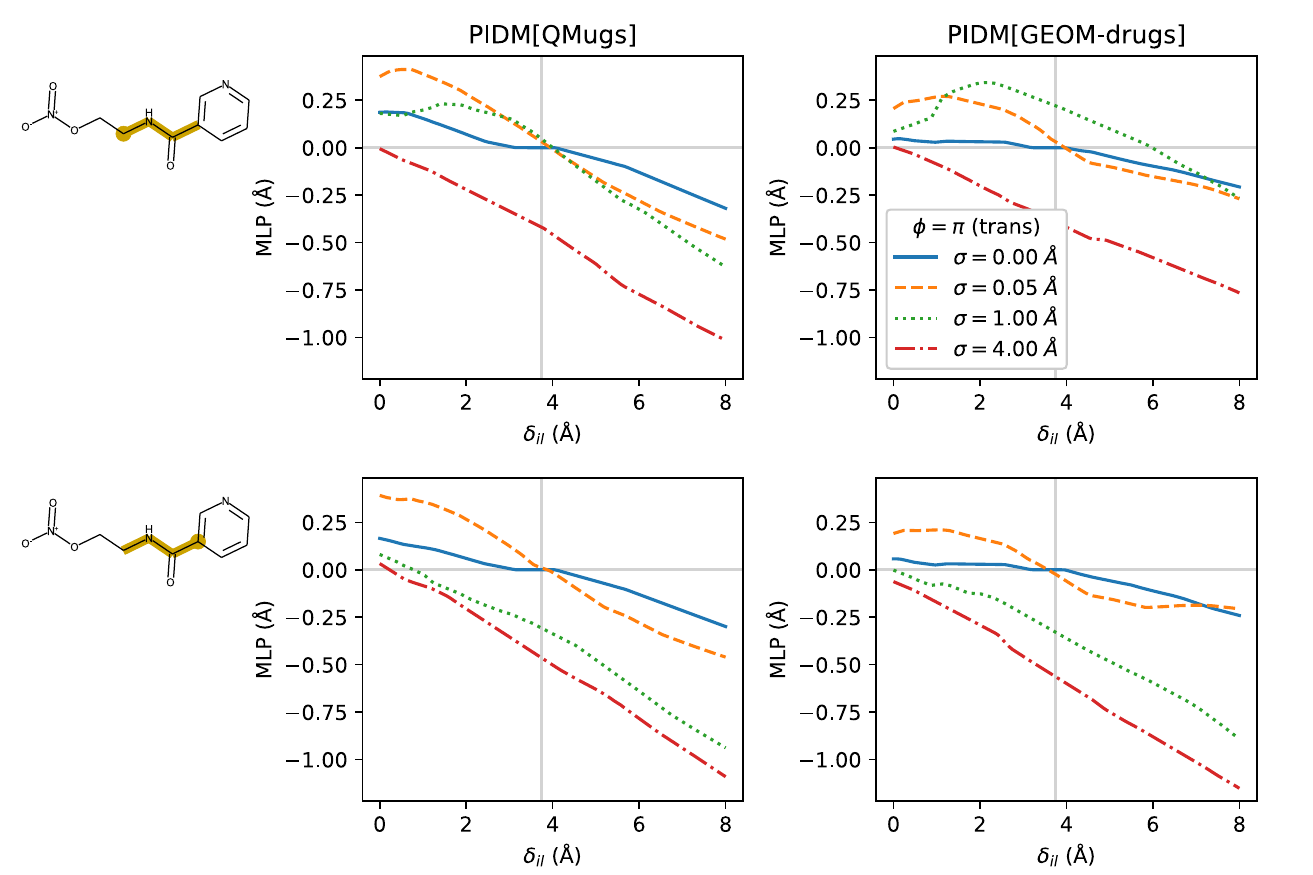}
    \caption{Example proper torsion angle correction for an amide 
    in nicorandil in the trans configuration. 
    }
\end{figure}

\begin{figure}[H]
    \centering
    \includegraphics[width=\textwidth]{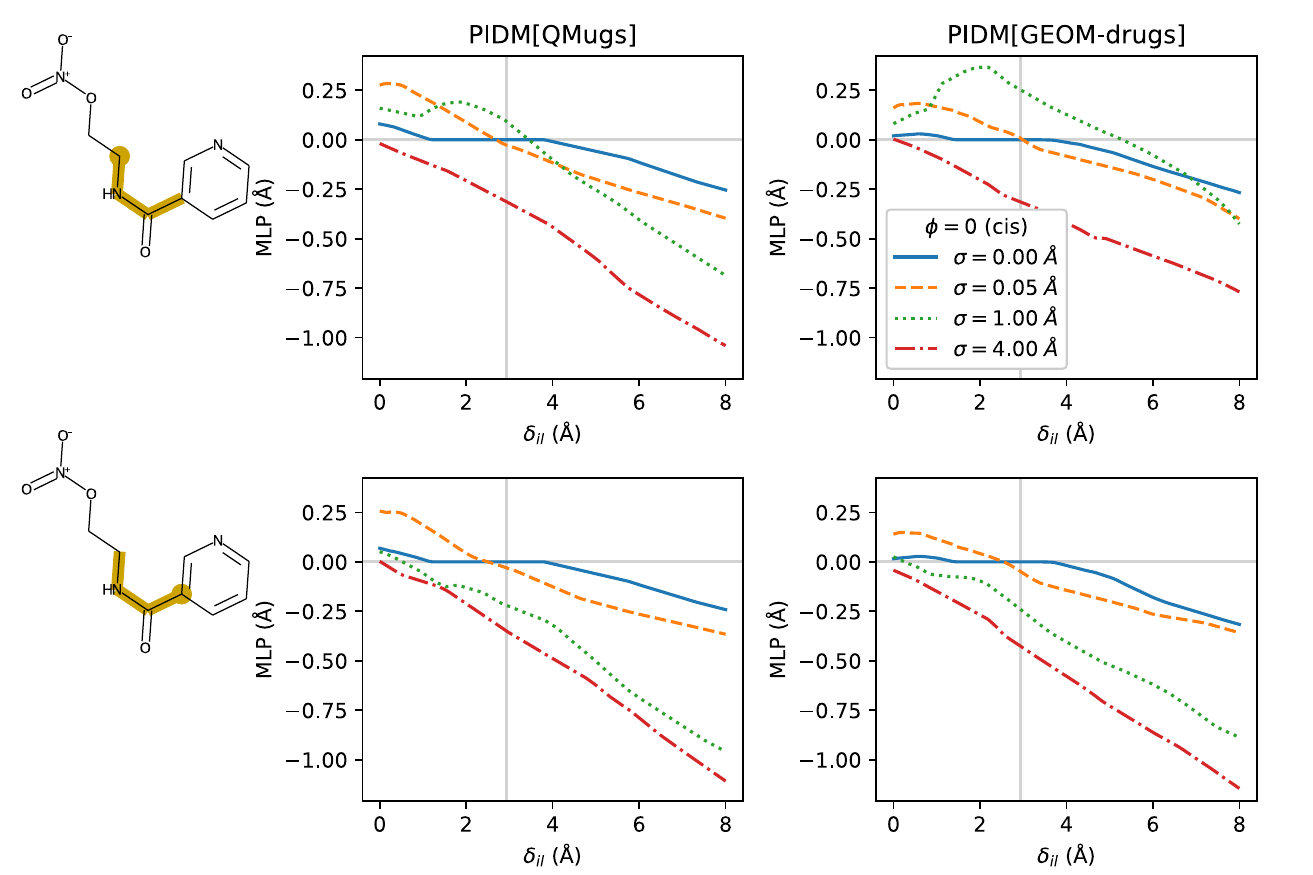}
    \caption{Example proper torsion angle correction for an amide 
    in nicorandil in the cis configuration.
    }
\end{figure}

\subsection{Chirality component}

\begin{figure}[H]
    \centering
    \includegraphics[width=\textwidth]{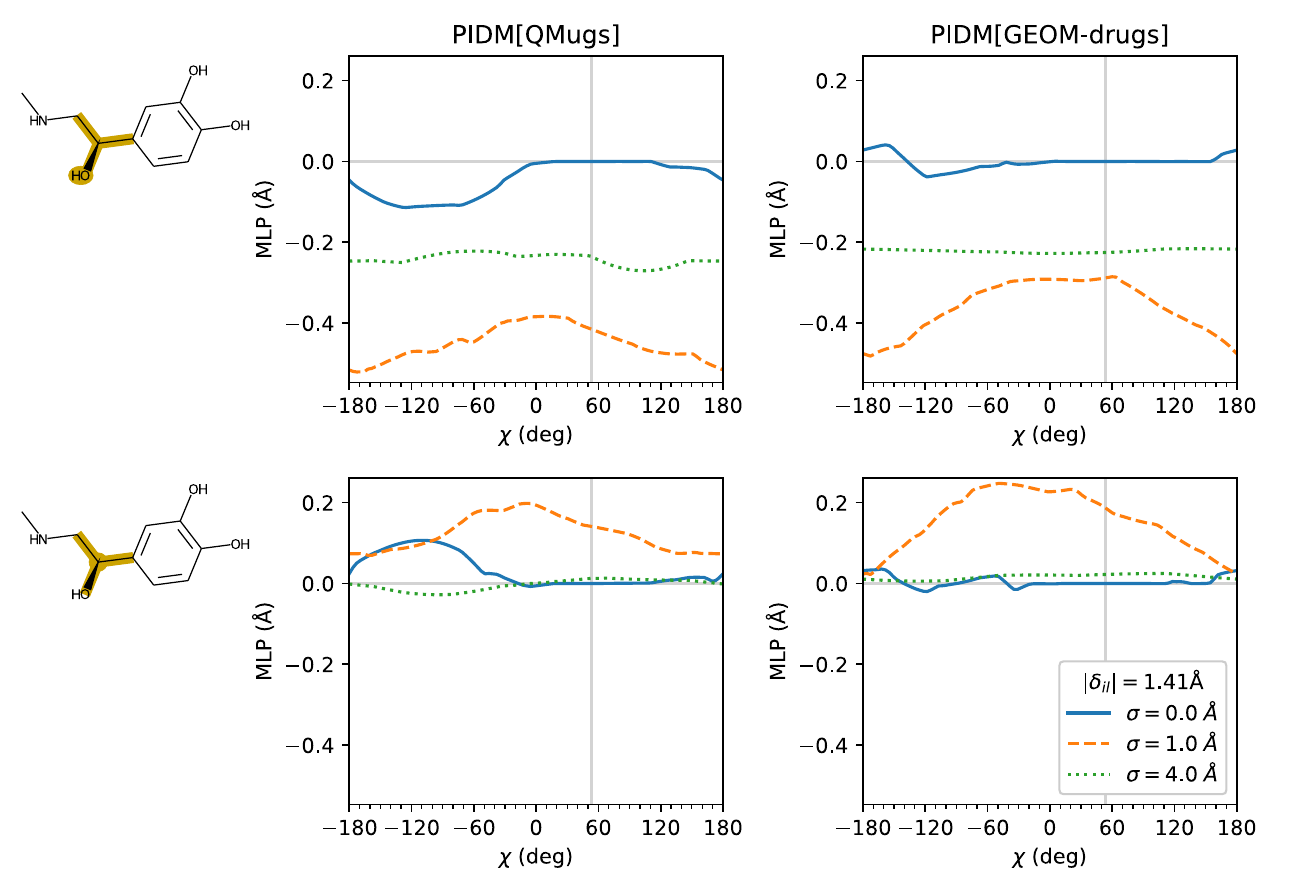}
    \caption{Example chirality correction for the hydroxyl on the 
    chiral center of adrenaline. The chirality angle $\chi$ is sampled at a
    bond length of 1.41\AA{}, corresponding to the GFN2-xTB prediction.
    }
\end{figure}

\subsection{Cis-trans component}

\begin{figure}[H]
    \centering
    \includegraphics[width=\textwidth]{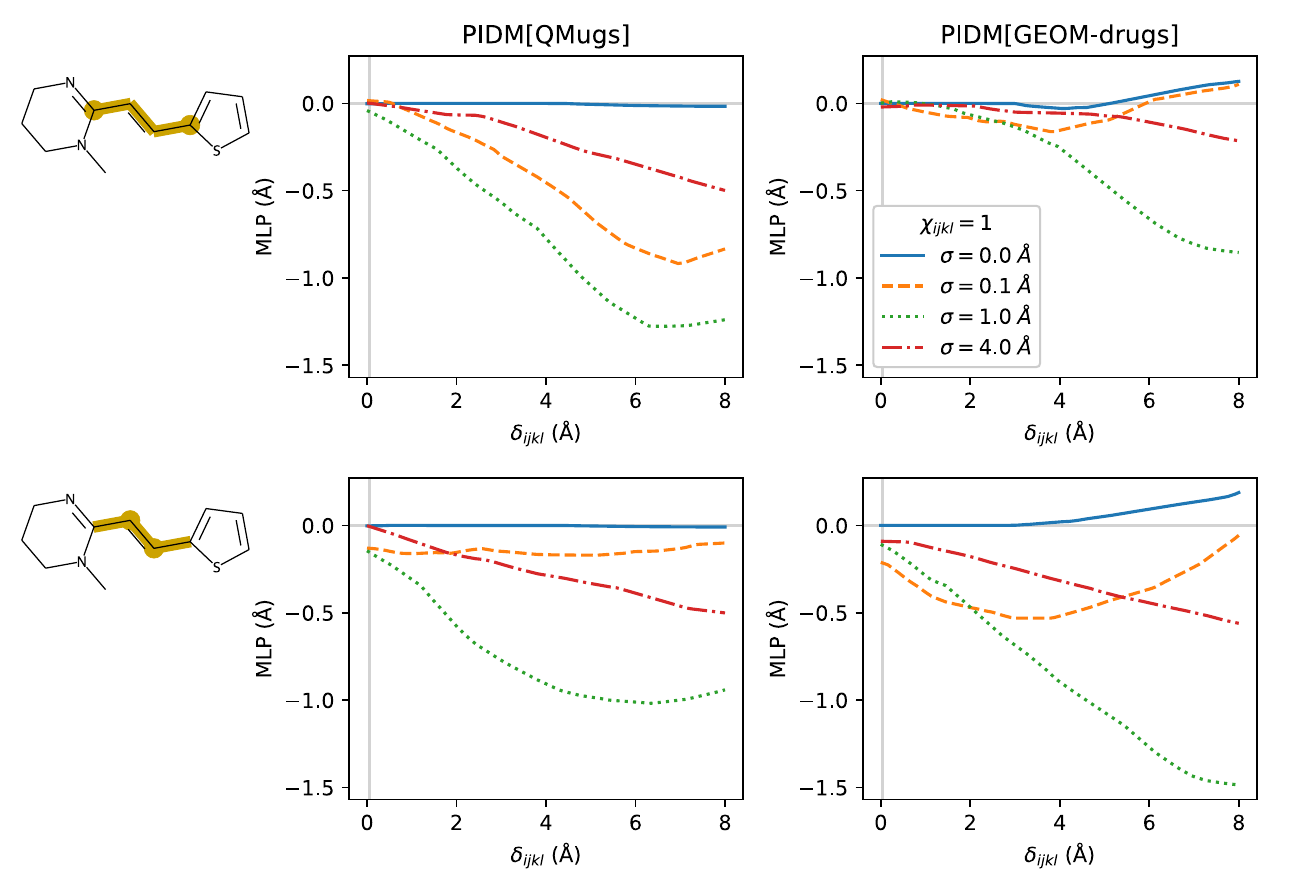}
    \caption{Example cis-trans correction for the carbons on the 
    double bond of pyrantel.
    }
\end{figure}

\begin{figure}[H]
    \centering
    \includegraphics[width=\textwidth]{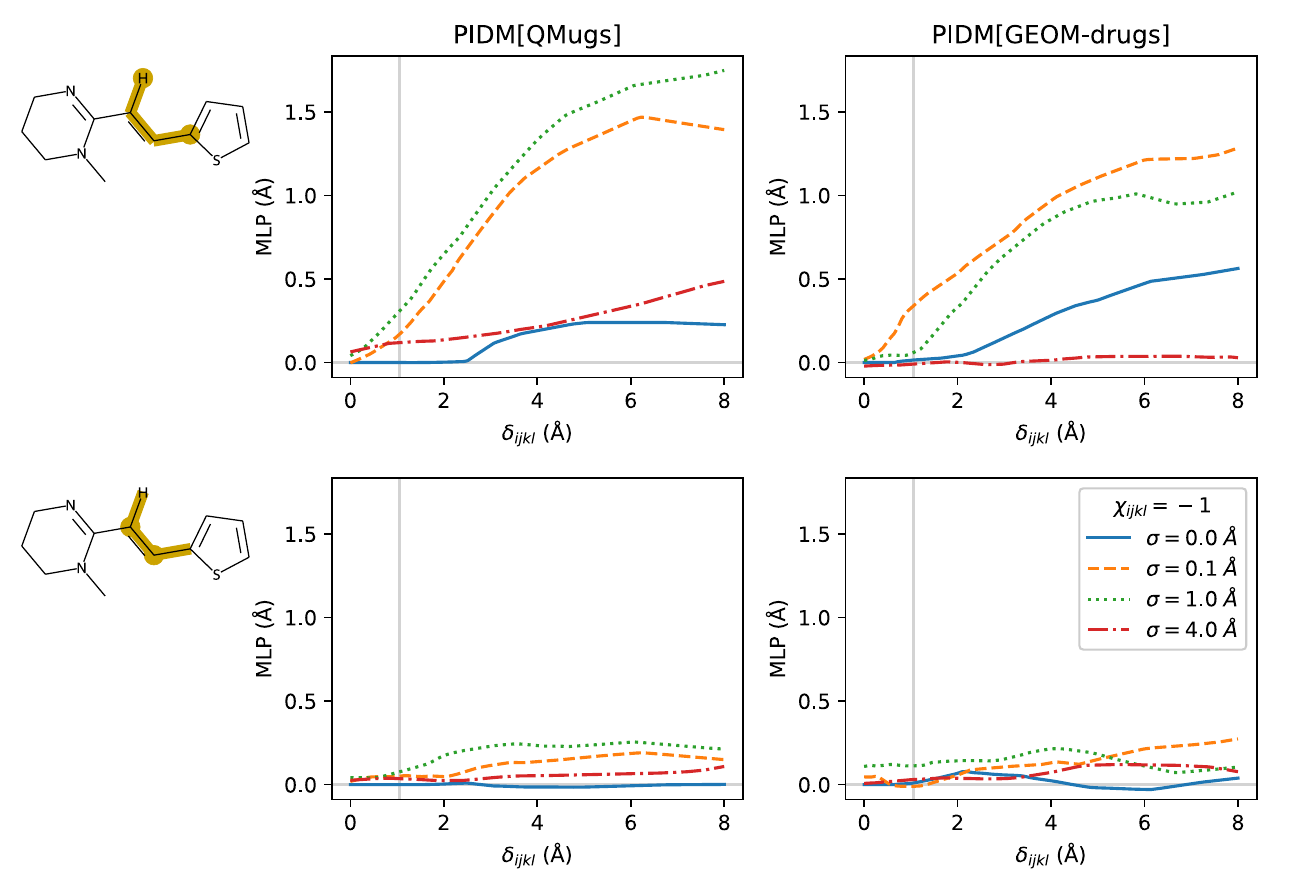}
    \caption{Example cis-trans correction for a hydrogen on the 
    double bond of pyrantel.
    }
\end{figure}

\section{Additional example output}

The following are conformers generated from molecules selected randomly
from the benchmark dataset. No filtering has been applied.
In all of these figures, the second column 
from left (grey background) is the first conformer from QMugs. Shown on
the right are the first four conformers generated by PIDM.
All molecule renderings are oriented by principal component.

\begin{figure}[H]
    \centering
    \centerline{\includegraphics[width=1.2\textwidth]{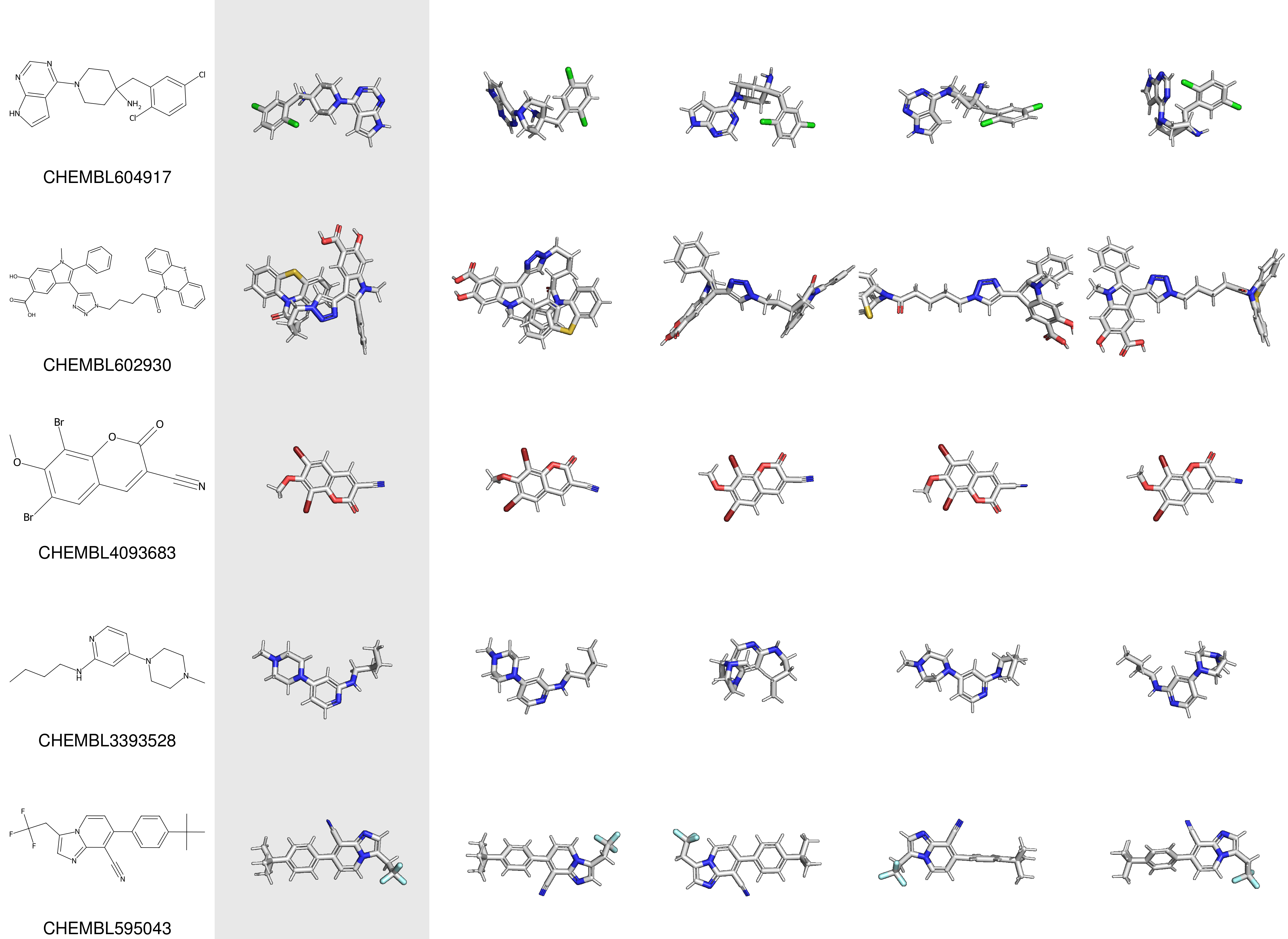}}
    \caption{Example conformer output, randomly selected, for PIDM[QMugs],
    using stochastic generation and 500 steps.
    }
\end{figure}

\begin{figure}[H]
    \centering
    \centerline{\includegraphics[width=1.2\textwidth]{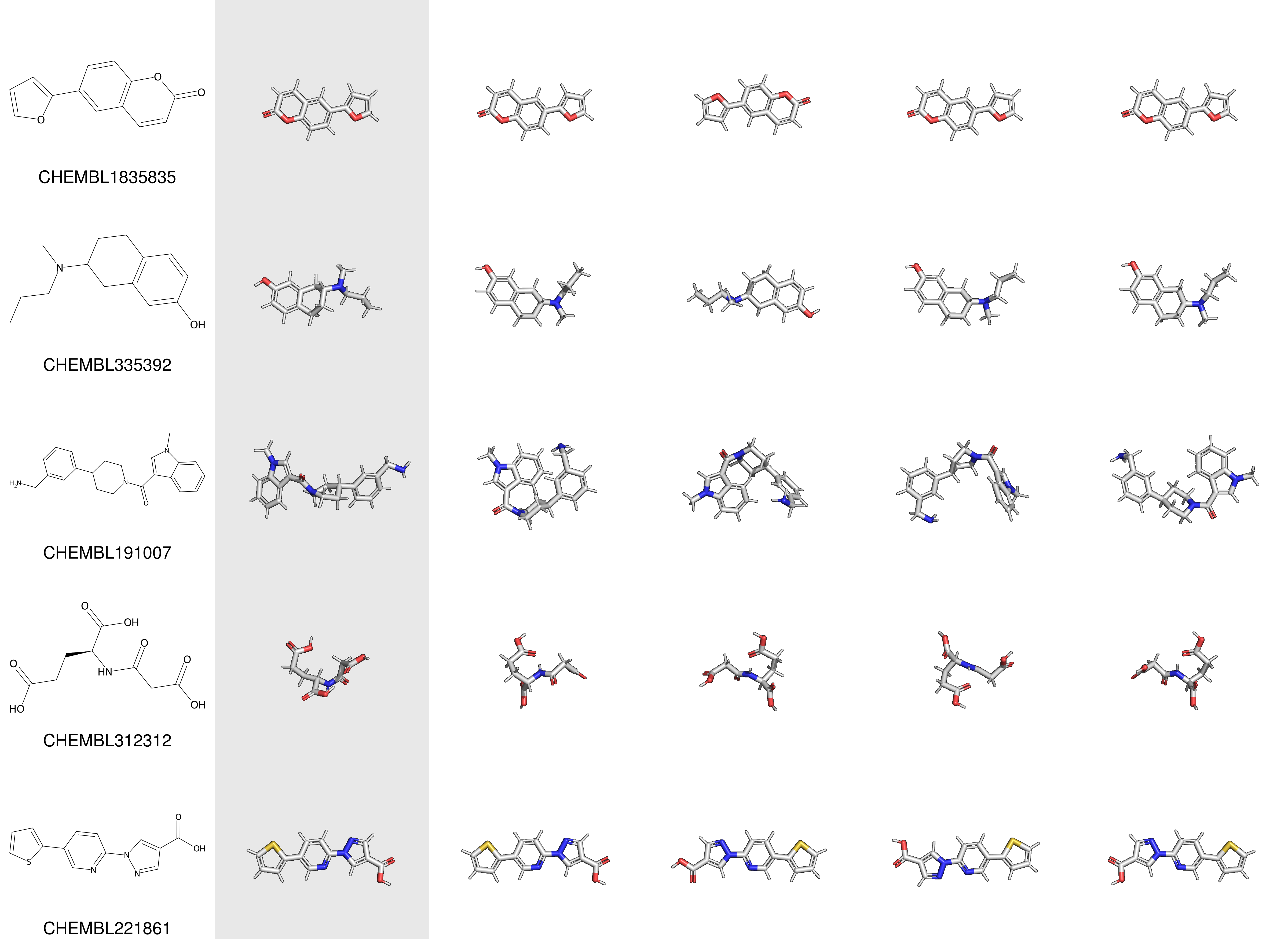}}
    \caption{Example conformer output, randomly selected, for PIDM[GEOM-drugs],
    using deterministic generation and 500 steps.
    }
\end{figure}

\begin{figure}[H]
    \centering
    \centerline{\includegraphics[width=1.2\textwidth]{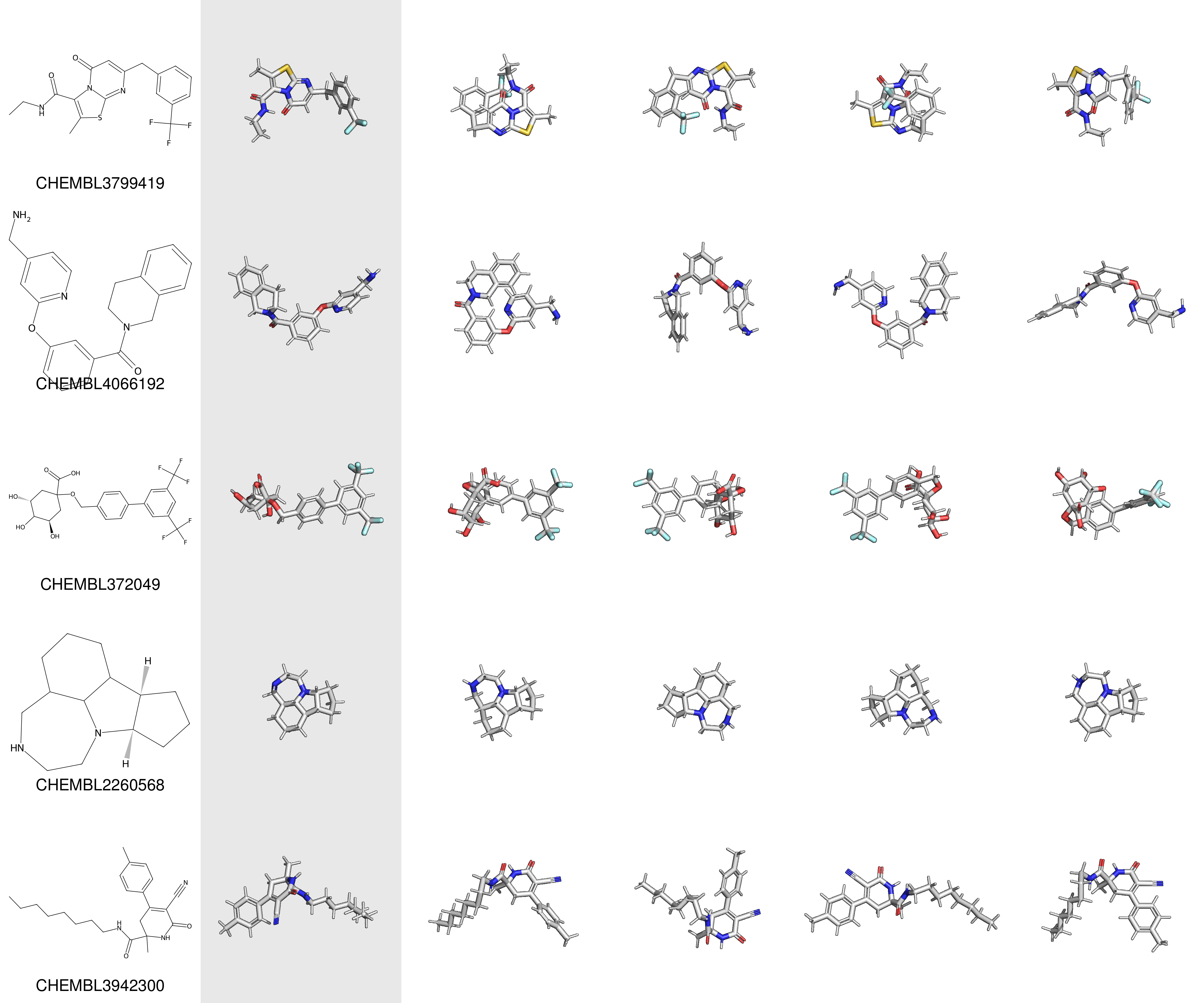}}
    \caption{Further example of conformer output, randomly selected,
    for PIDM[QMugs], using deterministic generation and 500 steps.
    }
\end{figure}

\begin{figure}[H]
    \centering
    \centerline{\includegraphics[width=1.2\textwidth]{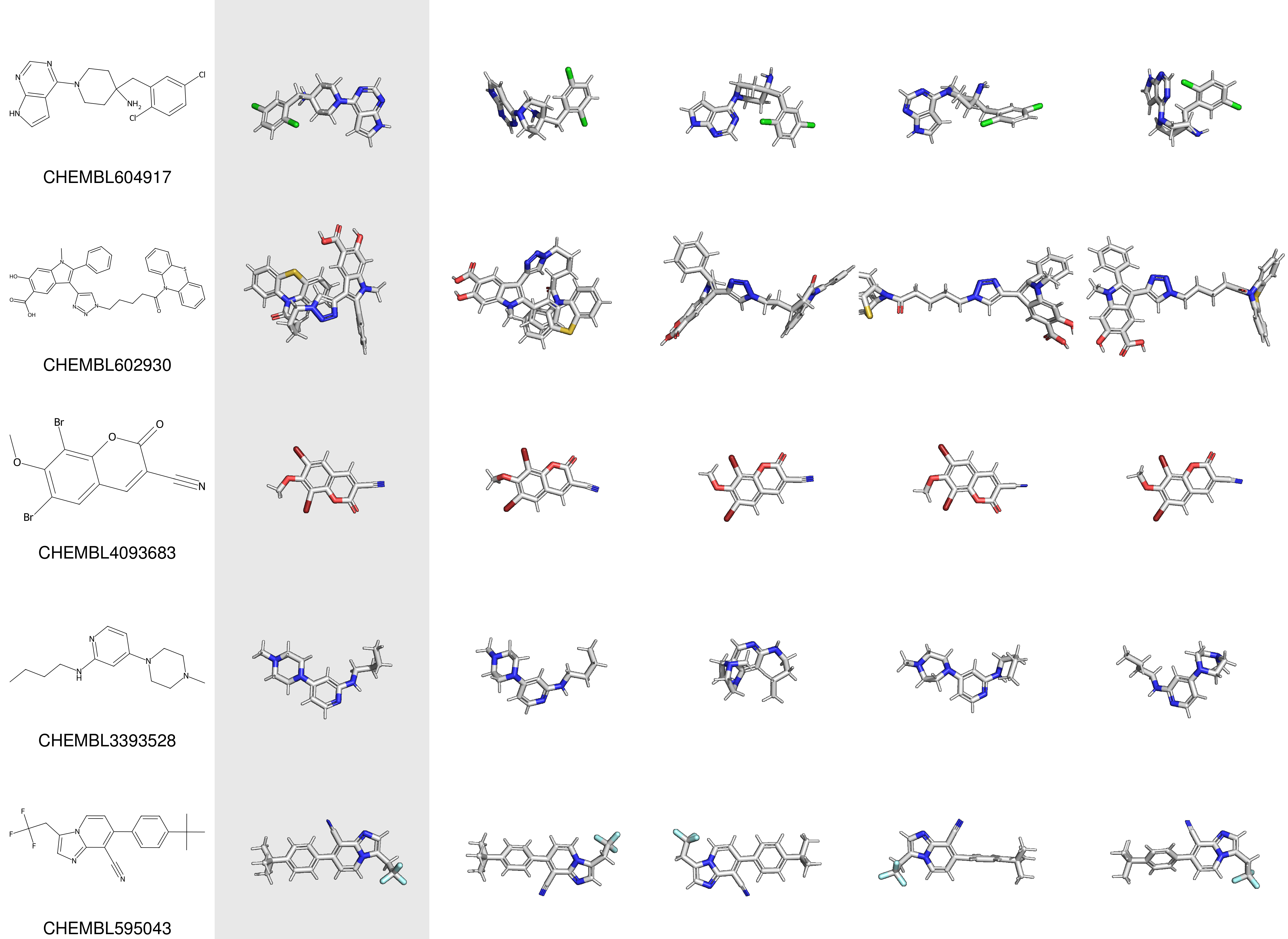}}
    \caption{Further example conformer output, randomly selected,
    for PIDM[QMugs], using deterministic generation and 500 steps.
    }
\end{figure}

\begin{figure}[H]
    \centering
    \centerline{\includegraphics[width=1.2\textwidth]{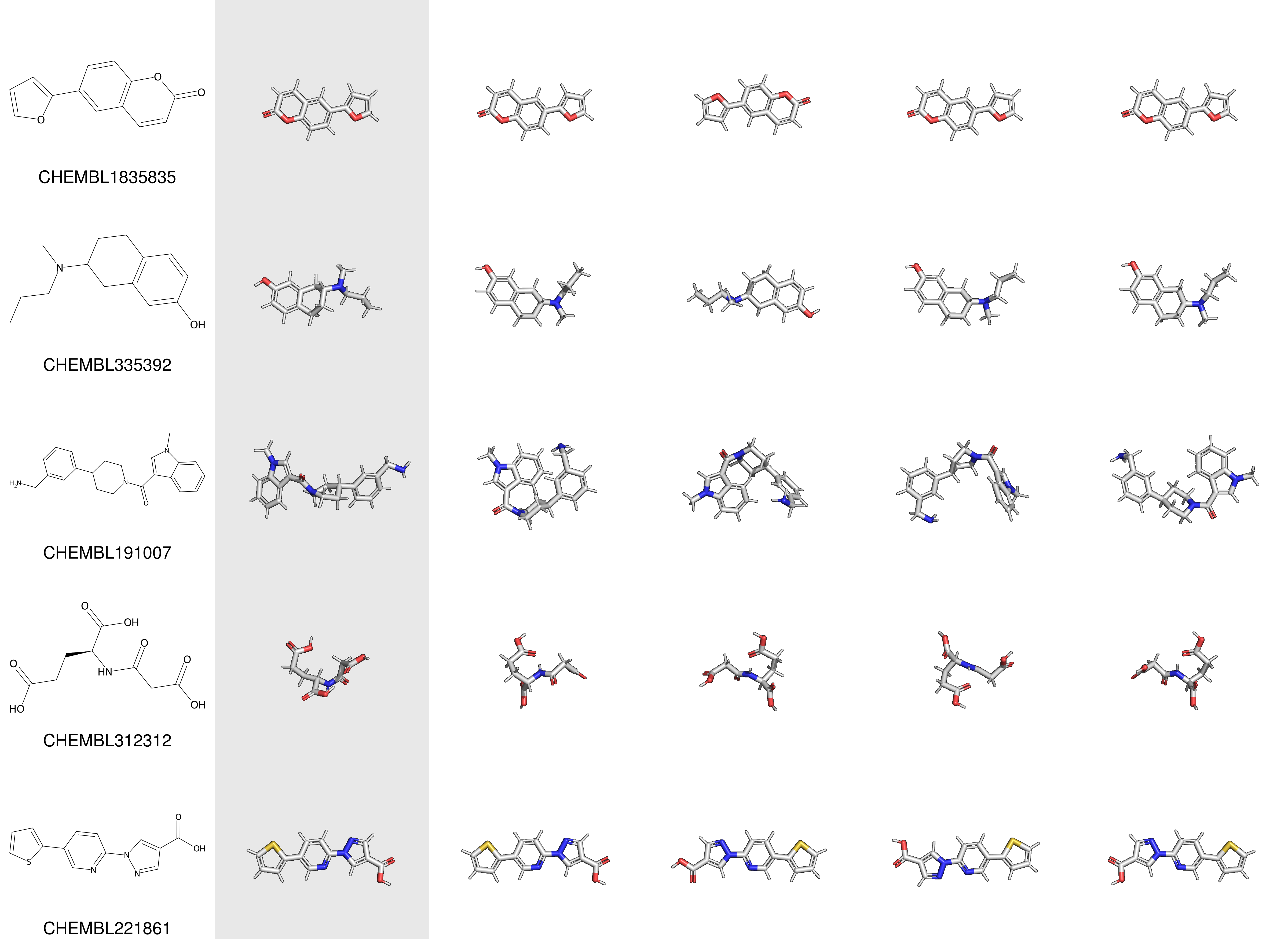}}
    \caption{Further example conformer output, randomly selected,
    for PIDM[QMugs], using deterministic generation, 500 steps, and repulsion of $\eta = 1$.
    }
\end{figure}

\begin{figure}[H]
    \centering
    \centerline{\includegraphics[width=1.2\textwidth]{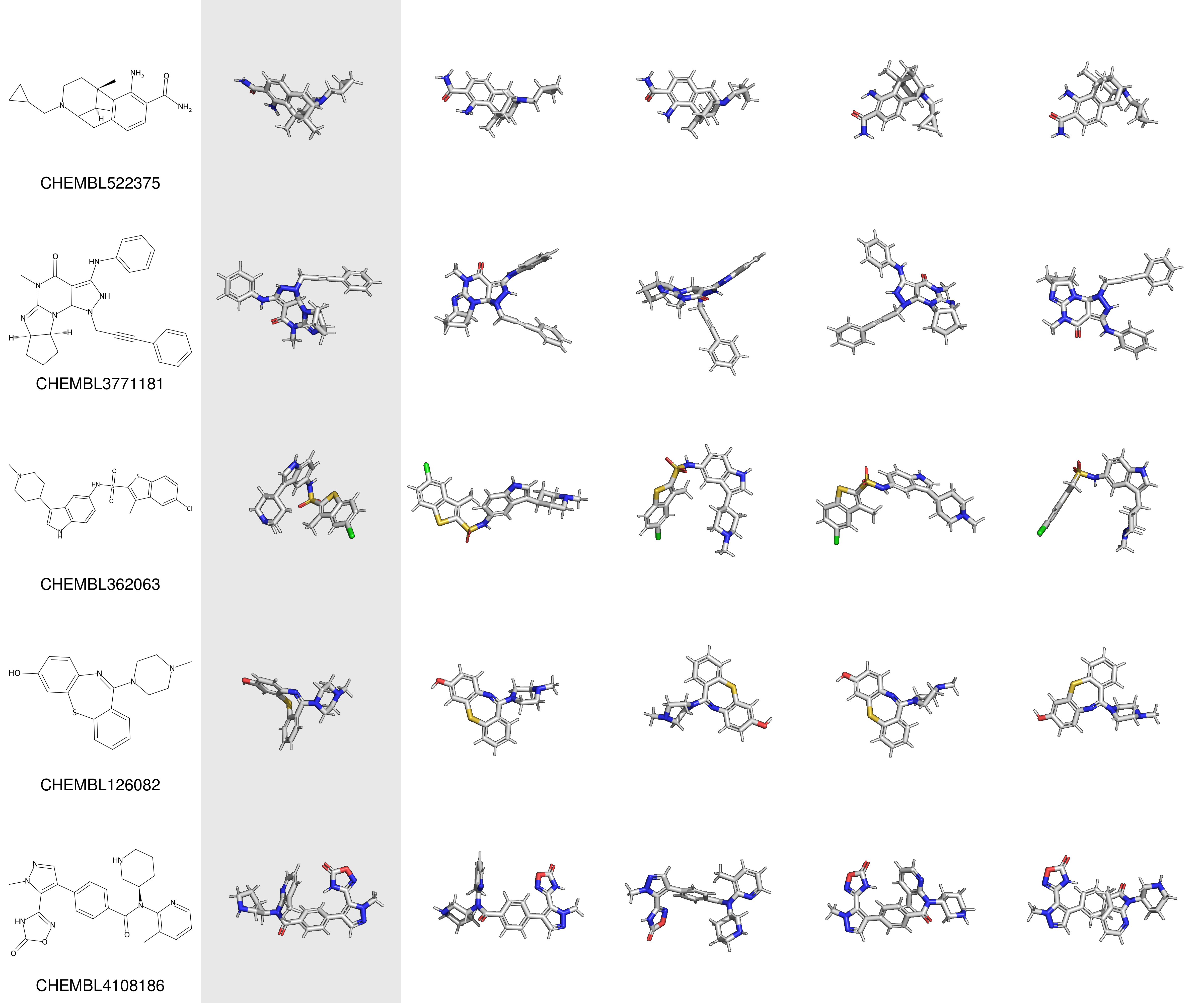}}
    \caption{Further example conformer output, randomly selected,
    for PIDM[QMugs],
    using deterministic generation, 500 steps, and repulsion of $\eta = 1$.
    }
\end{figure}

\end{document}